\documentclass[pdftex,twocolumn,epjc3]{svjour3}  
\usepackage[british,UKenglish,USenglish,english,american]{babel}
\RequirePackage[T1]{fontenc}
\smartqed 
\RequirePackage{graphicx}
\RequirePackage{mathptmx}      
\RequirePackage{flushend}
\RequirePackage[numbers,sort&compress]{natbib}
\RequirePackage[colorlinks,citecolor=blue,urlcolor=blue,linkcolor=blue]{hyperref}
\journalname{Eur. Phys. J. C}
\usepackage{xcolor}

\begin{document}

\title{Ergodic sampling of the topological charge using the density of states}

\author{ Guido Cossu \thanksref{braid,edin}
  \and
  David Lancaster\thanksref{e1,plym}
  \and
  Biagio Lucini\thanksref{e2,swan}
  \and
  Roberto Pellegrini\thanksref{edin,amozonwarning}
  \and
  Antonio Rago\thanksref{e3,plym}
}

\thankstext{e1}{e-mail: david.lancaster@plymouth.ac.uk}
\thankstext{e2}{e-mail: b.lucini@swansea.ac.uk}
\thankstext{e3}{e-mail: antonio.rago@plymouth.ac.uk}
\thankstext{amozonwarning}{Work done prior to joining Amazon.}

\institute{%
  Braid Technologies, Shibuya 2-24-12, Tokyo, Japan\label{braid}
\and
School of Physics and Astronomy, The University of Edinburgh,
Edinburgh EH9 3JZ, UK\label{edin}
\and
Centre for Mathematical Sciences, Plymouth University, 
Plymouth, PL4 8AA, United Kingdom\label{plym}
\and
Department of Mathematics, Computational Foundry, Swansea University
(Bay Campus), Swansea SA2 8EN, UK\label{swan}
}

\date{\today}

\maketitle

\begin{abstract}
In lattice calculations, the approach to the continuum limit is
hindered by the severe freezing of the topological charge, which prevents
ergodic sampling in configuration space. In order to significantly
reduce the autocorrelation time of the topological charge, 
we develop a density of states approach with a smooth constraint
and use it to study SU(3) pure Yang Mills gauge theory near the
continuum limit. 

Our algorithm relies on
simulated 
tempering across a range of couplings, which guarantees the
decorrelation of the topological charge and ergodic sampling of
topological sectors. Particular emphasis is placed on testing the
accuracy, efficiency and scaling properties of the method. In their
most conservative interpretation, our results provide firm evidence of
a sizeable reduction of the exponent $z$ related to the growth of the
autocorrelation time as a function of the inverse lattice
spacing. 

\end{abstract}

\section{Introduction}

In four spacetime Euclidean dimensions, gauge field configurations can be
classified according to their topological charge $Q$, defined as
\begin{equation}
\label{eq:topchargdef}
  Q =  \frac{1}{32 \pi^2} \int d^4 x \, \epsilon_{\mu \nu \rho
    \sigma} \mathrm{Tr} \left( G_{\mu \nu} G_{\rho \sigma} \right) \ , 
    \label{Qcontinuum}
\end{equation}
where $G_{\mu \nu}(x)$ is the field tensor. This integral can be re\-phra\-sed as a
mapping from a three-sphere to a $SU(2)$ group, exposing the fact that $Q$ is
related to the homotopy group $\pi_3: S^3 \mapsto SU(2)$.  With the
normalisation in equation~(\ref{eq:topchargdef}), $Q$ is integer-valued. 

$\chi$, the susceptibility of $Q$ per unit of volume,
is related to the mass of the $\eta^{\prime}$ particle by the
Witten-Veneziano formula~\cite{Witten:1979vv,Veneziano:1979ec}. Therefore, the
topological charge $Q$ and the topological susceptibility $\chi$ play
a phenomenologically relevant role in the theory of the strong
interactions and in any 
theoretically convenient generalisation of the latter such as the
large-$N$ limit of $SU(N)$ gauge theories~\cite{tHooft:1973alw}. It is hardly surprising
then that precise measurements of $Q$ and $\chi$ have been one
of the most active areas of activity of Lattice Gauge Theory since its
earliest days (see e.g.~\cite{Vicari:2008jw,Muller-Preussker:2015daa}
for recent reviews). In addition, more recent investigations have used
lattice determinations of topological quantities to estimate the mass of
axions from first
principles in order to determine a bound for their density in the Universe~\cite{Berkowitz:2015aua,Bonati:2015vqz,Borsanyi:2016ksw}.  

On a finite lattice with periodic boundary conditions in all
directions, gauge configurations split into topological sectors
labelled by the value of $Q$ and separated by free energy
barriers~\cite{vanBaal:1982ag}. Monte Carlo simulations have shown a
sharp growth of the height of these barriers as $a \to 0$, with a consequent
freezeout of the topological sector in numerical
calculations~\cite{Schaefer:2010hu}. The 
phenomenon becomes even more pronounced as the number of colours $N$
increases~\cite{DelDebbio:2002xa,Lucini:2004yh}. This dramatic slowing
down of $Q$ and the resulting very large autocorrelation times pose
serious problems of ergodicity in simulations 
and cause a very slow convergence of
physical observables such as hadron masses to their thermodynamic
limit values~\cite{Brower:2003yx,Aoki:2007ka}. Since the
toroidal topology of the domain plays a crucial role in the freezeout,
other setups have been proposed based for example on open boundary conditions
in time~\cite{Luscher:2011kk} (used at large $N$
in~\cite{Amato:2015ipe,Ce:2016awn}) or 
on non-orientable manifolds~\cite{Mages:2015scv}. While these setups
are very promising, their properties and potential drawbacks such as
the loss of translational invariance with open boundary conditions in
time, need further study.

In this work, we take an alternative approach to the problem of
topological freezing without abandoning the use of periodic
boundary conditions in all directions. For the theory formulated on
a torus, we build a sequence of couplings that interpolate
from a relatively strong value to a weak coupling value and we perform
numerical simulations using an algorithm in which the coupling is also
a dynamical variable taking values in the prebuilt set. More specifically, we use the 
density of states approach~\cite{Langfeld:2012ah,Langfeld:2015fua} formulated with a Gaussian
constraint~\cite{Pellegrini:2017iuy} and
tempering~\cite{Marinari:1992,Lucini:2016fid}. The idea behind
this approach is that the strong coupling values will act as an ergodic
reservoir of topological sectors, while the tempering
guarantees thermal equilibrium of the latter in the Monte Carlo at
weaker coupling. Preliminary results of our investigation have
appeared in~\cite{Cossu:2017sfu}.  

The rest of the work is organised as follows. In
Sect.~\ref{methodology}, we recall some key properties of the 
density of states approach.
Sect.~\ref{formalism} develops the formalism for a Gaussian constraint 
and discusses a simplifying approximation.
The lattice gauge theory model, the simulations and results are described in
Sect.~\ref{LGT}. This is a long section as it covers our discussion of the 
topological charge and demonstrates that the density of states approach leads 
it to couple more weakly
to the long timescale modes of the system. Techniques to analyse the
autocorrelation time in this situation are developed.
We work on three lattice sizes and  present our results concerning 
scaling   in Sect.~\ref{scaling}, before summarising
our findings in Sect.~\ref{conclusions}. Two appendices provide more
technical details:~\ref{appendixa} 
compares a multi-histogram calculation using our reference simulations 
with the density of states approach;
~\ref{appendixb} contains 
tables recording details of the simulations performed in this study.

\section{Density of States Method}\label{methodology}

The density of states technique has been shown to be useful in 
a variety of models with continuous variables.
For example: U(1)\cite{Langfeld:2015fua}, 
complex action\cite{Langfeld:2014nta},
Bose gas\cite{Pellegrini:2015dkk}.
Other versions of the technique have also been used to study SU(3) gauge theory
\cite{Gattringer:2017aa,Gattringer:2019aa}.
In this paper we further develop the technique and use it to 
make a detailed study of SU(3) lattice gauge theory that
extends the results of \cite{Cossu:2017sfu}.

The density of states approach developed in \cite{Langfeld:2012ah} 
reconstructs canonical expectation values from a 
sufficiently large and closely spaced 
set of
functional integrals each con\-strained to have central action
$S_i$. 
These constrained functional integrals provide an
approximate way to define observables on the micro-canonical ensemble at
 $S_i$ and while 
the original development of the subject employed
a sharp constraint, in this
paper we use a Gaussian constraint of width $\delta$.
The choice of an analytic Gaussian constraint  allows simulations to
use the global HMC algorithm which commonly underlies simulations of lattice gauge theory.
The constrained functional integrals are defined using double bracket notation as,

\begin{equation}
\langle \langle B[\phi]\rangle \rangle_i 
= {1\over {\cal N}_i}
\int {\cal D}\phi \, B[\phi] e^{-a_i S[\phi]} 
e^{-(S[\phi] - S_i)^2/2\delta^2} \ .
\label{eqn-constrained}
\end{equation}

Where $B$ and $\phi$ denote generic operators and fields
respectively.
The need for the term $e^{-a_i S[\phi]} $ is explained below.

${\cal N}_i$ is a normalisation factor written as a constrained functional
integral without the operator $B$,

\begin{equation}
{\cal N}_i
=
\int {\cal D}\phi \, e^{-a_i S[\phi]} 
e^{-(S[\phi] - S_i)^2/2\delta^2} \ .
\label{eqn-Nidefn}
\end{equation}

Noting the definition of the 
density of states,

\begin{equation}
\rho(S) =
\int {\cal D}\phi \, \delta(S[\phi]-S)  \ .
\label{rhodef}
\end{equation}

We see that since the expression for ${\cal N}_i$ involves only the action, 
 the functional integral defining it can be replaced
by an ordinary integral over the density of states,

\begin{equation}
 {\cal N}_i 
 = \int dS\, \rho(S) 
e^{-a_i S } 
e^{-(S - S_i)^2/2\delta^2} \ .
\label{Nidef1}
\end{equation}

So the ${\cal N}_i$ are closely related to the density of states
evaluated at $S_i$. This is most apparent
in the limit $\delta \to 0$, when the Gaussian constraints become
delta functions and the double bracket expressions truly represent
the evaluation of an observable on the micro-canonical ensemble. 
In this formal limit there would be no need for the $a_i$'s.
However, at finite $\delta$, the constraint is wide enough to require
the term $e^{-a_i S[\phi]} $ to compensate the growth of the density of states in
the measure and to ensure that the mean constrained action $\langle\langle S\rangle \rangle$
really is at $S_i$. The value of $a_i$ needed to accomplish this 
is determined by an iterative procedure described below.
The simulations which underly the method provide estimates of 
these double bracketed expectations (\ref{eqn-constrained}) for a set of 
closely spaced $S_i$ labelled by the index 
$i = 1\dots N_{TEMPER}$, which we refer to as ``tempers''
(we reserve the term ``replica'' to refer to repeats of the whole simulation
procedure with identical parameters, but different random number se\-quences). 
Having chosen an appropriate set of central action $S_i$ to 
characterise the tempers, the method consists of two phases.
\paragraph{Robbins Monro (RM) Phase} 
to determine $a_i$ through a procedure 
of minimising $\langle\langle S - S_i\rangle \rangle_i$
for an iterative sequence of $a_i$'s. 
This quantity
converges to zero when
$a_i$ becomes the derivative of the (log) density of states at $S_i$.
Knowledge of the final $a_i$'s is sufficient to construct a piecewise
approximation to the density of states that can immediately be used to 
estimate canonical expectation values of operators that depend only 
on the action, such as moments of the plaquette.
\paragraph{Measurement Phase (MP)} 
at fixed $a_i$ as determined by the RM phase.  This is necessary
to compute canonical expectation values of generic operators 
through a reweighting procedure. For example, the topological
charge requires more information than is contained in the density of states alone.
Moreover, the measurement phase provides dynamic information
about the efficiency of the method through the autocorrelation.

We are particularly concerned about the accuracy,
the relationship between expectation values computed using RM phase and
those based on the measurement phase (MP), the autocorrelation
time and the efficiency of the tempering method.
To achieve these aims we have repeated the whole simulation
procedure consisting of both phases,
$N_{REPLICA}$ times, as this provides a way of estimating
the errors inherent in the method.

While the density of states method is applicable to any model, 
and indeed has been used to study phase transitions \cite{Langfeld:2015fua}, 
the Yang Mills theory we study here is far from any phase transition and all
results are smooth functions of parameters. The conclusions
that rely on working in this context
are not restricted
to lattice gauge theory but are expected to hold for 
other models in which there is a single coupling multiplying
the action as this most closely mimics simple statistical mechanics systems.

\section{Formalism}\label{formalism}

According to the density of states approach, $\log \rho(S)$  is 
approximated as piecewise linear with coefficients given by the 
$a_i$ appearing in (\ref{eqn-constrained}),

\begin{equation}
\rho(S) =
\rho_{i} e^{a_{i} (S-S_{i})}  
,\  {(S_{i-1} + S_{i})\over 2} < S <  {(S_{i} + S_{i+1})\over 2}  \ .
\label{rho1}
\end{equation}

Where $\rho_i$ is shorthand for $\rho(S_{i}) $.
By convention we take $S_i$ to grow with $i$.
Piecewise continuity leads to,

\begin{equation}
\rho_{i+1}
= \rho_i e^{(a_i + a_{i+1}) (S_{i+1}-S_i)/2}  \ .
\label{rho2}
\end{equation}

This relation along with the $(S_i, a_i)$ is sufficient to compute
the piecewise approximation to the density of states
in terms of the value $\rho_{i_c}$ at some 
base index $i_c$. 
As $\rho$ varies over many orders of magnitude,
the base index is chosen depending on the particular problem.
When reweighting to coupling $\beta$ it is chosen so as 
to minimise $|a_{i_c} - \beta | $.

To avoid unduly long formulae, the expressions given here assume
 fixed spacing $S_{i+1} - S_{i} = \delta_S$ and also
 fixed Gaussian width $\delta_i = \delta$. However, motivated by the
desire to ensure 
effective mixing
across the range of tempers,
variable spacing is used in simulations.

Besides $\delta$ and $\delta_S$ another parameter 
provides a useful scale to the problem. This parameter is
model dependent and we define 
$\sigma$ to be the standard deviation of action fluctuations
in the unconstrained system. The strength of the Gaussian
constraint is determined by the dimensionless ratio
$\delta/\sigma$. Although $\sigma$ depends on the coupling, 
for this study it does not change very much over the 
narrow range of tempers  and thus provides a useful 
scale.
For the pure gauge SU(3) studied here, 
 $a_i$ turns out to be approximately linear: $a_i \approx a_{i_c} - (i-i_c)\delta_S/\sigma^2$ for indices close to  the base index $i_c$ where $\sigma$ can be
 regarded as constant.
Consequently for small $k$, $\rho_{i_c+k} \approx \rho_{i_c}e^{k a_{i_c} \delta_S}
e^{-k^2   \delta_S^2/2 \sigma^2}$.
However, we emphasise that this observation
relies on the smoothness of the model and limited range of tempers and
should not be regarded as any kind of modified version of the
density of states method.

\subsection{Canonical expectation values of the action and its variance: LLR}\label{llr}

The parameters $a_i$ obtained from the RM phase for each $S_i$ provide the
piecewise estimate of the log of the density of states, so without any measurement
phase
we have access to the full probability density of the action 
for a range of couplings $\beta$.
Estimates can be made of  
canonical expectation values of functions of the 
action, 
the most common being 
 $\langle S\rangle $ and $\langle(S-\langle  S\rangle)^2\rangle$ 
 which in the lattice gauge theory
 studied here are directly related to moments of the plaquette.
 We shall call this approach
 ``LLR'' as a shorthand to distinguish it from the
``reweighting'' approach discussed in the next subsection.

Direct application of the piecewise approximation (\ref{rho1}),(\ref{rho2})
leads to the following expression for the partition function.

\begin{eqnarray}
{\cal Z}_{LLR}
&=& \int dS \, \rho(S) e^{-\beta S}
\label{zeefulldos1}
\\
&=& \sum_i {\rho_i e^{-\beta S_i} \over a_i-\beta}
\left[e^{(a_i-\beta)\delta_S/2}-e^{-(a_i-\beta)\delta_S/2} \right]
\nonumber\\
&\approx&
{\cal Z}_{LLR'}
=
 {\delta_S}
\sum_i
\rho_i e^{-\beta S_i}  \ .
\label{zeefulldos}
\end{eqnarray}

And similar formulae for the expectation values
$\langle S\rangle $ and $\langle(S-\langle  S\rangle)^2\rangle$.
These results, because of their reliance on the piecewise approximation,
should be taken to be accurate to  order  $\delta_S^2/\sigma^2$.

The further approximation on the third line 
can be seen as an expansion
for $(a_i-\beta)\delta_S \ll1$ or can be regarded as a trapezium
evaluation of the integral.
In numerical work we always use the full expression
but the approximate one, which we denote as ``$\rm LLR'$'' approach 
is simpler yet remarkably accurate.
Because of its advantages in exposition we develop
some formalism for this $\rm LLR'$ approximation and
define weights $v_i$:

\begin{equation}
v_i =
\rho_i e^{- \beta S_i}
\label{weights_vi}
\end{equation}

then 

\begin{equation}
\langle  f(S) \rangle_{LLR'}
=
 {
\sum_i
v_i  f(S_i)
\over
\sum_i
v_i 
}  \ .
\label{vevdos}
\end{equation}

These weights can be normalised to $\sum_i v_i = 1$, but for numerical
reasons in the course of intermediate computations, 
 it is convenient to 
choose the weight at the central index $v_{i_c}$ to be unity by 
defining  $\rho_{i_c} = \rho(S_{i_c})=e^{\beta {S_{i_c}}} $.

The full LLR formulae are straightforward to derive from 
(\ref{zeefulldos1}), but ugly in comparison with (\ref{weights_vi})
and (\ref{vevdos}).

\subsection{Canonical expectation values of arbitrary operators:
  reweighting}\label{rewt}

Observables that are not simply functions of the action can 
not be computed using the density of states alone and a measurement
phase is needed 
The measurements are performed on a set of constrained
tempering runs with parameters  ($S_i,a_i$) and reweighting is then used to obtain 
the canonical expectation value:

\begin{equation}
\langle  B[\phi] \rangle 
= {1\over {\cal Z}}
\int {\cal D}\phi \, B[\phi] e^{-\beta S[\phi]}   \ .
\label{Canonical}
\end{equation}

To relate this to the constrained functional integrals we consider sums of spaced 
Gaussians.
We intend to use the functional integral equivalent of the
approximate identity presented below,

\begin{eqnarray}
K(x)& =& {1\over \sqrt{2\pi}}{\delta_x\over \delta}
\sum_i 
e^{-(x-i\delta_x)^2/2\delta^2}\nonumber \\
&=&1 + 2 \sum_{i>0} e^{-2 \left(\frac{i \pi
    \delta}{\delta_x}\right)^2} \cos(\frac{2 i \pi x }{\delta_x}) \nonumber \\
  &=&1  + 2 e^{-2 \left(\frac{ \pi
    \delta}{\delta_x}\right)^2} \cos(\frac{2 \pi x }{\delta_x}) +\dots \nonumber \\
    & \approx & 1
                \label{1DK}
\end{eqnarray}
where the omitted terms are exponentially suppressed in $1/\delta_x$.
This identity should be more properly understood in its integral form,

\begin{equation}
\int_\Gamma dx\, f(x)K(x) \approx \int_\Gamma  dx\, f(x) 
\label{1Dint}
\end{equation}

where $f(x)$
are smooth test functions of width exceeding $\delta$ or $\delta_x$
defined over the domain $\Gamma$.

This approximation is extremely accurate and
can be checked  in this one-dimensional context
either analytically for Gaussian test functions,
or numerically
to understand the errors introduced by a 
finite set of Gaussians, and by the choice of the ratio
 $\delta/\delta_x$.
In the limit $\delta \to 0$ the sum of Gaussians becomes a
comb of delta functions, 
the right hand
side of (\ref{1Dint})
becomes a trapezium approximation
 and the accuracy of the approximation in
this limit is order $\delta_x^2$.

Returning to (\ref{Canonical})
we introduce the functional integral generalisation of
the sum (\ref{1DK})
into the expression to obtain:

\begin{eqnarray}
\langle  B[\phi] \rangle 
&\approx& {1\over {\cal Z}}
{\delta_S\over \sqrt{2 \pi} \delta}
\sum_i
\int {\cal D}\phi \, B[\phi] e^{-\beta S[\phi]} 
e^{-(S[\phi] - S_i)^2/2\delta^2}  \ .\nonumber\\
\label{sumgaussian}
\end{eqnarray}

In the light of the discussion above, the accuracy of the approximation is
rather better than 
anticipated from the delta function limit, but nonetheless
still contains an order  $\delta_S^2/\sigma^2$ term when there are
only a finite set of Gaussians.

Notice that from the definition of the double bracket quantities
it is possible to reweight in the usual sense, that is, to
include part of the exponential term as a measured operator 
rather than use for importance sampling in the Monte Carlo simulation. 

\begin{eqnarray}
\langle \langle B[\phi] e^{(a_i-\beta)S}\rangle \rangle_i  
&=&
{1\over {\cal N}_i}
\int {\cal D}\phi \, B[\phi] e^{-\beta S[\phi]}  
e^{-(S[\phi] - S_i)^2/2\delta^2}   \ , \nonumber\\
\end{eqnarray}

this resembles the terms in the sum (\ref{sumgaussian}), 
so the canonical expectation value can be written
in terms of the double bracket quantities as:

\begin{equation}
\langle  B[\phi] \rangle 
\approx
 {1\over {\cal Z}}
{\delta_S\over \sqrt{2 \pi} \delta}
\sum_i
{\cal N}_i 
\langle \langle B[\phi]e^{(a_i-\beta)S}\rangle \rangle_i   \ .
\label{Bphi}
\end{equation}

With,

\begin{equation}
{\cal Z}
=
 {\delta_S\over \sqrt{2 \pi} \delta}
\sum_i
{\cal N}_i 
\langle \langle e^{(a_i-\beta)S}\rangle \rangle_i    \ .
\label{Zphi}
\end{equation}

The normalisation factors, ${\cal N}_i$, 
only involve the action  (\ref{Nidef1}), so they
can be evaluated using the piecewise approximation.
This introduces an additional approximation of order  $\delta_S^2/\sigma^2$.

\begin{eqnarray}
 {\cal N}_i 
&\approx& 
\sum_j \rho_j
\int_{(S_{j-1} + S_j)/2}^{(S_{j+1} + S_j)/2}dS \,
e^{a_j(S - S_j)}\times\nonumber\\
&&\hspace{1cm}e^{-a_i S }
   e^{-(S - S_i)^2/2\delta^2}  \ .
\end{eqnarray}

So the ${\cal N}_i$ can be analytically evaluated in terms of truncated
Gaussian integrals or error functions. 
Using this expression in (\ref{Bphi}) and (\ref{Zphi}), we obtain 
the reweighting formula for fixed separation between Gaussians:

\begin{equation}
\langle  B[\phi] \rangle 
\approx
 {
\sum_i
 {\cal N}_i  
\langle \langle B[\phi]e^{(a_i-\beta)S}\rangle \rangle_i  
\over
\sum_i
 {\cal N}_i  
\langle \langle e^{(a_i-\beta)S}\rangle \rangle_i 
}  \ .
\label{expectation1}
\end{equation}

We define the following weights

\begin{equation}
 w_i 
 =
{\delta_S \over  \delta}
{ {\cal N}_i \over \sqrt{2\pi} \delta}  
 e^{(a_i - \beta)S_i}  
 \label{wt-wi}
 \end{equation}

where the prefactors are motivated by the simple relation
between $w_i$ and the $\rm LLR'$ weights, $v_i$ (\ref{weights_vi}),
in various limits explored in section (\ref{limits}) below.

The reweighting formula takes its final form, which
overall
is accurate to order  $\delta_S^2/\sigma^2$.

\begin{equation}
\langle  B[\phi] \rangle 
\approx
 {
\sum_i
 w_i  
 \langle \langle B[\phi]e^{(a_i-\beta)(S-S_i)}\rangle \rangle_i  
\over
\sum_i
 w_i  
 \langle \langle e^{(a_i-\beta)(S-S_i)}\rangle \rangle_i 
}  \ .
\label{expectation_dynamic}
\end{equation}

This expectation value 
is defined in terms of the ratio of sums and
it is important to avoid bias in estimating the ratio.
Bootstrap techniques are used as explained later.

\subsubsection{Quenched Approximation}\label{quenched}

If the fluctuations of the action are ignored and we
approximate (\ref{expectation_dynamic})
by neglecting the exponential factors in the
double brackets, we obtain a simpler formula

\begin{equation}
\langle  B[\phi] \rangle 
\approx
 {
\sum_i
 w_i  
 \langle \langle B[\phi]\rangle \rangle_i  
\over
\sum_i
 w_i  
}  \ .
\label{expectation_static}
\end{equation}

At the risk of confusion in the lattice gauge theory community, we
term this approximation ``quenched'', following the more general usage 
in the theory of disordered systems.

In an attempt to justify
this approximation 
we first consider the
term   $(a_i-\beta)$.
Provided that the range
of reweighting is sufficiently narrow to allow the approximate linear
relation
$a_i = a_{i_c} - (i-i_c)\delta_S/\sigma^2$,  and noting that
$(a_{i_c}-\beta) < \delta_S/\sigma^2$, we have:

\begin{equation}
a_i-\beta \approx -(i-i_c)\delta_S/\sigma^2 \ .
\end{equation}

The double bracket expectation of the second term in the exponent
vanishes,
$\langle \langle  S - S_i \rangle \rangle_i  = 0$, according to the
tuning of $a_i$ during the RM phase.
Of course, because this second term appears in the exponent,
fluctuations of the action cause higher moments to contribute.
However, the role of the Gaussian constraint is
precisely to limit the size of these fluctuations
and we expect even moments to scale as:

\begin{equation}
\langle \langle  (S - S_i )^{2n}\rangle \rangle_i  \sim \delta^{2n}  \ .
\end{equation}

Therefore, provided the exponential can be expanded, we
 may anticipate the overall result: 

\begin{equation}
 \langle \langle e^{(a_i-\beta)(S-S_i)}\rangle \rangle_i  \sim 1 +  {\cal O}(\delta \delta_S/\sigma^2)   \ .
\end{equation}

If, by ignoring possible correlations between $S$ and $B[\phi]$,
we make a similar assumption for the numerator 
of (\ref{expectation_dynamic}), we
may expect that the quenched approximation differs
from the full formula by terms of order 
$\delta \delta_S/\sigma^2$. Since, in practical work, we
set $\delta/\delta_S$ to be a value of order 1, 
the quenched approximation is plausibly accurate to 
the same order $ \delta_S^2/\sigma^2$ as was the full
reweighting formula (\ref{expectation_dynamic}).
However, it will become clear in the numerical work that 
while the quenched approximation is often good, it 
can lead to very different predictions in some circumstances.

The quenched approximation has a much simpler structure
than the full expression and it
can be interpreted as a reweighting at each time step or
configuration of the Monte Carlo simulation in the measurement phase.
This allows for a straightforward understanding  
of the quenched autocorrelation as that of the 
evolution of the reweighted quantity. 
Indeed, this is the basis for the autocorrelation time reported in \cite{Cossu:2017sfu}.

In the full expression, (\ref{expectation_dynamic}), 
the expectation value 
is defined
in terms of the ratio of sums.
The signal to evaluate the autocorrelation time will be dominated by the larger of the autocorrelators of 
the numerator and denominator, in practice always the numerator.
It is then important to avoid bias in estimating the ratio
and bootstrap techniques are used as explained later.

\subsubsection{Limits}\label{limits}

Further intuition about the reweighting formula comes from considering 
various limits.
Firstly, consider the limit $\delta/\sigma \to 0$
holding other quantities, including $\delta_S$, fixed.
The weights simplify, $w_i = v_i$, and moreover in this limit  
fluctuations of the action are suppressed,
so the reweighting formula becomes 
the quenched version (\ref{expectation_static})
resembling the
density of states result in the $\rm LLR'$ approximation 
(\ref{vevdos}) extended to operators
that do not solely depend on the action. 

For fixed spacing $\delta_S$, without taking any limit, 
the weights can be written as:
\begin{eqnarray}
w_i & = &
{1\over \sqrt{2\pi}}
{\delta_S\over \delta}
v_i 
\sum_j 
{\rho_j\over \rho_i} e^{a_j(S_i-S_j)}
\nonumber
\\
&{1\over \sqrt{2\pi}\delta}&
\int_{S_j-S_i-\delta_S/2}^{S_j-S_i+\delta_S/2}
dS e^{(a_j-a_i)S}e^{-S^2/2\delta^2}  \ .
\label{wforlineara}
\end{eqnarray}

When the range of reweighting is sufficiently narrow to allow the
the approximate linear behaviour
$a_i = a_{i_c} - (i-i_c)\delta_S/\sigma^2$, 
the term 
${(\rho_j/ \rho_i)} e^{a_j(S_i-S_j)} = 1$, and the integral becomes
a function of $j-i$, so the sum is independent of $i$ and the
weights $w_i$ are proportional to the $LLR'$ weights $v_i$.

Numerically, because the ranges over which we reweight are small,
the $a_i$ are linear and
the weights
obey $w_i \propto v_i$ rather accurately. When the sum of weights 
is normalised to unity, $w_i$ and $v_i$ differ by at most $4\times 10^{-3}$.

The limit  $\delta \to \infty$, corresponding to no constraint,
resembles the situation usually treated with the multi-histogram 
approach (see~\ref{appendixa} for a general comparison). 
Relying on the approximate linearity of the $a_i$, the
expression (\ref{wforlineara}) can be  simplified by expanding in $\sigma^2/\delta^2$.
However, fluctuations are not suppressed in this limit and the
reweighting formula (\ref{expectation_dynamic}) does not
reduce to the multi-histogram formula. 

A more interesting limit is to take $\delta/\sigma \to 0$, $\delta_S/\sigma \to 0$ holding the
ratio  $\delta_S/\delta$ fixed. 
Again we assume a narrow range and rely on linearity of the $a_i$ to obtain,

\begin{equation}
w_i = 
{\delta_S\over \delta}
v_i 
{\sigma \over 
\sqrt{\delta^2 + \sigma^2}} 
\to 
{\delta_S\over \delta}
v_i   \ .
\end{equation}

Since fluctuations are suppressed in this limit
the reweighting formula again becomes 
the quenched version (\ref{expectation_static}).

\section{Lattice Gauge Theory}
\label{LGT}

Pure Yang Mills lattice gauge theory has been used as a testing ground
for the density of states method since 
\cite{Langfeld:2012ah} studied SU(2) with a sharp ``top hat'' constraint 
and either Metropolis or Heat Bath algorithm. 
The smooth Gaussian constraint allows the HMC method to be used
as described for SU(2) in \cite{Pellegrini:2017iuy}.
The tempering method was introduced for SU(3) in \cite{Cossu:2017sfu}.

We work with SU(3) in four dimensions and employ a standard Wilson action,
on $16^4, 20^4$ and $24^4$ lattices. The HMC algorithm was used 
with the same trajectory length and integrator as \cite{Cossu:2017sfu}
and
these parameters allow 
us to validate our simulations in
comparison with results in \cite{Schaefer:2010hu}.

\begin{figure}
\begin{center}
\includegraphics[scale=0.65]{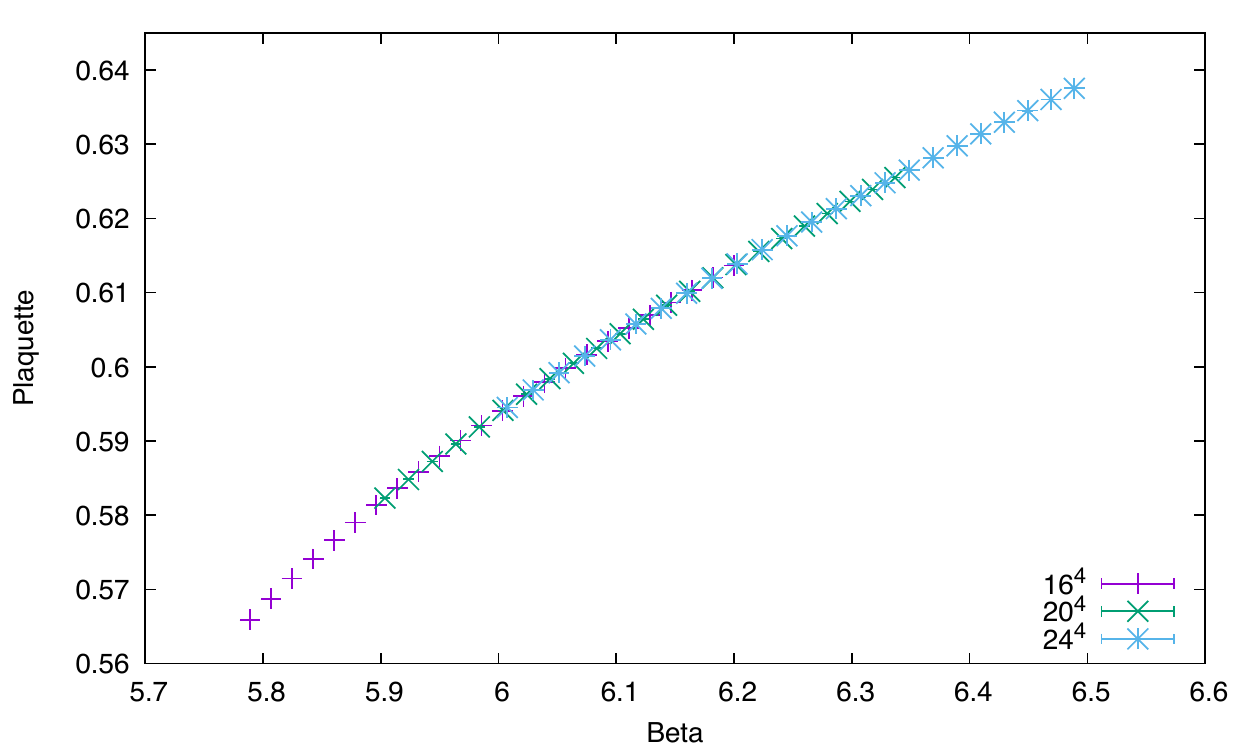}
\includegraphics[scale=0.65]{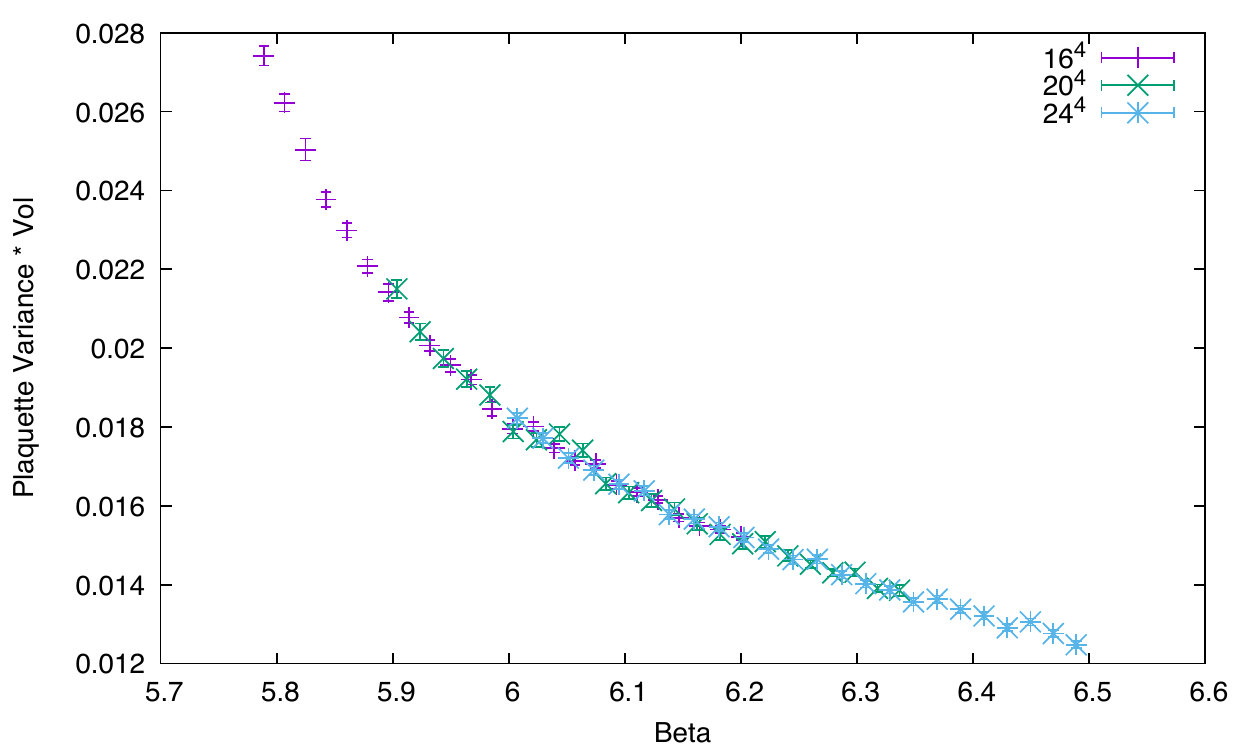}
\caption{
Reference simulations using an unconstrained approach
illustrating the range of parameters under consideration
over the lattice sizes 
$16^4$, $20^4$, $24^4$.
Top: Plaquette expectation.
Bottom: Plaquette variance, including a volume factor
for scaling.
}
\label{fig-overall}
\end{center}
\end{figure}

\begin{table}[!htb]
\begin{center}
\caption{
Parameters of canonical reference simulations. 
For each lattice size, the range of $\beta$ 
was spanned by 24 simulations.
}
\begin{tabular}{|l|c|c|}
\hline
Size & $\beta$ range  & Number of sweeps\\
\hline
$16^4$ &  $5.79 < \beta < 6.20$ &  $10^5$\\
$20^4$ &  $5.90 < \beta < 6.35$ &  $5\times 10^4$\\
$24^4$ &  $6.00 < \beta < 6.49$ &  $10^5$\\
\hline
\end{tabular}
\label{ReferenceSim}
\end{center}
\end{table}

Preliminary simulations 
generated for the purposes of comparison
used the canonical unconstrained approach rather than the density of states method.
These simulations consider several
size lattices and a range of couplings among which many trajectories at constant
boxsize can be identified to define a continuum limit.
Figure \ref{fig-overall} illustrates the range of parameters considered
and these are given in table \ref{ReferenceSim}.
At larger values of $\beta$ the dynamics of the topological charge
becomes extremely slow.
All simulations, both these unconstrained comparison runs and later
constrained runs, use the
HiRep code\footnote{Available at
  https://github.com/claudiopica/HiRep.}
\cite{DelDebbio:2008zf,DelDebbio:2009fd} and are 
initialised by thermalising a random configuration.

As can be seen in figure \ref{fig-overall}, the $\beta$ values for 
each lattice size all overlap in the approximate range $6.0 < \beta < 6.2$.
A comparison of the predictions for plaquette observables
from simulations 
in this range confirms the absence of finite size effects for this observable
to the accuracy of these results.
This conclusion matches the claim in
\cite{Schaefer:2010hu}
that  finite size effects are small for the parameters we consider
both for these local observables and later on for the more extended ones.

A table of parameter values for the constrained runs is given in~\ref{appendixb}, but
some rough observations are given here as a prelude to a discussion
of the method of selecting them. 
The action is of order  $1.5\times 10^5$ at  $16^4$, 
$3.6\times 10^5$ at $20^4$
and
$7.3\times 10^5$ at $24^4$.
The standard deviation
of the action in the unconstrained case  is about
$\sigma =196$ at $16^4$,
$\sigma =286$ at  $20^4$
and $\sigma =391$ at  $24^4$
 in the centre of the reweighting range.
This parameter provides the scale for the spacing 
$\delta_S$ of order 
$210$ at $16^4$, $310$ at $20^4$
and $420$ at $24^4$.
The width of the Gaussian constraints were set to be about 
1.2 times the spacing in each case,
$\delta/\delta_S = 1.2$.


\subsection{Tempering Method}

Constrained functional integrals are estimated from 
HMC simulations with action 
$H(a_i,\phi,S_i)$ which incorporates
the Gaussian constraint (see equation (\ref{eqn-constrained})).
\begin{equation}
H(a_i,\phi,S_i) =  a_i S[\phi] + (S[\phi] - S_i)^2/2\delta^2 \ .
\end{equation}

Because the constraint might allow energy
barriers to block the simulation dynamics
there is concern that  density of states simulations
will be slower than unconstrained simulations.
For pure gauge theory even the
unconstrained simulations are notorious for the slow dynamics 
apparent in
topological modes. In an attempt to overcome this problem we
use tempering  \cite{Marinari:1992}
(this was called the replica exchange method in \cite{Cossu:2017sfu,Lucini:2016fid}).
Tempering is a technique that considers an ensemble of
simulations over a range of $\beta$, or in our case $S_i$, 
and allows configurations to migrate over this range.
Configurations suffering from slow dynamics at large $\beta$ can 
diffuse to small $\beta$ where they evolve more quickly before
returning to  small $\beta$.
We chose a set of $N_{TEMPER}=24$ tempers each with a fixed central action $S_i$.
The central energies are listed in table \ref{tab-params} in~\ref{appendixb} and
correspond to a range of coupling given in table \ref{TemperSim}.
These ranges are much narrower than the ranges of the unconstrained
reference  simulations.

\begin{table}[!htb]
\begin{center}
\caption{
Parameters of measurement phase of density
of states simulations. 
There were 24 tempers, or simulations at specific
values of $S_i$. To aid comparison, the range
of central action
is given in terms of a coupling range below.
}
\begin{tabular}{|l|c|c|}
\hline
Size & $\beta$ range  & Number of sweeps \\
\hline
$16^4$ &  $6.07 < \beta < 6.16$  & $2 \times 10^4$\\
$20^4$ &  $6.26 < \beta < 6.32$  & $2 \times 10^4$\\
$24^4$ &  $6.42 < \beta < 6.46$  & $2 \times 10^4$\\
\hline
\end{tabular}
\label{TemperSim}
\end{center}
\end{table}

At regular intervals (every 15 HMC steps in the measurement phase), 
the tempering method swaps configurations 
between pairs of tempers
with the following probability
\begin{eqnarray}
{\rm min}(&1&,exp\lbrace H(a_1,\phi_1,S_1) + H(a_2,\phi_2,S_2) \nonumber\\
&-& H(a_1,\phi_2,S_1) - H(a_2,\phi_1,S_2) \rbrace ) \ .
\end{eqnarray}

This preserves detailed balance of the entire system of multiple
tempers and aids ergodicity.

\begin{figure}
\begin{center}
\includegraphics[scale=0.65]{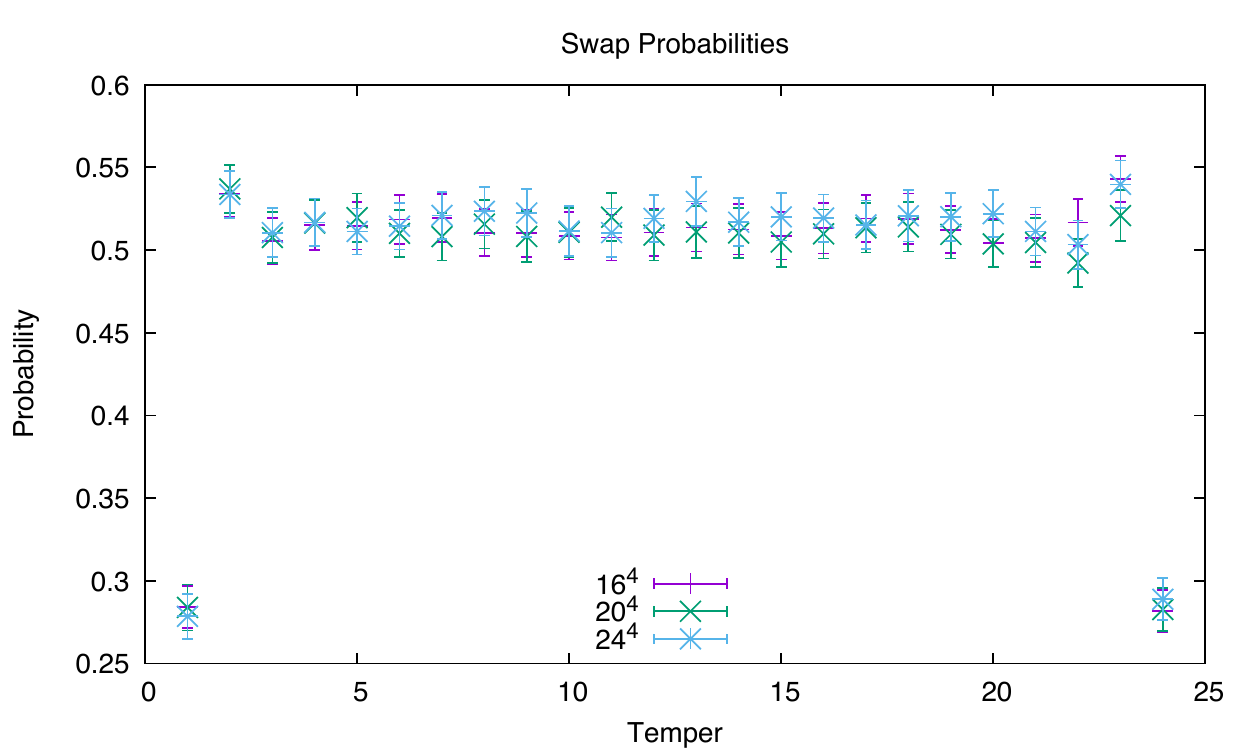}
\caption{
Average swap probabilities during the course of the 
measurement phase. Averaged over all replicas.
Errors are a combination of statistical and from the variation between replicas.
The overlapping symbols indicate that the probabilities for  $16^4$, $20^4$ and $24^4$  are indistinguishable.
}
\label{fig-swapprob}
\end{center}
\end{figure}

In order to ensure that swap probabilities remain
constant across the range of tempers, 
the values of the set of central energies $S_i$
and the width of the Gaussians
 $\delta_i$ were chosen so that the overlap
 of probability distributions remain fixed.
This amounts to a requirement on the ratio of the spacing $\delta_S$ 
and the standard deviation of the constrained action
(given by $\delta_i/\sqrt{1 + \delta_i^2/\sigma^2}$).
The absolute values of $\delta_i/\sigma$ and $\delta_S/\sigma$ determine
the potential accuracy of the method.
The effect of this tuning was monitored through the swap
probabilities as shown in figure \ref{fig-swapprob}
where the aim was for a swap probability of 0.5.
Effects from the edge of the temper range are apparent, but
do not extend far into the main range.

\begin{figure}
\begin{center}
\includegraphics[scale=0.65]{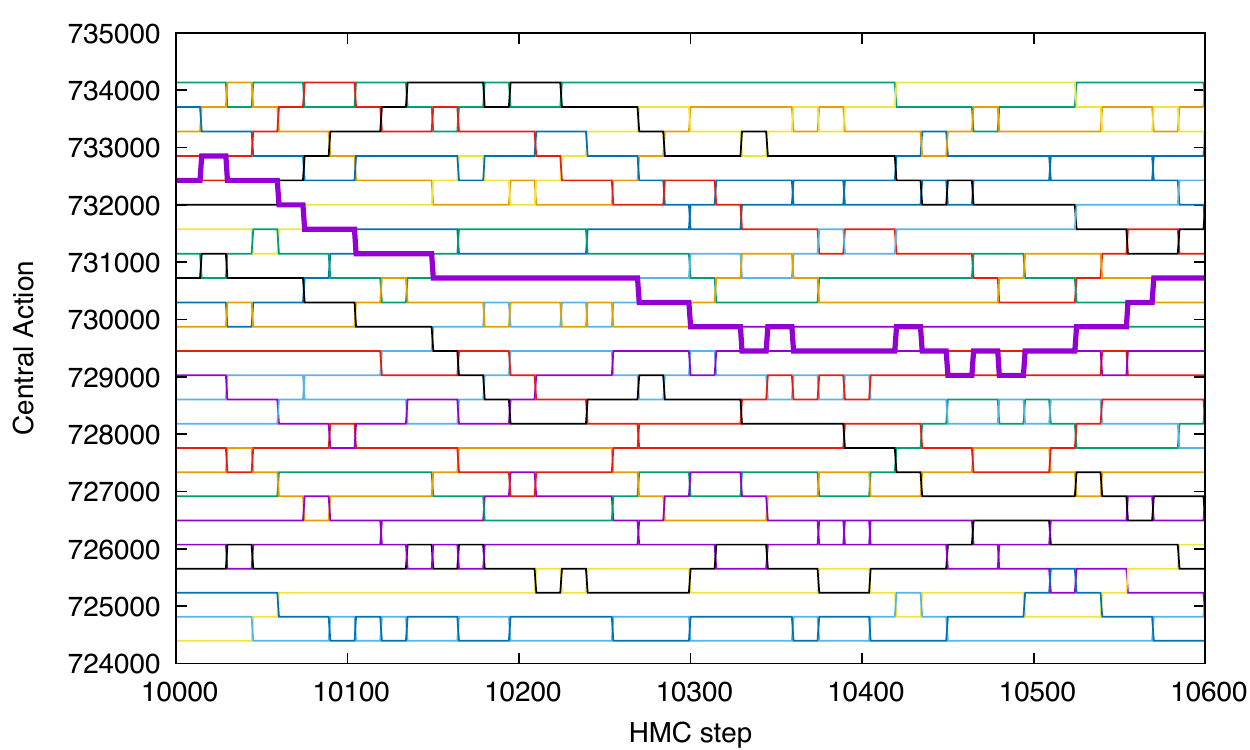}
\caption{
Evolution of configurations showing how swaps  between
adjacent tempers lead to global mixing.
Example for $24^4$ over a short
part of the run for one replica near the middle of the measurement
phase. The bold line shows the history of a certain configuration.
}
\label{fig-swapmix}
\end{center}
\end{figure}

Figure \ref{fig-swapmix} shows the swap history over a short part
of the measurement phase of the simulation illustrating how 
the pair swaps enable mixing though the whole space.
For all lattice sizes, more than half the 
configurations appear at some point in every temper. In other words, they
appear at every possible central energy. Moreover, every configuration
always appears in at least half the tempers.


\subsection{Observables}
\label{observables}

Besides the plaquette expectation value which is an action dependent
quantity and hence can be computed using either 
of the techniques outlined in section 2,
it turned out that
the variation of the plaquette provides a sensitive test of the
accuracy of the method. 

Other observables were defined after configurations had 
been evolved according to the Wilson flow \cite{Luscher:2010iy}.
The simplest such observables 
were the action density $E$ and its clover version $E_{sym}$
which have been studied before in the context of
Wilson flow \cite{Luscher:2010iy}.

At $16^4$ the flow time was set to $t=2$ 
and this was scaled for the simulations at $20^4$ to $t=3.125$
and at $24^4$ to $t=4.5$ so as to preserve the 
extent of the smoothing radius
as the same fraction of the lattice in each case, keeping the physical
length fixed.
These choices correspond to a requirement on the dimensional parameter $t$:
\begin{equation}
\sqrt{2\, d\,  t} = L/4 \ .
\label{eqn-WFt}
\end{equation}

Where $d$ is the dimension ($d=4$) and $L$ is the linear size of the box in
lattice units. Fixing the extent of the smoothing to $1/4$ of the lattice
will allow us to study scaling  in section \ref{scaling}.

Topological charge was the observable of most interest as it provides a test of the
efficiency of the tempering method in improving autocorrelation
times. The charge was computed after Wilson flow of the operator
to a value of $t$ that was scaled according to the size of the lattice
as described above.
The charge was defined in terms of plaquette-like quantities on the
smoothed configuration \cite{Luscher:2010iy,Luscher:2005rx}
that
amount to the continuum expression 
given in the introduction (\ref{Qcontinuum}).


\subsection{RM Phase}

With a single coupling constant and far from
any phase transition,
the value of $a_i$ for a given central energy $S_i$ will be close to the
value of the coupling $\beta$ at which $\langle S-S_i\rangle = 0$.
The preliminary simulations allow a fit to the dependence of $\langle S \rangle$
on $\beta$ which provides a value that is used
as a seed for the RM phase.
Since the seed is known so accurately, the role of the RM phase
is simply to improve rather than 
to search for the solution in a complex landscape. 
The literature \cite{Spall:2003}
indicates that the efficiency of the Robbins Monro technique can depend sensitively
on the relaxation parameters. 
This may be true for high dimension problems common 
in AI but is not the case for this improvement of a one dimensional problem.
Indeed,  a simple Gaussian model for the RM process allows the effect of
different RM schedules to be explored for much longer convergence
times than are possible for the real system. 
The model consists of sampling $\langle\langle S-S_i\rangle\rangle$
from a Gaussian of width $\delta/\sqrt{1+\delta^2/\sigma^2}$
and mean $-(a_i - a_i^{*})\delta^2/({1+\delta^2/\sigma^2})$
where $a_i^{*}$ is the assumed solution.
It was found that 
there was little advantage in more elaborate
tuning than the choice of parameters $n_0$ and $c$ in the 
RM iteration:

\begin{equation} 
a_i(n) = a_i(n-1) - {c\over n_0 + n} \langle\langle S- S_i\rangle\rangle_i \ .
\label{RMiteration}
\end{equation} 

In the RM phase, 
$ \langle\langle S- S_i\rangle\rangle_i$
was computed from
$50$ measurements on consecutive HMC configurations
with fixed  $a_i$. Test runs which made an update 
according to (\ref{RMiteration}) every measurement
had the same behaviour and were barely any faster.
These simulations were performed without any
replica swaps during the evaluation of $\langle\langle S- S_i\rangle\rangle_i$
because, while these do not usually affect final convergence,
they do prevent it at the very edges of the temper range. Moreover
they lead to larger early fluctuations which would require a different
tuning regime. Swaps were allowed between 
evaluations of $\langle\langle S- S_i\rangle\rangle_i$.

The value of $c$ was chosen to be rather smaller than the
supposed optimal one as larger values tended to drive the system
too far from the accurate seed value. This choice does not compromise
the convergence properties of the algorithm.
The starting iteration $n_0$ was set to $20$ as a compromise between
excess change over the first few iterations and speed of convergence
later (the value was slightly raised to $n_0=25$ for $24^4$).

In contrast to the details of the RM schedule, starting and 
following independent RM simulations
was useful both to test the convergence and to provide a bootstrap
mechanism for computing errors  arising from 
inaccuracy in this phase. 
The whole simulation procedure is repeated for a set of $N_{REPLICA}$
replicas. 
During the RM phase, each replica or RM iterate starts from a different random configuration 
separately thermalised with a different sequence of random numbers,
but with the same seed value $a_i(0)$.
We distinguish the replicas with an upper index $a = 1\dots N_{REPLICA}$ and in
this work there are $N_{REPLICA}=8$ different replicas. 
The mean over replicas at each RM phase iteration step $n$, is 
denoted with an overbar as,

\begin{equation}
{\bar a_i(n)} = {1\over N_{REPLICA}}  \sum_{a=1}^{N_{REPLICA}} a_i^a(n)   \ .
\end{equation}

According to the theory of the Robbins Monro 
technique \cite{Spall:2003}, different iterates will
asymptotically  converge to a Gaussian
distribution of width proportional to 
$1/\sqrt{n}$. 
The variance in replica space is therefore a convenient
convergence parameter,

\begin{equation}
C_i(n) = {1\over N_{REPLICA}}  \sum_{a=1}^{N_{REPLICA}} 
(a_i^a(n)-{\bar a_i(n)} )^2   \ .
\end{equation}
\begin{figure}
\includegraphics[scale=0.65]{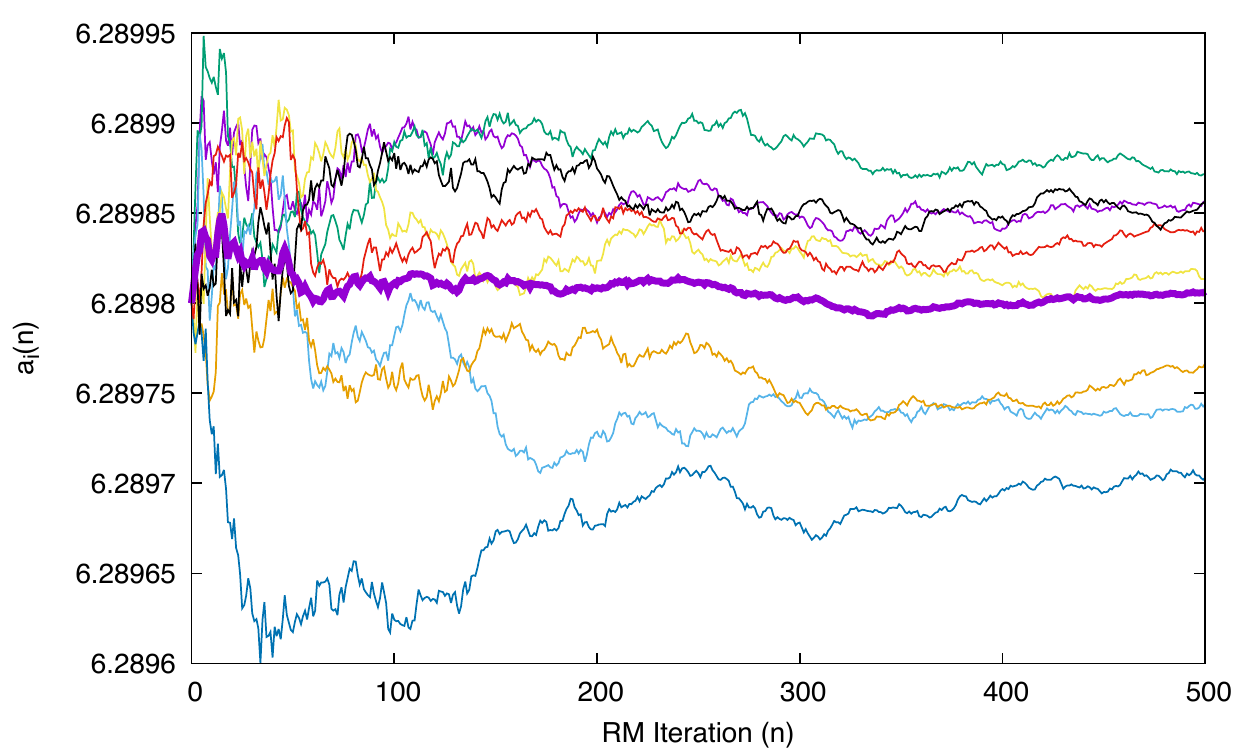}
\includegraphics[scale=0.65]{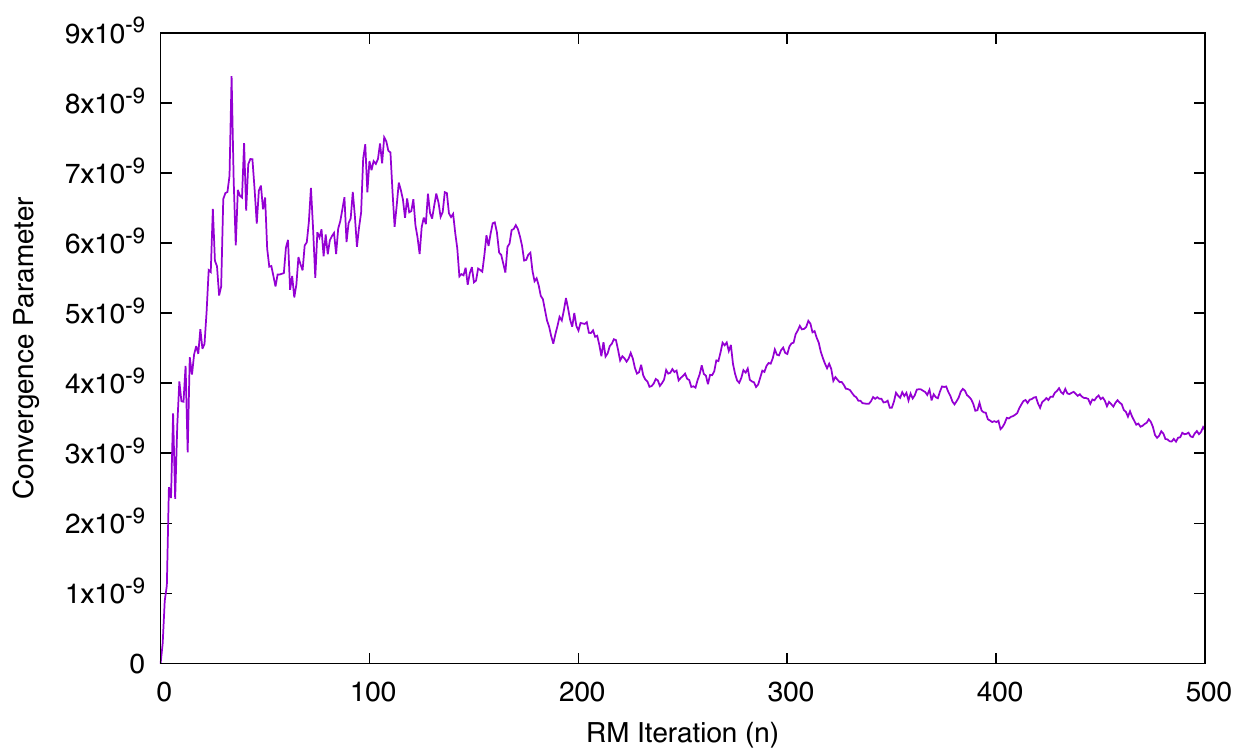}
\caption{
Convergence during the RM phase at $20^4$
for the temper with central action 363310. 
Top: history of $a_i(n)$ 
showing 8 independent replicas all starting from the 
seed 6.28980. The mean is also shown as the thick line. 
Bottom: history of convergence parameter $C_i(n)$ defined in the text.
}
\label{fig-convergence20}
\end{figure}

\subsubsection{Convergence}

Examples of typical convergence are shown in figure
\ref{fig-convergence20}. 
Ideal behaviour is for the convergence parameter to decrease as
$1/n$.

During the RM phase the behaviour of the convergence parameter is characterised by a noisy initial phase after
which the scaling regime
is reached. The shape of these curves varies considerably between
replicas and tempers but all eventually display the asymptotic
behaviour.

The primary convergence criterion is to ensure that convergence 
parameter has reached 
the asymptotic regime, but this still leaves some flexibility in
exactly when to stop.
In the present work 
we first ensure that the convergence has reached the asymptotic phase
and terminate 
after a fixed number of iterations $n_{max}$.
$n_{max}=442$ iterations for $16^4$ and $500$ in the case of $20^4$ and $24^4$. 

The error in the final value of $a_i$ may be estimated from the 
standard deviation, 
$\sqrt{C_i({n_{max}})/N_{REPLICA}} $,
at the end of the RM phase.
When averaged over tempers,  this was
$2.7\times 10^{-5}$, $1.7\times 10^{-5}$, $1.3\times 10^{-5}$ for $16^4$, $20^4$
and $24^4$ respectively.
Note that these values are similar to the difference between the final mean 
value  ${\bar a_i}(n_{max})$  and the initial seed value. 
As is clear from the behaviour of the convergence parameter of the RM phase
convergence shown in figure \ref{fig-convergence20}
it takes considerable effort to improve  the accuracy of 
the $a_i$.

\begin{figure*}
\centering
\includegraphics[scale=0.5]{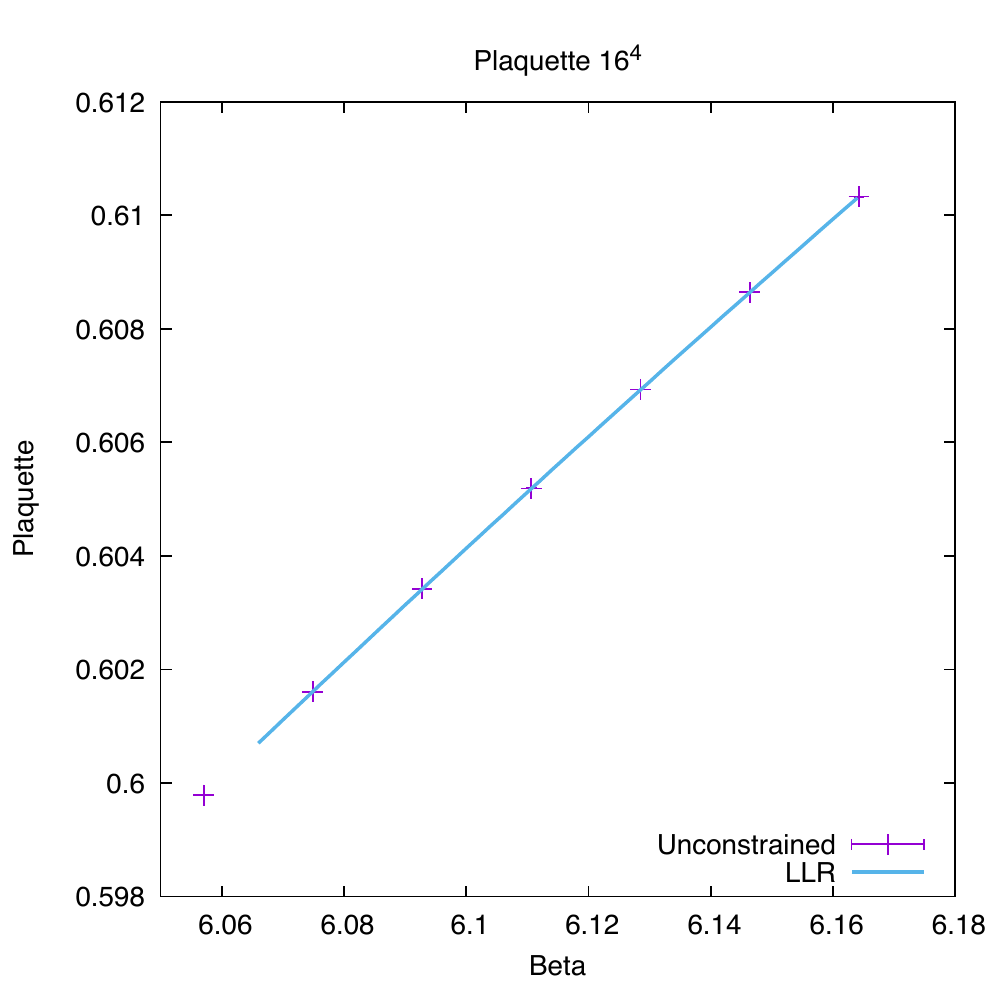}\hfil
\includegraphics[scale=0.5]{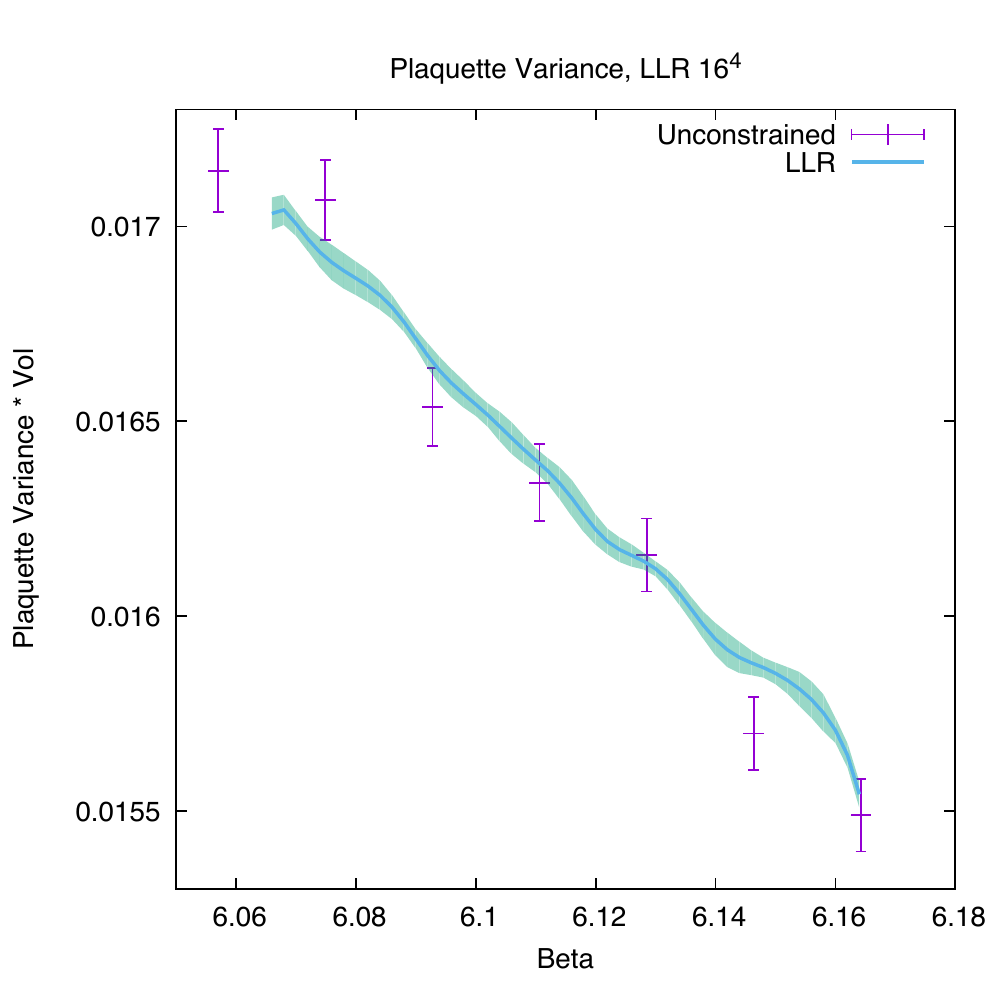}\hfil
\includegraphics[scale=0.5]{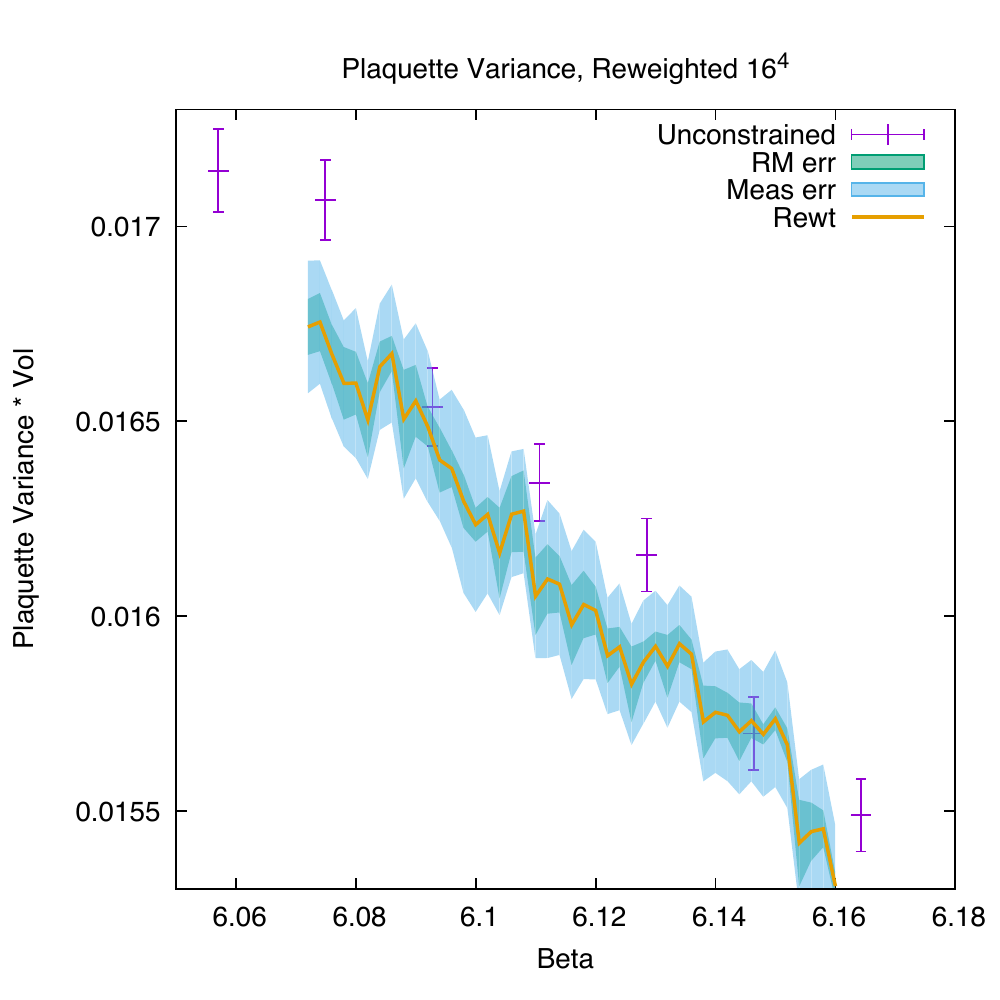}\par\medskip
\includegraphics[scale=0.5]{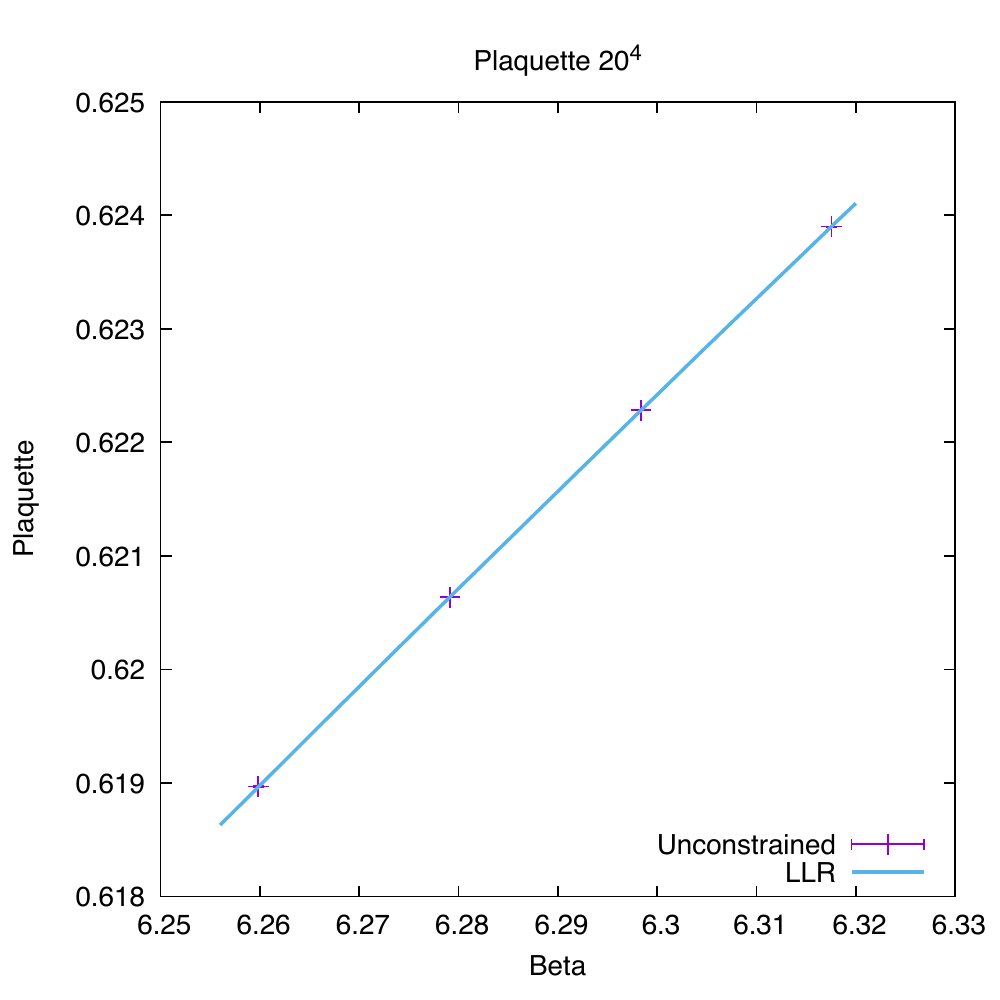}\hfil
\includegraphics[scale=0.5]{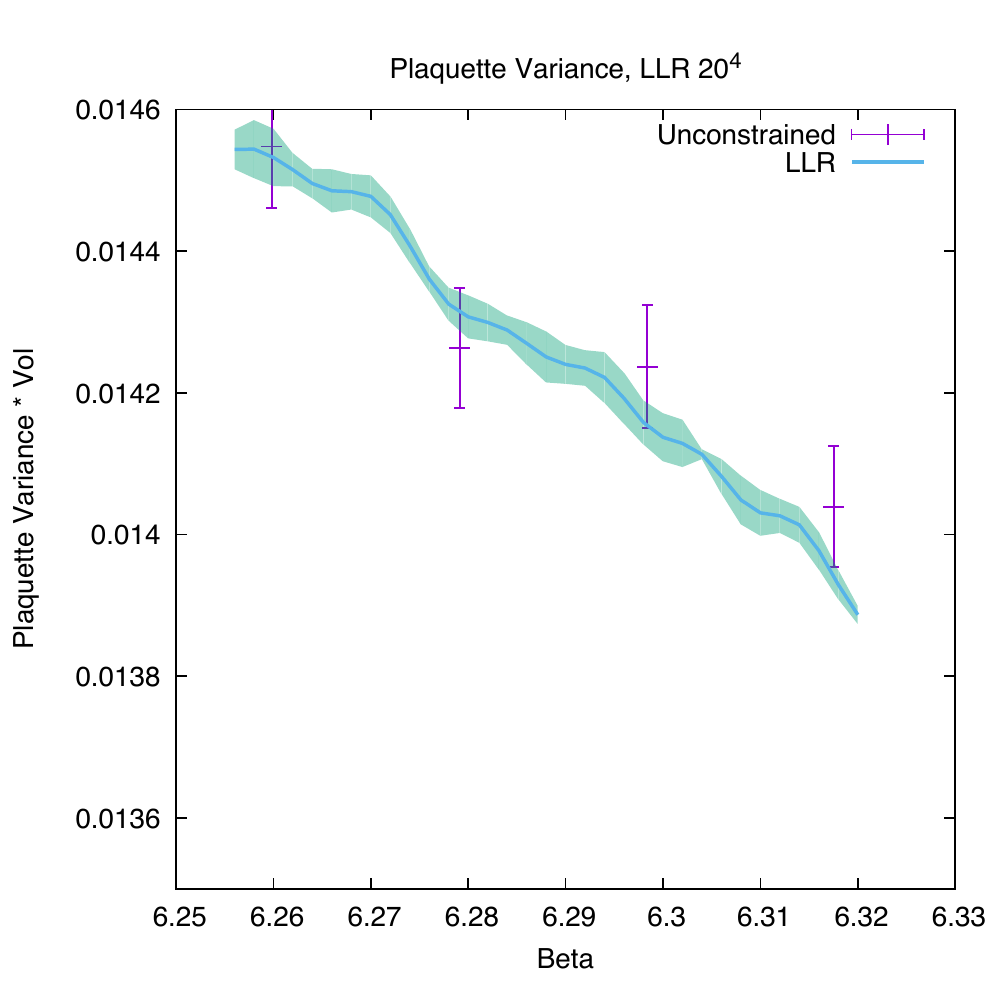}\hfil
\includegraphics[scale=0.5]{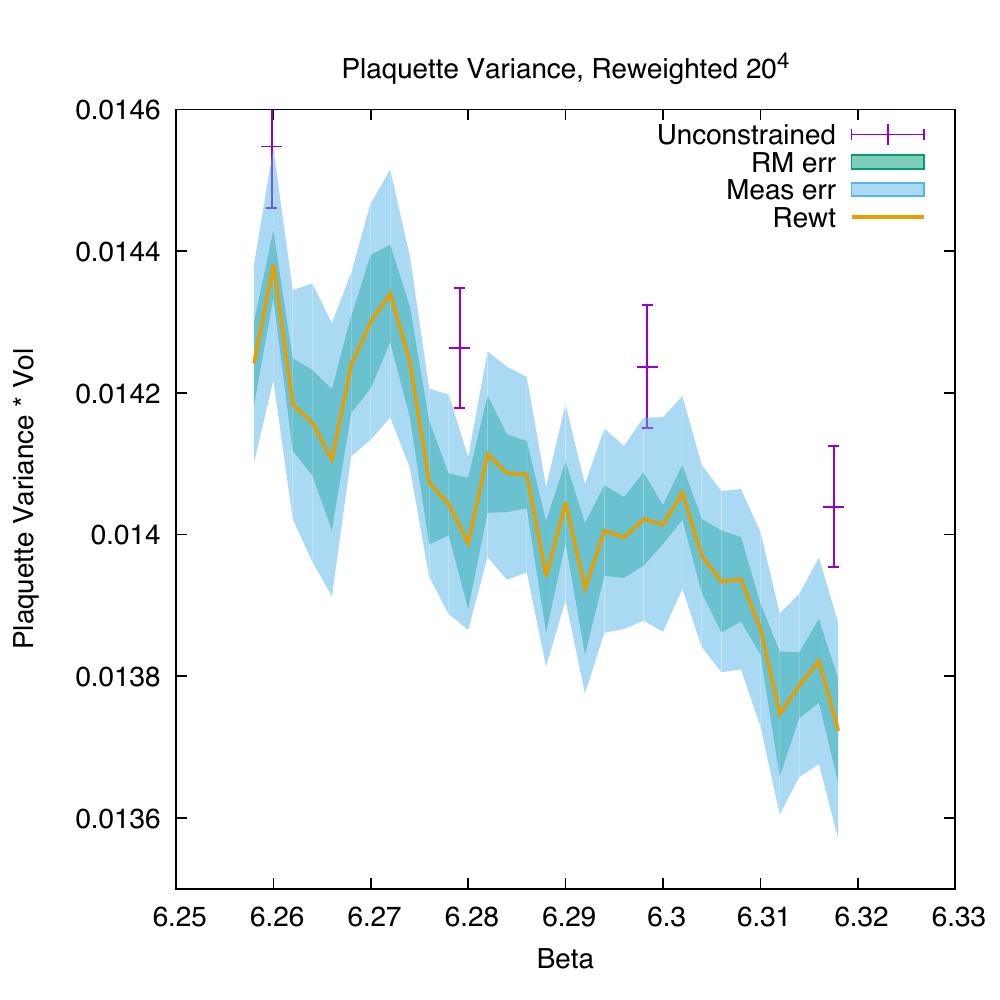}\par\medskip
\includegraphics[scale=0.5]{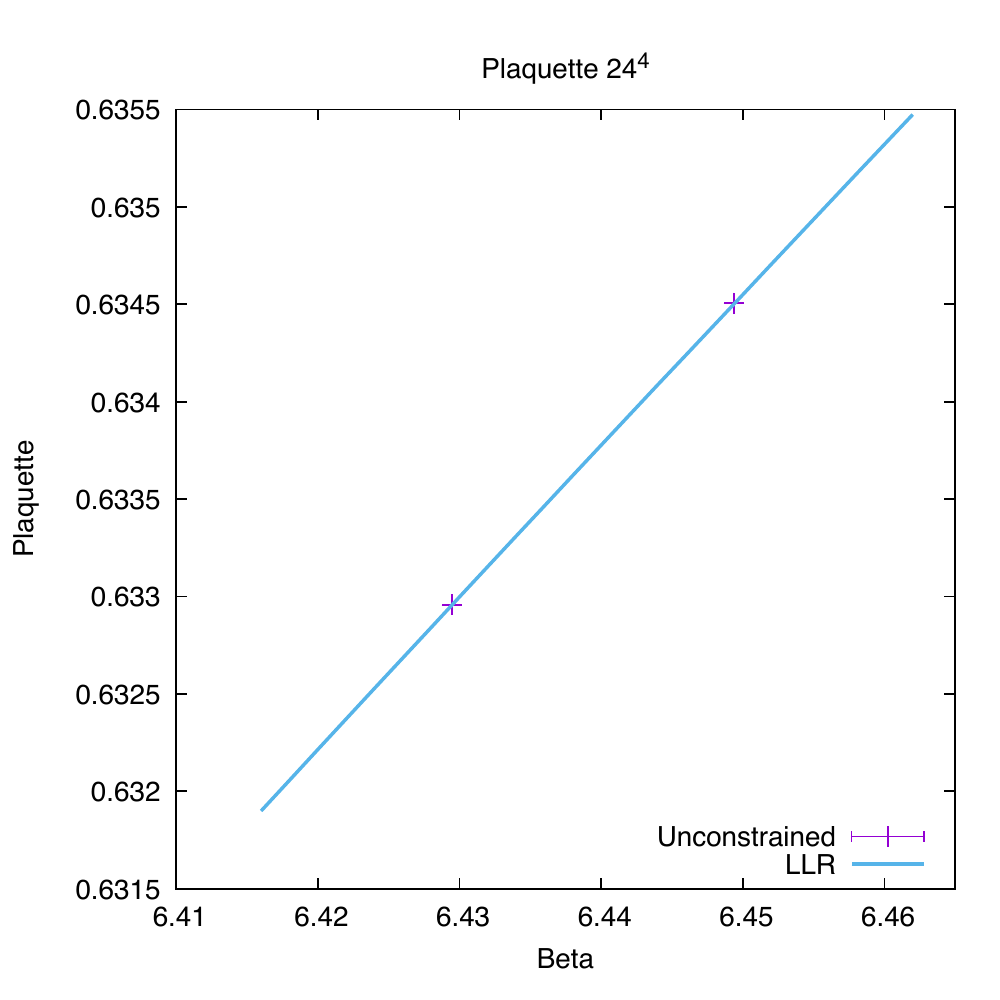}\hfil
\includegraphics[scale=0.5]{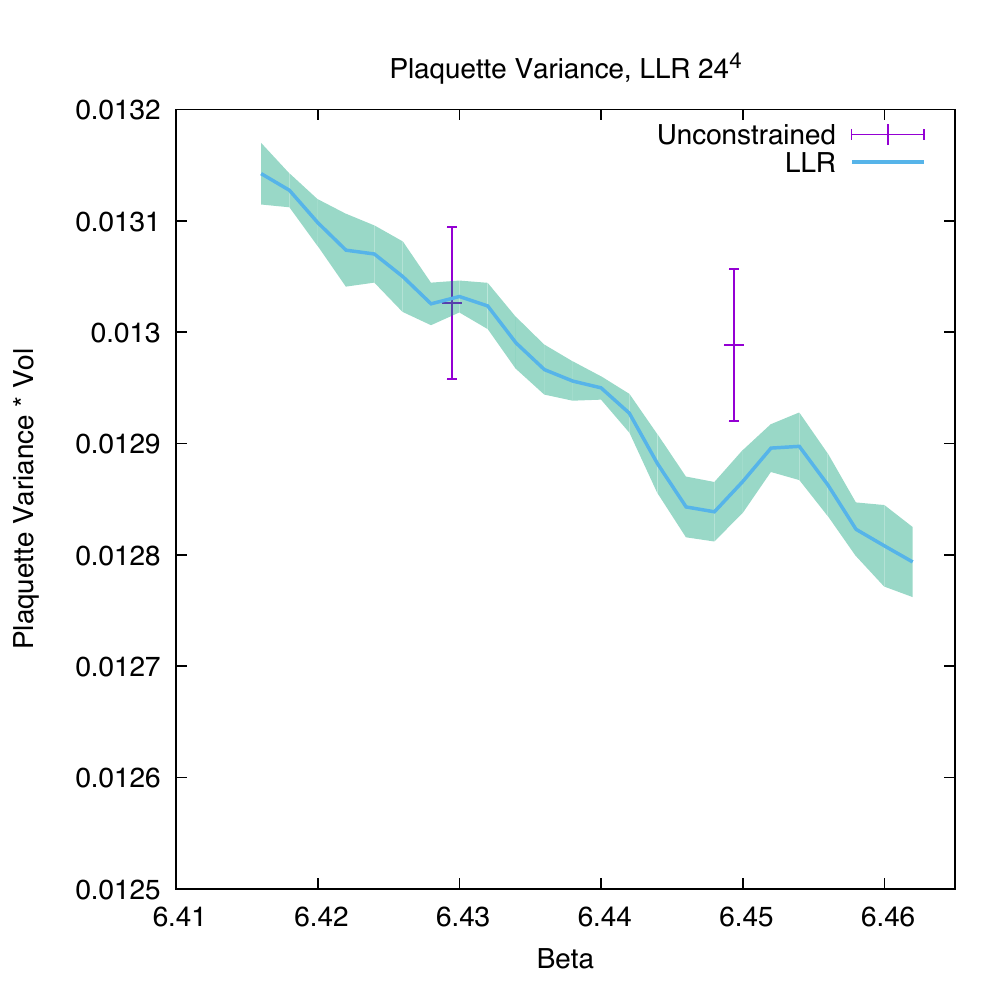}\hfil
\includegraphics[scale=0.5]{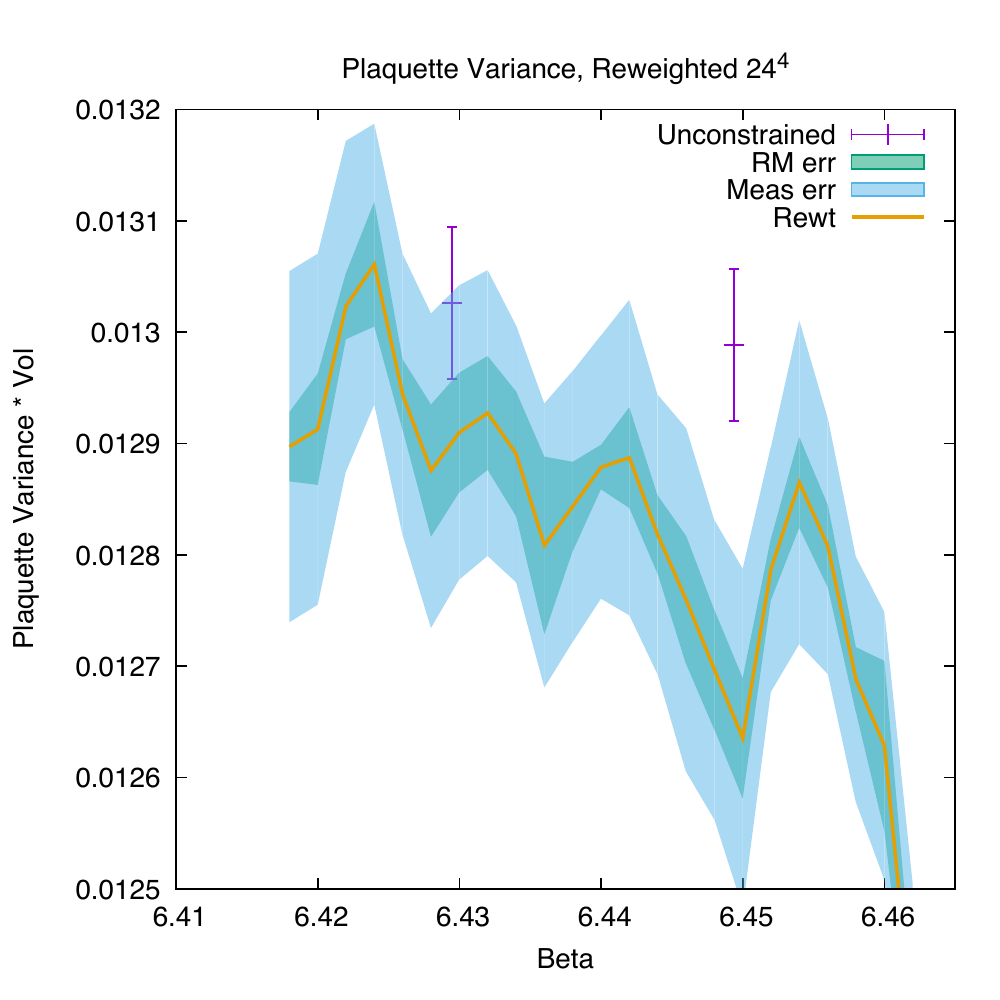}
\caption{
Predictions for the mean plaquette and its variance. 
Rows show $16^4$, $20^4$ and $24^4$.
The first column shows the mean plaquette with errors that are
too small to be visible  and this
appears the
same whether computed using 
information from the RM phase (LLR)
or reweighting results from the measurement phase.
The second and third columns show the plaquette variance
computed using LLR and reweighting respectively.
Errors are shown with shaded regions.
All plots also show data points obtained
using traditional unconstrained methods. 
}
\label{fig-llr}
\end{figure*}

\subsubsection{Predictions from RM phase}

The LLR procedure of section \ref{llr} was followed to compute the 
plaquette and its variation based solely on the $S_i$ and the $a_i$
determined by the RM phase.
This was done separately for each of the $8$ replicas
and at the final stage errors are evaluated using bootstrap.
Note that the scales are different for each size shown in figure \ref{fig-llr}.
For the values of  $\beta$ shown in these plots, the original runs using 
traditional HMC were supplemented by additional unconstrained runs 
at $20^4$ and $24^4$ intended to improve statistics.

The range of $\beta$ considered for these measurements 
was limited because deviations from a smooth plot
appear near the edges of the temper range. 
The full LLR version of the weights (\ref{weights_vi}) changes
abruptly from an approximately linear dependence on $\beta$ 
and it is straightforward
to choose
a cutoff that amounts to discarding 3 tempers at each boundary.
The resulting ranges are given in table \ref{LLRanges}.

\begin{table}[!htb]
\begin{center}
\caption{
Ranges of $\beta$ for LLR predictions based on the RM phase.
}
\begin{tabular}{|l|c|}
\hline
Size & $\beta$ range   \\
\hline
$16^4$ & $6.066 \le \beta \le 6.164$\\
$20^4$ & $6.256 \le \beta \le 6.320$\\
$24^4$ & $6.416 \le \beta \le 6.462$\\
\hline
\end{tabular}
\label{LLRanges}
\end{center}
\end{table}

Although the plaquette can easily be measured using traditional
techniques, it is reassuring that we obtain the same results,
with high accuracy, using the density of states method.
The mean plaquette shown in the first column of 
figure \ref{fig-llr} appears the
same whether computed using 
density of states information from the RM phase
or reweighting results from the measurement phase.
The plaquette variance 
shown in the second column of figure \ref{fig-llr} 
provides a more sensitive test.
Indeed, an early $20^4$ run with shorter RM phase and 
values of $a_i$ only known to accuracy of 
around $1\times 10^{-4}$ (compared to $3 \times 10^{-5}$ for
the data shown in figure \ref{fig-llr}), 
led to errors in the variance about 5 times bigger then in the
data presented here.
The third column of figure \ref{fig-llr} is discussed in the next
section.

\subsection{Measurement Phase}

The measurement phase has a duration of $2 \times 10^4$ HMC steps
with potential swaps every 15 steps
and was repeated for each of the 8 RM replicas
and also for all three lattice
sizes.
There are two distinct contributions to the error: from uncertainly
in the $a_i$ arising from the RM phase and from statistical fluctuation 
in the measurement phase due to limited length runs. 
We refer to these as the RM error and MP error respectively.

The set of $\beta$ at which to compute canonical expectation
values by the reweighting procedure of section \ref{rewt} 
as adapted for simulations below,
are chosen sufficiently close
to give a good approximation to a continuous curve in the plots.
We choose a $\beta$ spacing of $0.002$ irrespective of lattice size.

\begin{figure}
\centering
\includegraphics[scale=0.65]{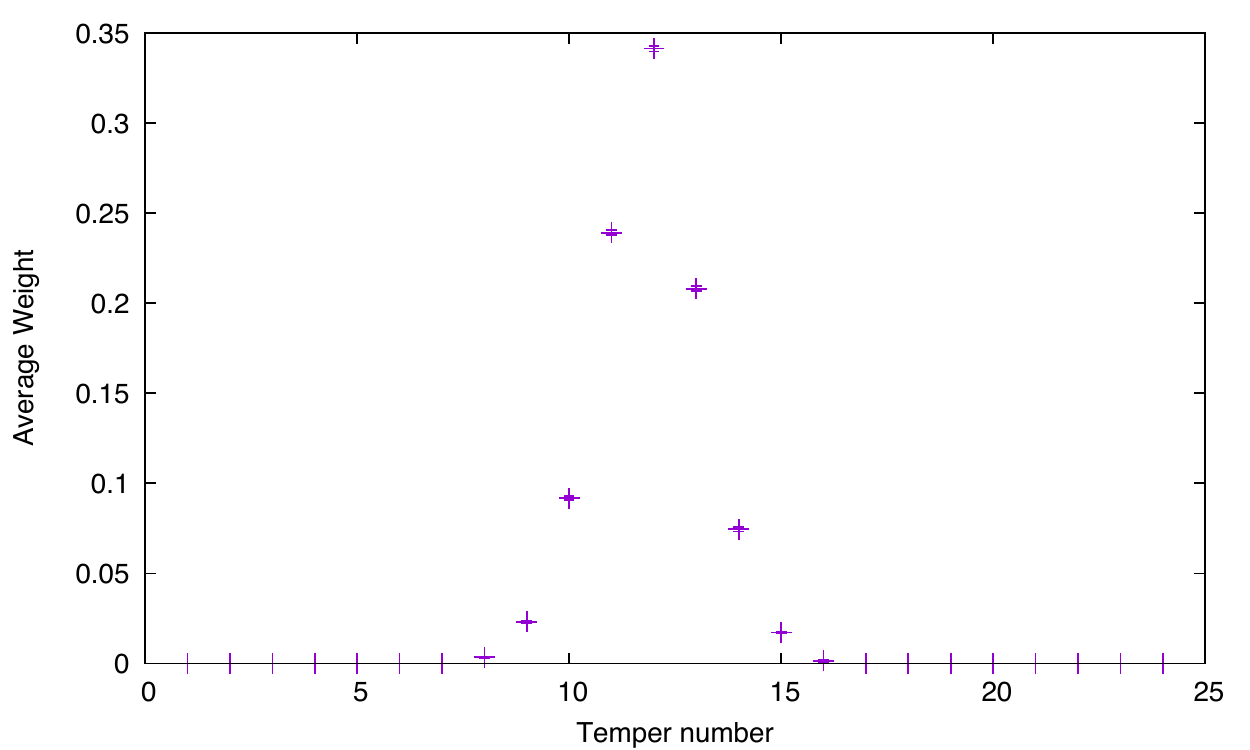} 
\caption{
The average normalised weight for $\beta$ near the centre of the temper range.
Shown for $16^4$, $\beta=6.118$; other sizes and $\beta$ are very similar.
The time dependent weight (\ref{wt-wit}) is averaged over all measurements
and all RM replicas. Errors are too small to be visible.
}
\label{fig-wts}
\end{figure}

\subsubsection{Reweighting}

In order to relate the formalism developed in section \ref{formalism}
to the output of simulations 
it is helpful to explicitly recognise that 
in the measurement phase of the constrained Monte Carlo, 
the double bracket quantities appearing in (\ref{expectation_dynamic})
are identified 
as averages over HMC configurations or time-steps $t$ for a particular temper $i$.
The usual estimator is:

\begin{equation}
\langle \langle B[\phi]e^{(a_i-\beta)(S-S_i)}\rangle \rangle_i 
= 
{1\over T} \sum_t
e^{(a_i-\beta)(S_i[t]-S_i)} B_i [t]  \ .
\end{equation}

Where $S_i[t]$ and $B_i [t]$ are the measurements of the respective
observables at that configuration or time-step.

For each HMC configuration 
and separately for each RM replica labelled by index $a$, we
perform the following temper sums 
appearing in the numerator and denominator of
 equation (\ref{expectation_dynamic}).
Usually, the variance of the operator is also a quantity of interest
so computations involving $B_i^2[t]$ are made at the same time.

 \begin{eqnarray}
B^a[t] &=& \sum_i 
 w^a_i e^{(a^a_i - \beta)(S^a_i[t]-S_i)} B^a_i[t] \\
& = &\sum_i 
 w^a_i[t] B^a_i[t]  \\
w^a[t]&=& \sum_i 
 w^a_i e^{(a^a_i - \beta)(S^a_i[t]-S_i)}
 = \sum_i 
 w^a_i[t]  \ .
\end{eqnarray}

Where we have defined the time dependent weights,

\begin{equation}
w^a_i[t]
= w^a_i e^{(a^a_i - \beta)(S^a_i[t]-S_i)}  \ .
\label{wt-wit}
\end{equation}

Which in the quenched approximation 
reduce to the ordinary static weights $w^a_i$
(\ref{wt-wi}).

Using these definitions the reweighting expression 
(\ref{expectation_dynamic}) for a particular RM replica, $a$, becomes

\begin{equation}
\langle B[\phi]\rangle =
{\sum_t B^a[t] \over \sum_t w^a[t] }  \ .
\label{simulation_dynamic}
\end{equation}

The integrated autocorrelation time is defined as the maximum of the autocorrelation
time in numerator and denominator and  is computed according to
the prescription in 
\cite{Wolff:2003sm}.
In order to  estimate the ratio that appears in the
reweighting formula (\ref{simulation_dynamic})
without
bias, a bootstrap technique is used over the $2\times 10^4$ measurements
divided by the autocorrelation time.
The same bootstrap selection is
employed for both numerator and denominator. This procedure 
also yields an estimate of the error of the ratio due to the
fluctuations in the measurements. We call this statistical error the ``measurement
phase (MP) error''. 

The procedure above is repeated for the $N_{REPLICA}$ different 
RM replicas. 
A separate analysis over these reweighted measurements gives the error
associated with imprecision in knowledge of the value
of the    $a_i$'s that were derived in the RM phase.
Depending on the observable,
the relative sizes of the errors arising from each source, RM or measurement,
is different and
in plots the two different contributions to the error
are represented by distinct shaded regions.

The static weights, named  $w_i$ 
and defined by equation (\ref{wt-wi})  
in the formalism above,
are computed
in such a way as to avoid danger of numerical overflow,
based on the parameters
 $(S_i, a_i)$ and the desired $\beta$ for reweighting using
generalisations of the formulae  (\ref{wt-wi})  adapted for variable action spacing.
When the reweighting parameter $\beta$ happens to be similar 
to a value of $a_i$ in the middle of the temper range, 
the static weights are largest for tempers that are near that central part of the range.
For time varying weights (\ref{wt-wit}) there are fluctuations 
and 
 the  weights of more distant
tempers contribute. 
Although these  weights vary for each step of the measurement 
simulation, their average is very stable.
Figure \ref{fig-wts} 
shows how the average normalised time dependent weight varies across
the temper index for $\beta$ chosen near the centre 
of the temper range. 
Plots for other size lattices or other values of $\beta$ near the
centre of the tempering range
look very similar indicating that the range of tempers that
contribute appreciable weight is fairly constant at about 7 
tempers for our parameters.

As $\beta$ approaches an edge of the temper range, reweighting
becomes less accurate since tempers not simulated could have 
significant weights. 
The reweighting range is determined by requiring that the average
normalised weight for edge tempers be less than 1\%.
A different criterion based on the limiting the fraction of weights
that are greater than 0.01 for example, leads to the same range.
The resulting ranges are apparent from the plots, but are explicitly given in table \ref{RewtRanges}.
There appears to be a slight decrease in the useful fraction of the temper range 
as the lattice size increases.

\begin{table}[!htb]
\begin{center}
\caption{
Ranges of $\beta$ for reweighting predictions based on the measurement phase
according to criteria in the text.
}
\begin{tabular}{|l|c|}
\hline
Size & $\beta$ range   \\
\hline
$16^4$ & $6.072 \le \beta \le 6.160$\\
$20^4$ & $6.258 \le \beta \le 6.318$\\
$24^4$ & $6.418 \le \beta \le 6.462$\\
\hline
\end{tabular}
\label{RewtRanges}
\end{center}
\end{table}


\begin{table}[!htb]
\begin{center}
\caption{
Plaquette observables for $16^4$ at $\beta = 6.118$.
Entries are computed using LLR, $\rm LLR'$ or reweighted measurements
with quenched (QRewt) or varying weights (Rewt).
Multi-histogram results based on unconstrained simulations are given
for comparison (see~\ref{appendixb}).
}
\begin{tabular}{|l|c|c|c|c|}
\hline
Obs & method  & value & RM-err & MP-err \\
\hline
Mean                & MultiHist       & 0.605906 & 3.6e-06  & - \\
Mean                & LLR'        & 0.605904 &  1.8e-06  & - \\
Mean                & LLR         & 0.605904 & 1.9e-06  & - \\
Mean                & QRewt     & 0.605904 & 1.7e-06  & 3.4e-06 \\
Mean                & Rewt        & 0.605905 & 1.9e-06  & 4.2e-06 \\
Var 		& MultiHist               & 2.4761e-07 & 12.2e-10 & - \\
Var 		& LLR'                & 2.4820e-07 & 6.3e-10 & - \\
Var		 & LLR               & 2.4813e-07 & 6.9e-10 & - \\
Var		 & QRewt             & 3.9087e-07 &  8.6e-10 & 31.7e-10 \\
Var		 & Rewt              & 2.4446e-07 & 13.2e-10 & 29.2e-10 \\
$\tau_{int}$   & QRewt             &  2.17& 0.05 & 0.13\\
$\tau_{int}$   & Rewt             & 0.73& 0.05 & 0.03\\
\hline
\end{tabular}
\label{tab-errs16}
\medskip

\caption{
Plaquette observables for $20^4$ at $\beta = 6.290$.}
\begin{tabular}{|l|c|c|c|c|}
\hline
Obs & method & value & RM-err & MP-err \\
\hline
Mean               & MultiHist          & 0.621560  & 10.6e-06  & - \\
Mean               & LLR'          & 0.621569  & 0.9e-06  & - \\
Mean                & LLR          & 0.621569  & 1.0e-06  & - \\
Mean                & QRewt      & 0.621569 & 0.7e-06  & 1.9e-06 \\
Mean                & Rewt         & 0.621570 & 1.0e-06  & 2.4e-06 \\
Var		 & MultiHist              & 0.8806e-07 & 3.9e-10 & - \\
Var		 & LLR'              & 0.8902e-07 & 1.6e-10 & - \\
Var		 & LLR                & 0.8900e-07 & 1.7e-10 & - \\
Var		 & QRewt          & 1.4034e-07 & 2.0e-10 & 10.8e-10 \\
Var		 & Rewt            & 0.8778e-07 &  3.6e-10 &  8.6e-10 \\
$\tau_{int}$   & QRewt             & 1.95& 0.02 & 0.11\\
$\tau_{int}$   & Rewt                & 0.90& 0.06 & 0.04\\
\hline
\end{tabular}

\label{tab-errs20}
\medskip

\caption{
Plaquette observables for $24^4$ at $\beta = 6.438$. }
\begin{tabular}{|l|c|c|c|c|}
\hline
Obs & method  & value & RM-err & MP-err \\
\hline
Mean                & MultiHist      & 0.633594 & 5.6e-06  & - \\
Mean                & LLR'       & 0.633621 & 0.6e-06  & - \\
Mean                & LLR        & 0.633621  & 0.6e-06  & - \\
Mean                & QRewt    & 0.633622 & 0.6e-06  & 1.4e-06 \\
Mean                & Rewt       & 0.633620 & 1.0e-06  & 1.4e-06 \\
Var		 & MultiHist       & 0.3257e-07 & 2.34e-9 & - \\
Var		 & LLR'       & 0.3906e-07 & 0.5e-10 & - \\
Var		 & LLR        & 0.3905e-07 & 0.5e-10 & - \\
Var		 & QRewt    & 0.6199e-07 & 1.6e-10 & 4.9e-10 \\
Var		 & Rewt       & 0.3871e-07 & 1.2e-10 & 3.7e-10 \\
$\tau_{int}$   & QRewt     & 1.91& 0.06 & 0.11\\
$\tau_{int}$   & Rewt        & 0.89& 0.02 & 0.03\\
\hline
\end{tabular}
\label{tab-errs24}
\end{center}
\end{table}

\subsubsection{Plaquette Predictions}

The expectation
of the mean plaquette and variance can be computed via either of the
techniques, LLR based on the RM phase or reweighting measurements, and it
is interesting to compare the results.
A plot of the mean is indistinguishable from 
the left column of figure \ref{fig-llr} though as is apparent
from tables \ref{tab-errs16}-\ref{tab-errs24}, 
the LLR approach leads to greater accuracy.
The plaquette variance according to the reweighting method is shown in the
right column of figure \ref{fig-llr}. 
The RM errors for each approach are similar but the
reweighting approach is also subject to MP errors
which for the run length chosen, are a few times larger that the RM errors.

In order to provide more detail about the relative size of the 
RM errors from uncertainly
in the $a_i$ from the RM and the MP errors due to limited length runs, 
tables \ref{tab-errs16}-\ref{tab-errs24}
show data for a representative value of $\beta$ 
chosen in the centre of the range for each lattice size.
Overall, because it is only subject to RM and not MP errors, the
LLR approach that only uses the density of states 
is more accurate than
reweighting.
In this case it is possible to see 
that the difference between 
measurements using the full formulae  based on (\ref{zeefulldos})
and those
using the approximation denoted $\rm LLR'$ are always smaller than
the error due to uncertainty in $a_i$.
The tables also show results from unconstrained simulations 
re\-weight\-ed to the reference $\beta$ using multi-histogram \cite{Ferrenberg:1989ui}
(see \ref{appendixb}). Note that the error from this approach reflects 
both the intrinsic statistical error of the unconstrained simulations 
and also how
close the reference  $\beta$ happens to be to 
one of the widely spaced set of  $\beta$  used in
those simulations, so the way it changes with lattice size is not
significant.

The tables also allow us to  compare 
the use of quenched and fluctuating weights
for reweighted measurements. 
Both lead to very similar
results for the mean plaquette and its errors.
However there is a clear difference in the result for the
plaquette variance and comparison with either density of states or
traditional simulations favours the approach derived in 
(\ref{expectation_dynamic}), that is, the one with fluctuating weights.
This result clearly indicates that the quenched approach
is incorrect, we will continue to study it to illuminate the autocorrelation of
the topological charge, but in all further plots of expectation values we
use the fluctuating weights. 
For both the plaquette mean and variance, the MP error for the
chosen run length is
a few times larger than the RM error.


\begin{figure*}
\centering
\includegraphics[scale=0.65]{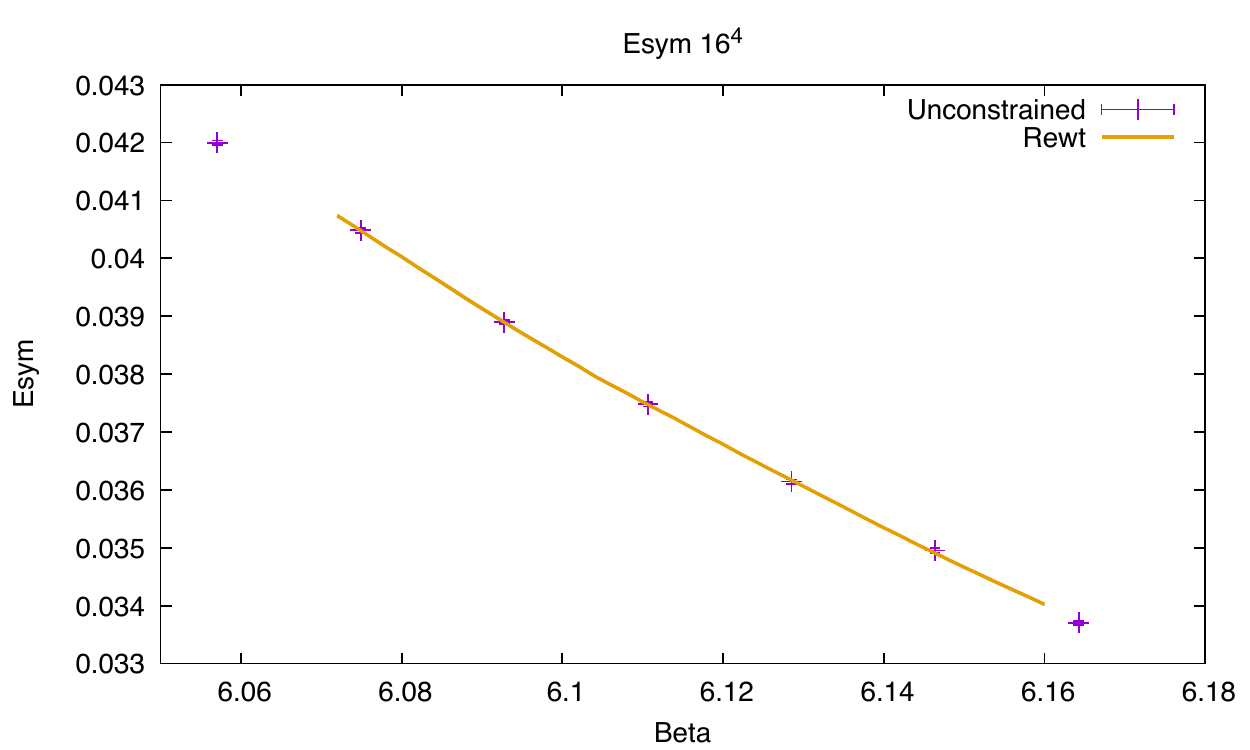}\hfil
\includegraphics[scale=0.65]{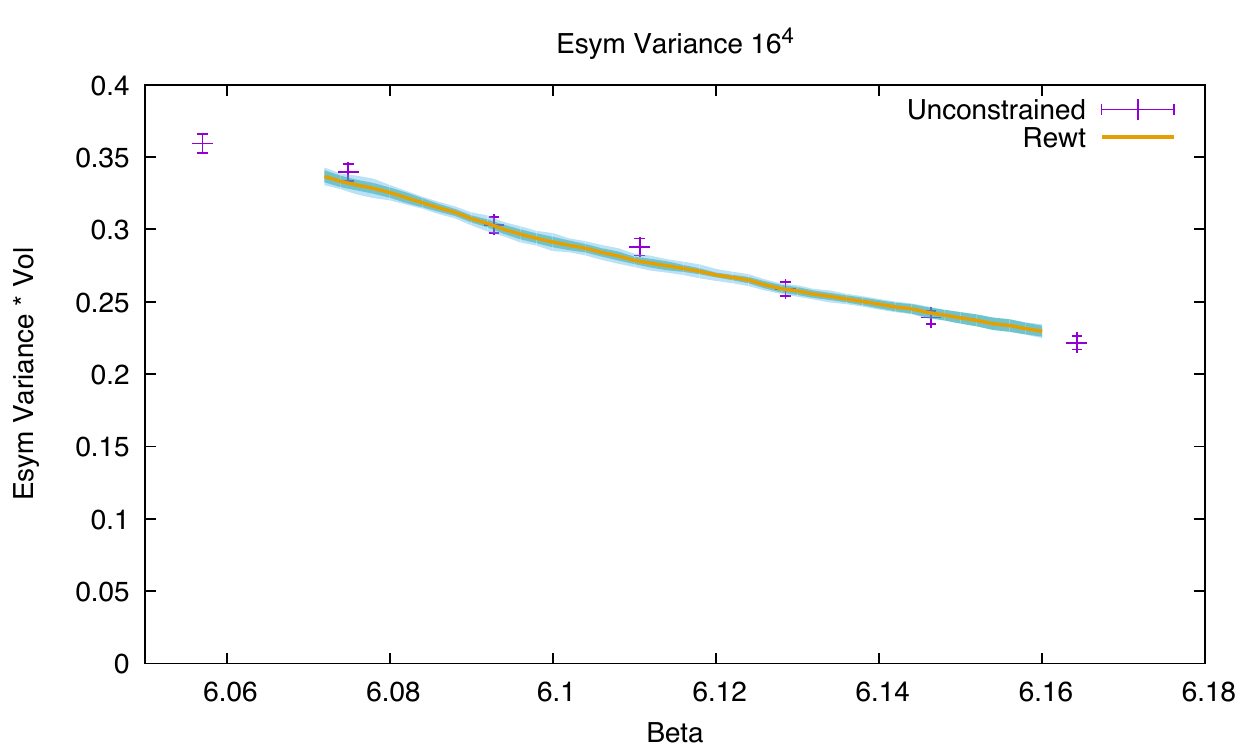}\par\medskip
\includegraphics[scale=0.65]{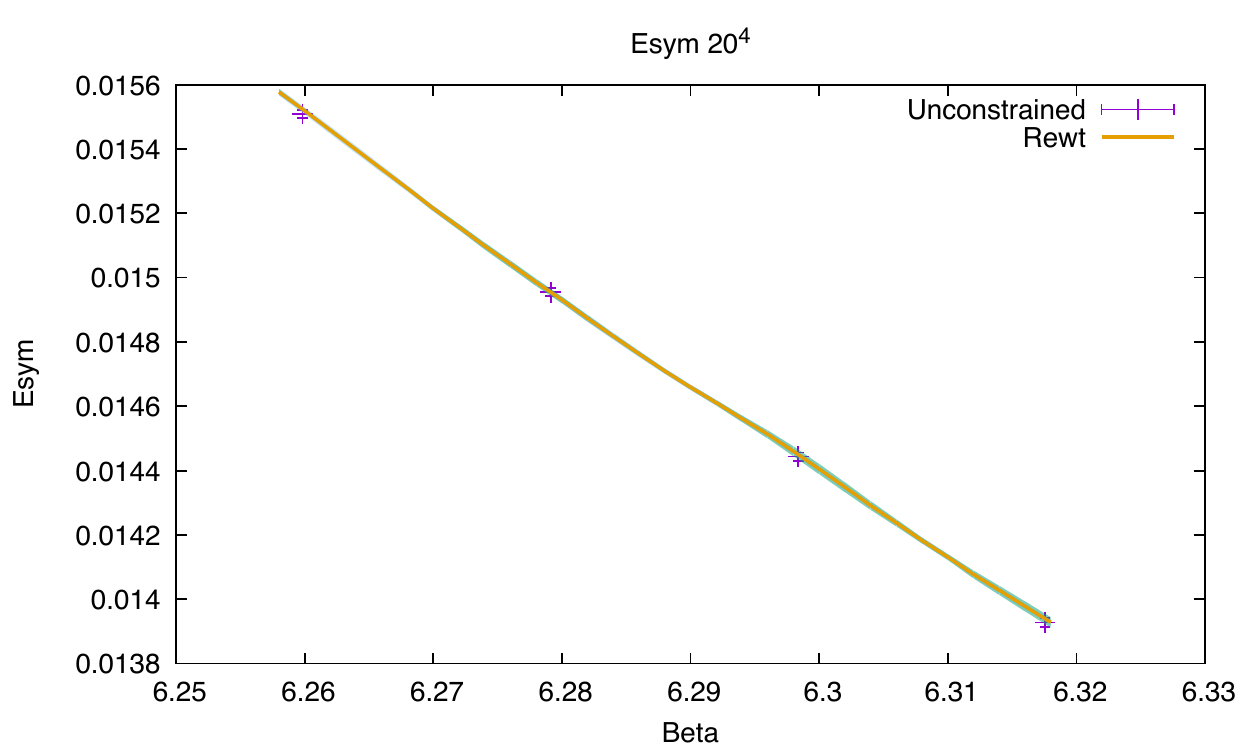}\hfil
\includegraphics[scale=0.65]{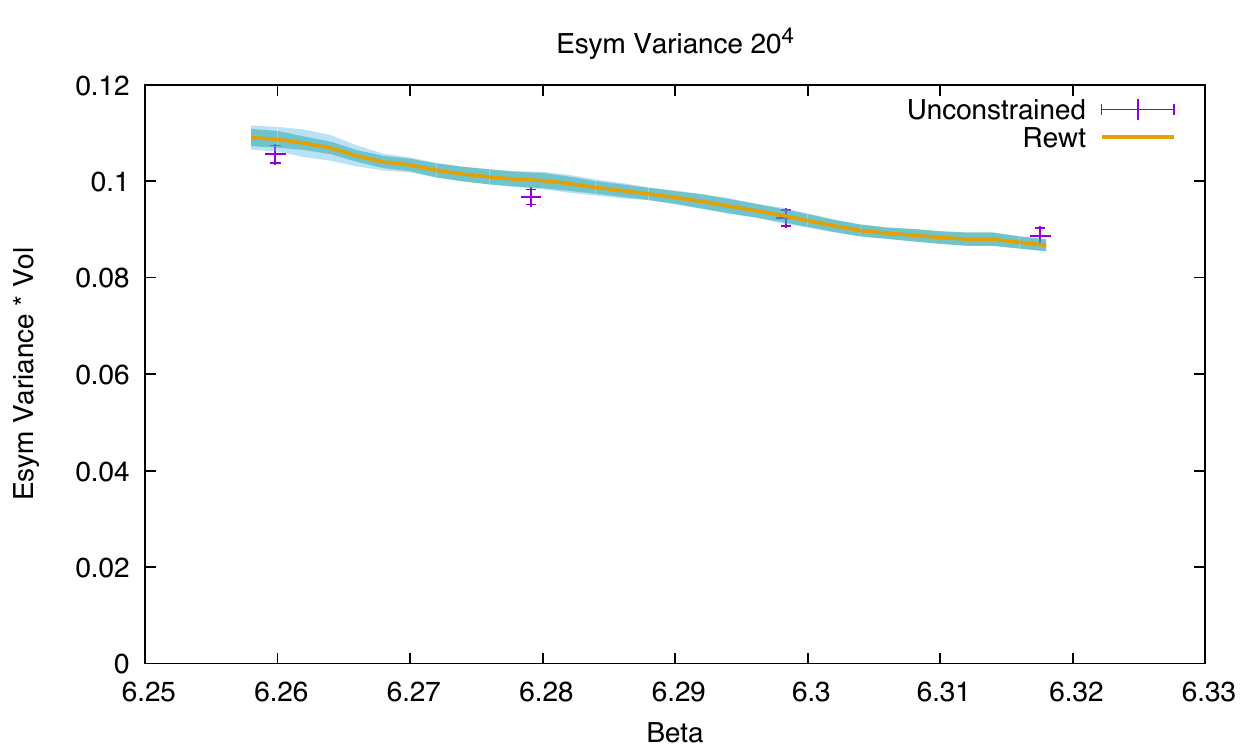}\par\medskip
\includegraphics[scale=0.65]{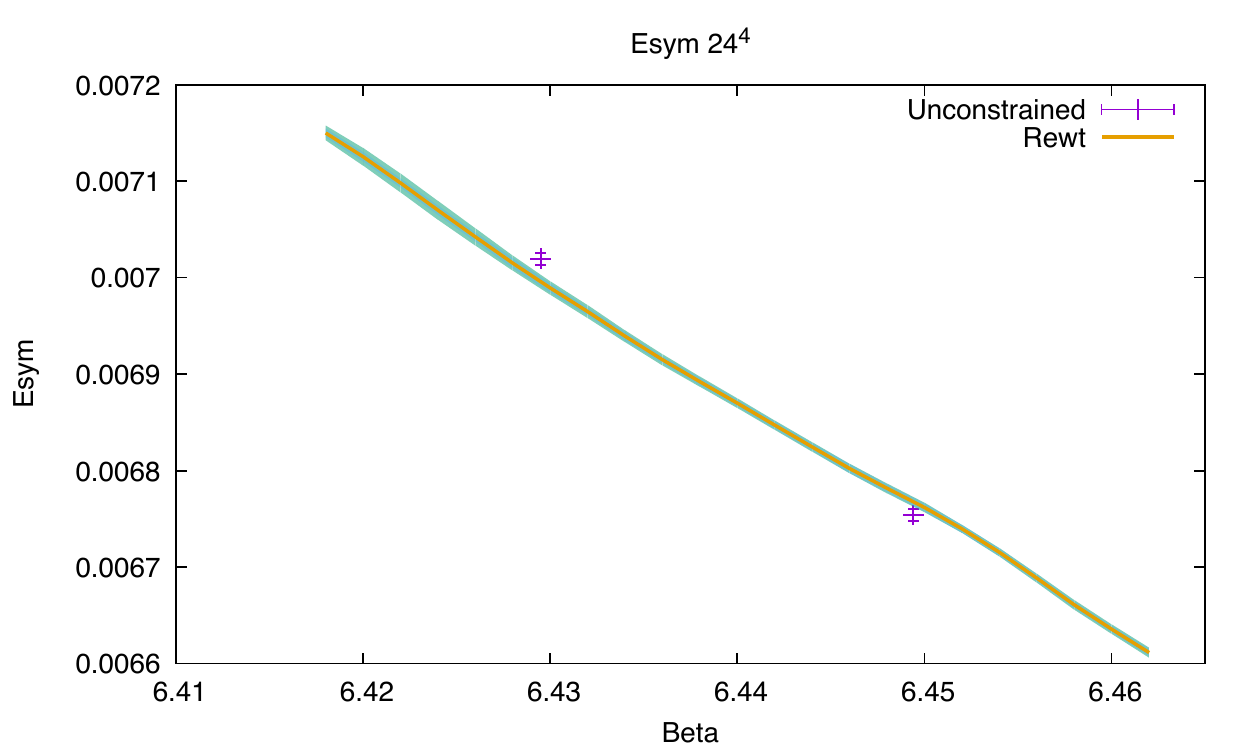}\hfil
\includegraphics[scale=0.65]{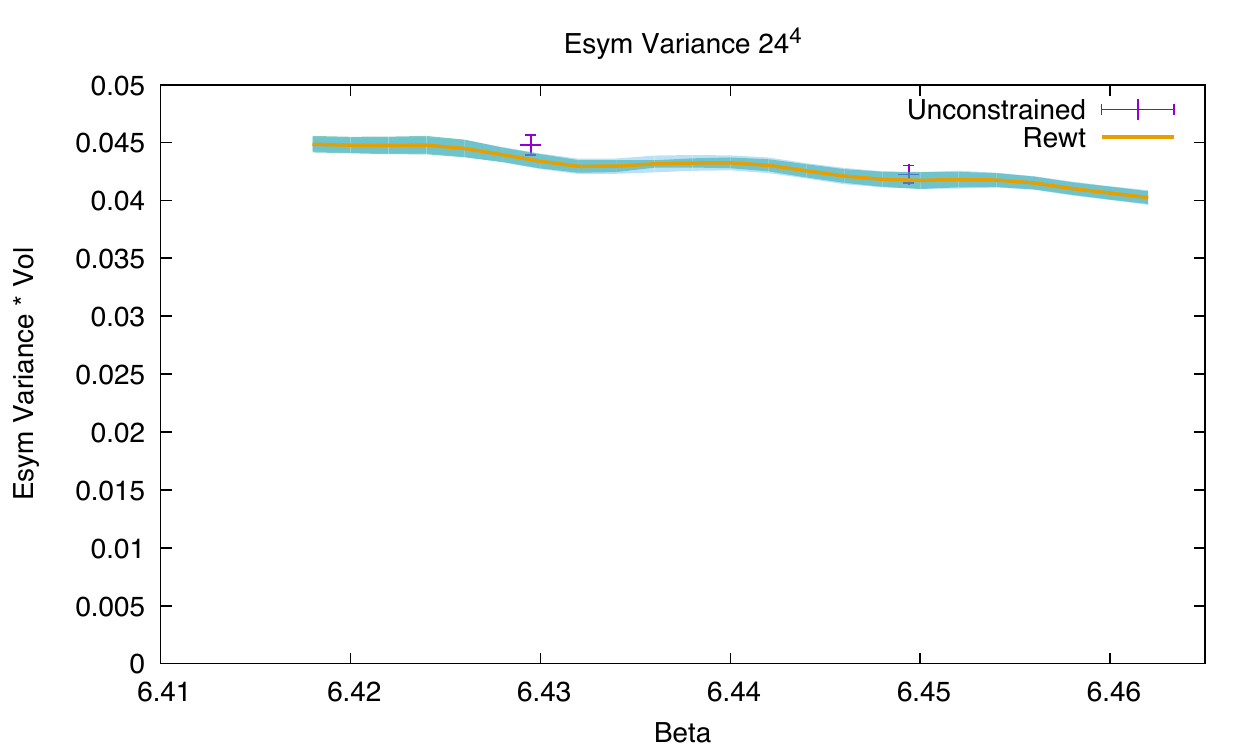}
\caption{
The symmetric cloverleaf energy density
after Wilson flow using 
reweighted measurements.  
Rows for $16^4$, $20^4$, $24^4$. 
Also shown for comparison
are results from traditional simulations supplemented by additional runs.
Left: the mean value. Right: the variance (with a volume factor).
Errors from the measurement and the RM phase are 
similar but small and are shown as the shaded region
barely visible in the case of the mean.
}
\label{fig-Esym}
\end{figure*}

The autocorrelation times displayed in the tables are 
the integrated version computed using the Wolff criterion as described in
section \ref{autocorr} and are order 1 HMC step
remaining fairly constant as
$\beta$ varies. 
The consequences of the use of fluctuating weights are most apparent
in the integrated autocorrelation time which becomes half the value 
obtained in a quenched analysis.
The equivalent value for the traditional simulation dropped to a plateau of 
about 3 HMC steps for $\beta > 6$.
For varying weights, the MP
and RM errors are of similar size while for quenched weights the
MP error is larger.

\subsubsection{Observables defined after Wilson Flow}

Wilson flow corresponds to an operation 
that cannot be 
treat\-ed without using reweighting. 
Expectation values for the symmetric cloverleaf energy density, $E_{sym}$, after Wilson flow 
are shown in figure \ref{fig-Esym}
and the equivalent results for the energy density after Wilson flow
are very similar. 
For the values of $\beta$
 shown, the traditional unconstrained simulations 
have been supplemented with additional
runs to improve statistics.
As we have come to expect for the 
plaquette, the mean value of $E_{sym}$ has small error and 
agrees closely with the reference simulations for all  sizes. 
The variance is much better behaved  than 
it was for the original plaquette variance without Wilson flow
and has closer agreement with the unconstrained simulations.
Errors arising from RM and measurement are of similar size
and both less than 0.1\% for the mean and less than 2\% for
the variance.

A discussion of the autocorrelation time for these observables
is postponed until after we discuss the computation of autocorrelation in the
case of  the topological charge.

\begin{figure*}
\centering
\includegraphics[scale=0.65]{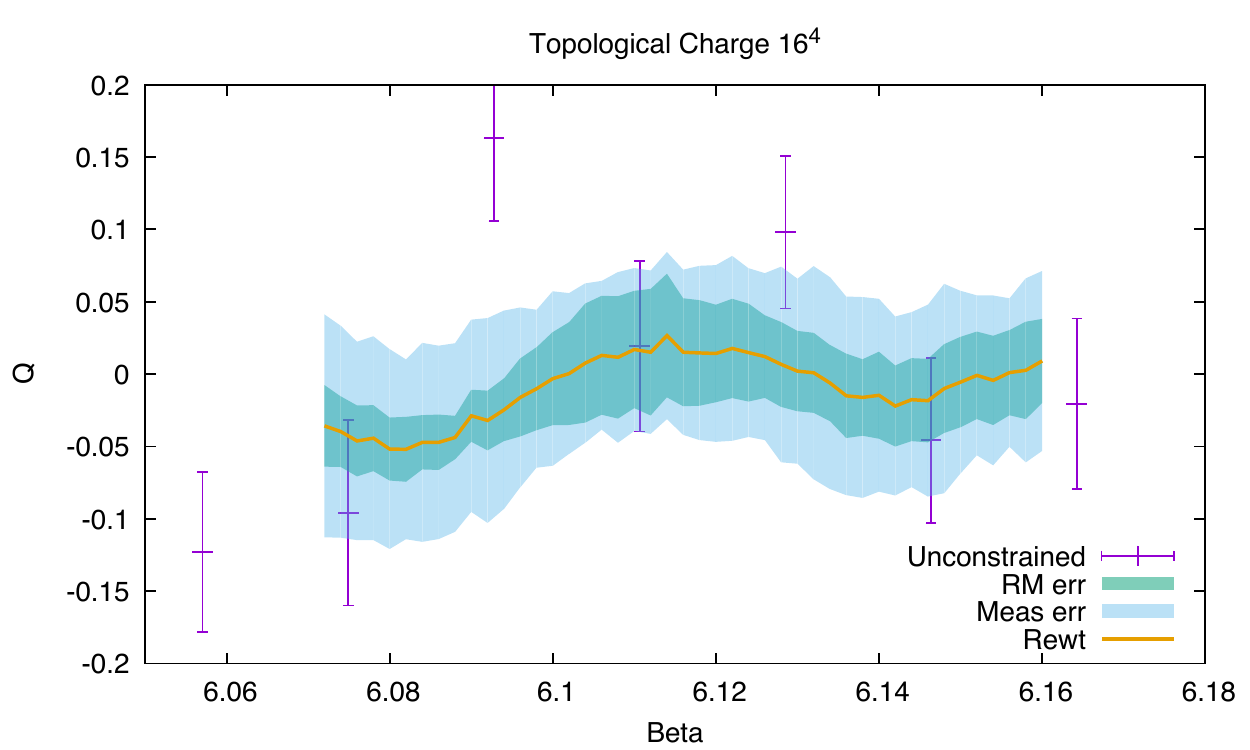}\hfil
\includegraphics[scale=0.65]{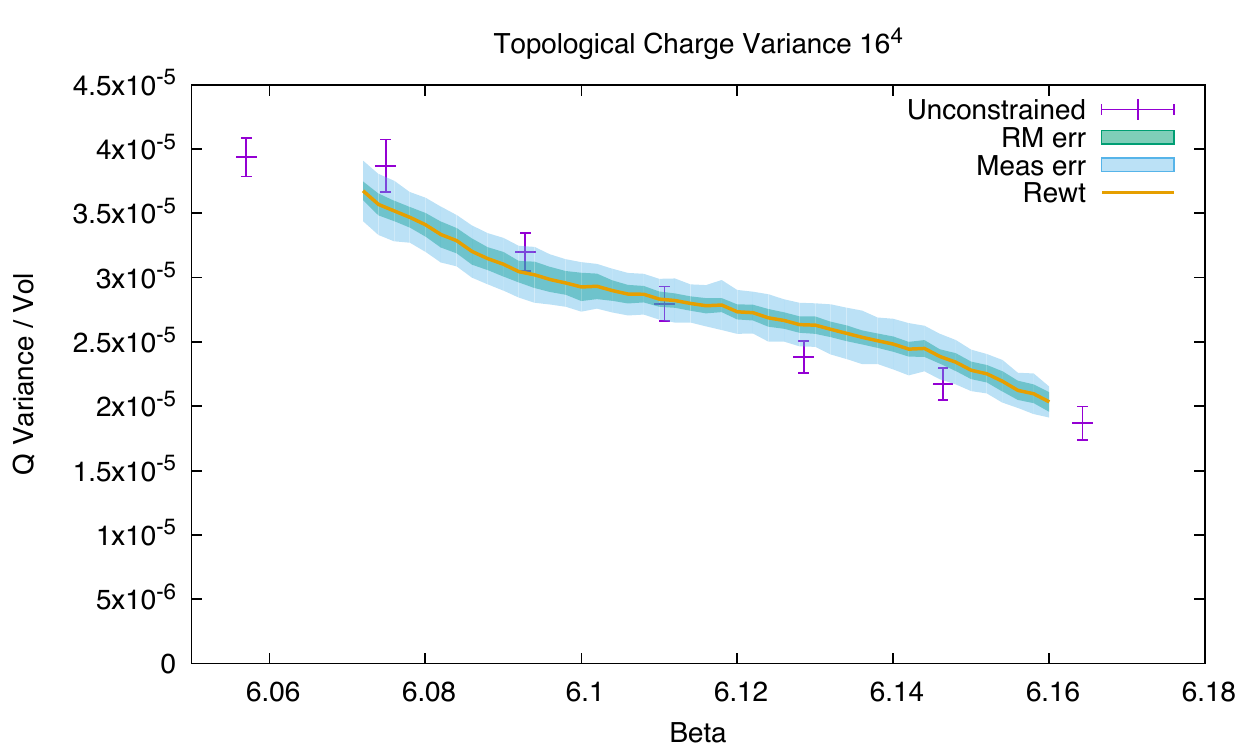}\par\medskip
\includegraphics[scale=0.65]{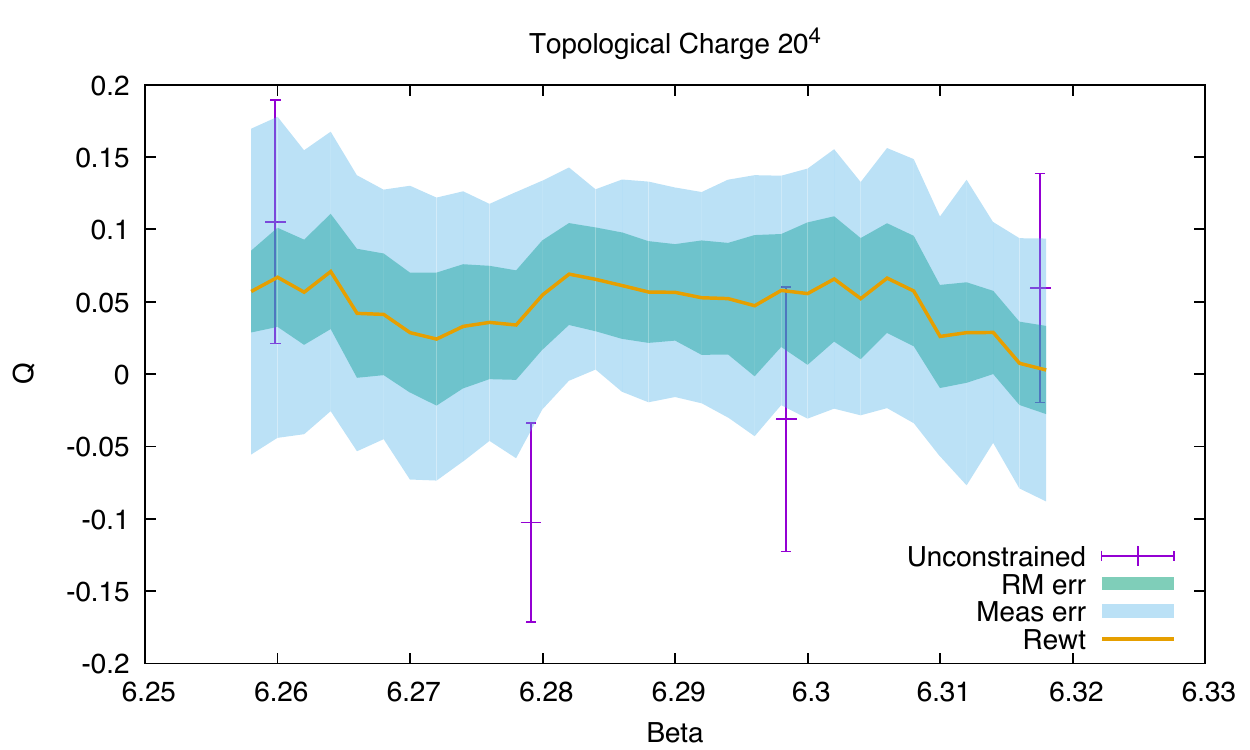}\hfil
\includegraphics[scale=0.65]{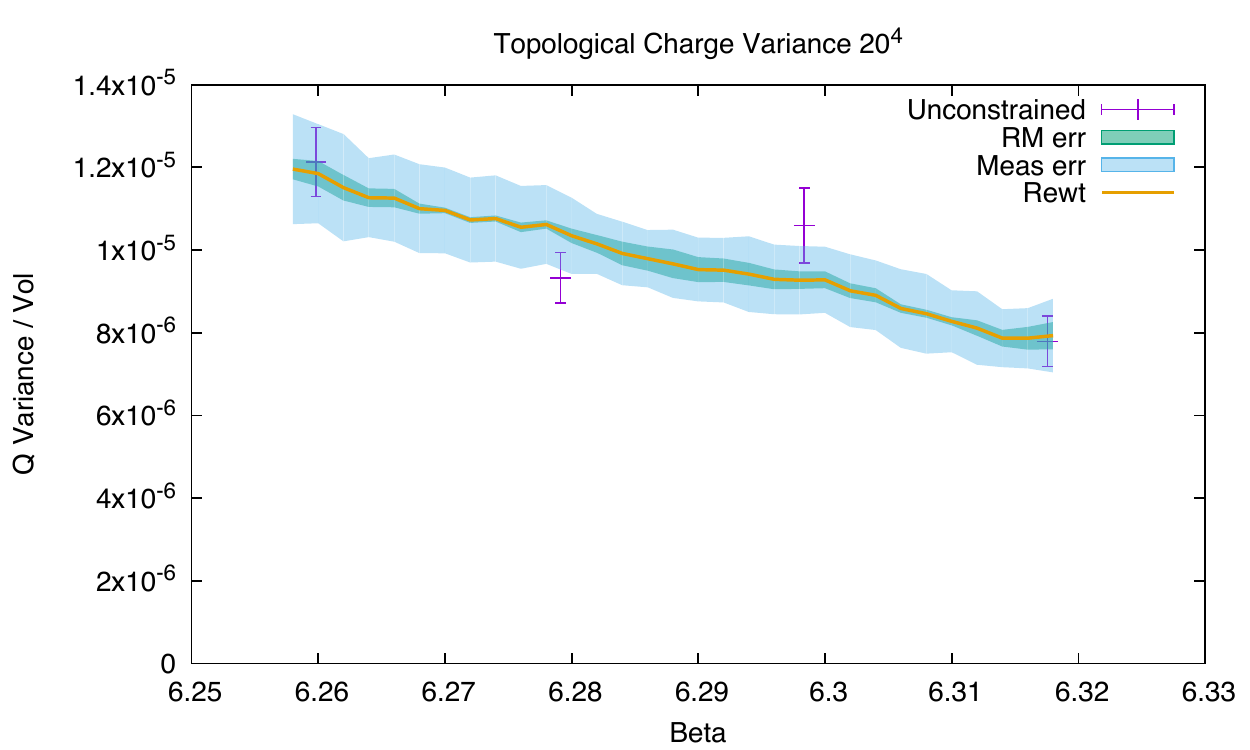}\par\medskip
\includegraphics[scale=0.65]{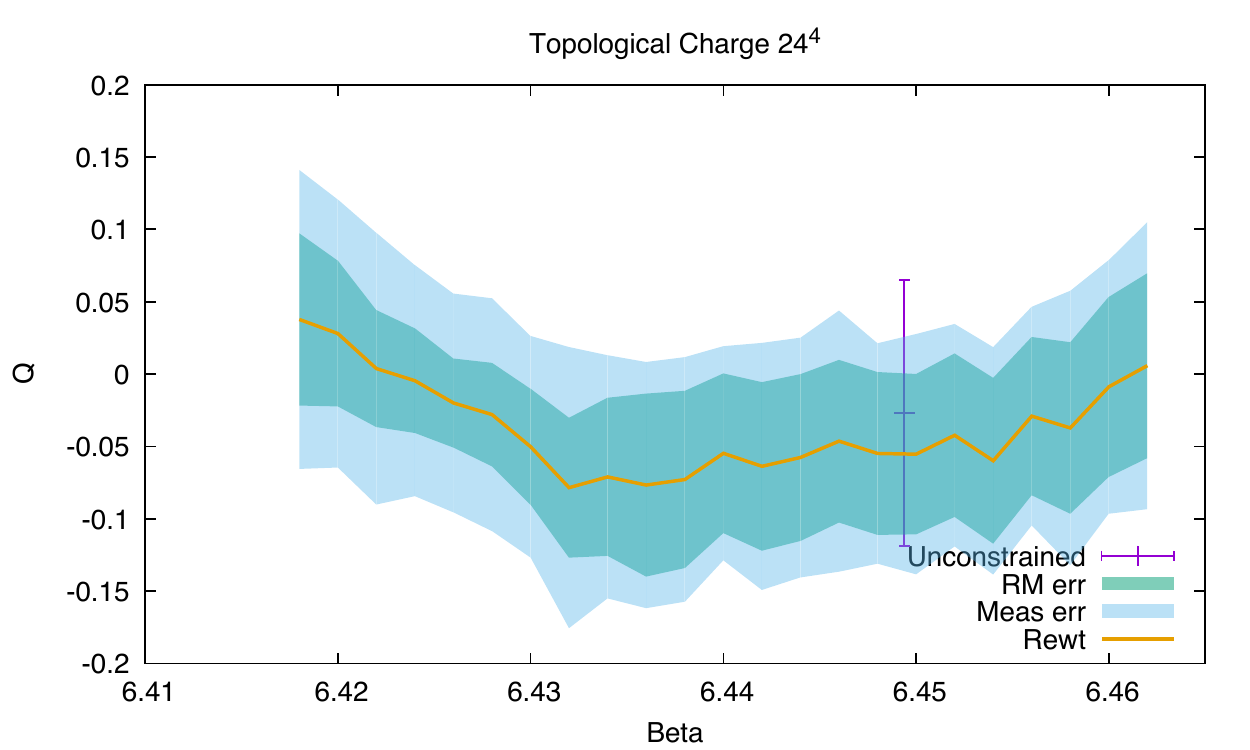}\hfil
\includegraphics[scale=0.65]{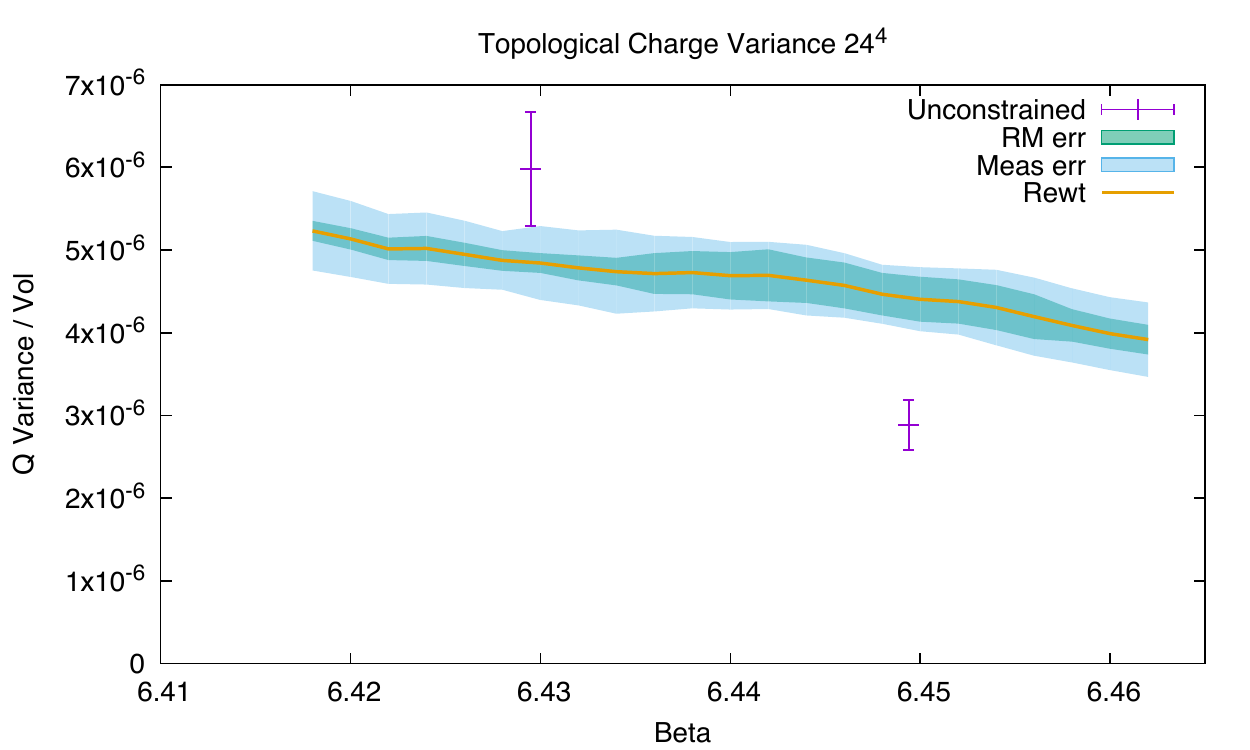}
\caption{
The topological charge based on
reweighted measurements. Also shown for comparison
are results from unconstrained simulations with additional runs.
Left: the charge itself which must be zero. Right: the variance of the charge.
From top to bottom: $16^4$, $20^4$, $24^4$.
Errors from the measurement and the RM phase are 
shown as shaded regions.
The Q data point for $24^4$ at $\beta=6.43$ is off plot.
}
\label{fig-TC}
\end{figure*}

\subsection{Topological Charge}

We devote a whole subsection to this topic because 
a strong motivation for studying alternative approaches such as density
of states is the freezing of the topological charge in the continuum
limit of traditional simulations.

The expectation of the topological charge 
must of course vanish
and the left column of figure \ref{fig-TC} shows that
 the reweight\-ed measurements always contain zero within the error range, 
while the rather longer traditional simulations have not yet met this requirement.
Vanishing charge is often regarded as a check that simulations are sufficiently
long and the widely varying sizes of the error bars for the variance 
of the traditional simulations at larger sizes,
even with the help of supplemental runs,
 indicates that these simulations are
not yet ergodic. 
Even the configurations of the measurement phase simulations
rarely change charge at large size, but the swaps lead
to much more rapid change of fixed central energy trajectories.
These qualitative considerations are more formally studied with the
autocorrelator below.

The effect of Wilson flow on the topological charge is well 
known \cite{Luscher:2010iy}  and for measurements
on either traditional unconstrained or constrained simulations
the flowed charge bunches close to integer values.
Several tempers that can have different 
(each almost integer) values of the charge contribute to the reweighted
charge which is no longer expected to be close to an integer.

The variance or topological susceptibility is shown in the
right  column of figure \ref{fig-TC}. Note that the scale varies 
and that the size of the errors of the reweighted results
actually decreases as the lattice size increases.

Tables  \ref{tab-TCerrs16}, \ref{tab-TCerrs20} and \ref{tab-TCerrs24},
show detailed results for a single value of $\beta$ from the centre of the
range at each lattice size to illustrate the relative size of the
RM and MP error
and to expose the relation between reweighting with quenched 
or fluctuating weights.
The values predicted 
for mean or variance of the topological charge
by either type of reweighting is close.
The unconstrained results reweighted according to
multi-histogram always have large errors and
 become poorer at larger lattice size.
Note that, as described in the next subsection, 
$\tau_{int}$ is computed using a different criterion 
for fluctuating weight than
for quenched weights.
Both RM and MP errors for each quantity are fairly similar
for quenched and unquenched. MP errors are in all cases
larger than RM errors.

The autocorrelation time and the dramatic difference between the
its value for quenched and fluctuating weights is 
the subject of the next subsection. 

\begin{table}[!htb]
\begin{center}
\caption{
Topological charge observables for $16^4$ at $\beta = 6.118$.
All tables are computed using reweighted measurements with 
quenched weights (QRewt) or fluctuating weights (Rewt).
Entries for reference simulations reweighted according to multi-histogram
(see~\ref{appendixb}) are given for comparison.}
\begin{tabular}{|l|c|c|c|c|}
\hline
Obs & Method & value & RM-err & MP-err \\
\hline
TC                & QRewt    & 0.03 & 0.04  & 0.07 \\
TC                & Rewt       & 0.02  & 0.04  & 0.06 \\
TC Variance & MultiHist   & 1.72 & 0.06 & - \\
TC Variance & QRewt    & 1.82 & 0.05 & 0.13 \\
TC Variance & Rewt       & 1.83 & 0.04 & 0.13 \\
TC   $\tau_{int}$   & QRewt     & 91 & 8& 27\\
TC   $\tau_{int}$   & Rewt        & 26   & 5 & 9\\
\hline
\end{tabular}
\label{tab-TCerrs16}

\caption{
Topological charge observables for $20^4$ at $\beta = 6.290$.}
\begin{tabular}{|l|c|c|c|c|}
\hline
Obs & wts & value & RM-err & MP-err \\
\hline
TC                & QRewt    & 0.05 & 0.04  & 0.08 \\
TC                & Rewt      & 0.06  & 0.03  & 0.07 \\
TC Variance & MultiHist    & 1.44 & 0.39 & - \\
TC Variance & QRewt    & 1.53 & 0.06 & 0.15 \\
TC Variance & Rewt       & 1.53 & 0.05 & 0.12 \\
TC   $\tau_{int}$   & QRewt     & 170 & 11& 63\\
TC   $\tau_{int}$   & Rewt      & 50   & 5 & 22\\
\hline
\end{tabular}
\label{tab-TCerrs20}

\caption{
Topological charge observables for $24^4$ at $\beta = 6.438$.}
\begin{tabular}{|l|c|c|c|c|}
\hline
Obs & wts & value & RM-err & MP-err \\
\hline
TC                & QRewt    & -0.08 & 0.07  & 0.10 \\
TC                & Rewt       & -0.07 & 0.06  & 0.09 \\
TC Variance & MultiHist    & 2.57 & 0.94 & - \\
TC Variance & QRewt    & 1.59 & 0.10 & 0.20 \\
TC Variance & Rewt        & 1.59 & 0.09 & 0.14 \\
TC   $\tau_{int}$   & QRewt     & 264 & 83& 128\\
TC   $\tau_{int}$   & Rewt       & 89   & 30 & 52\\
\hline
\end{tabular}
\label{tab-TCerrs24}

\end{center}
\end{table}

\subsection{Autocorrelation}
\label{autocorr}

\begin{figure}
\includegraphics[scale=0.65]{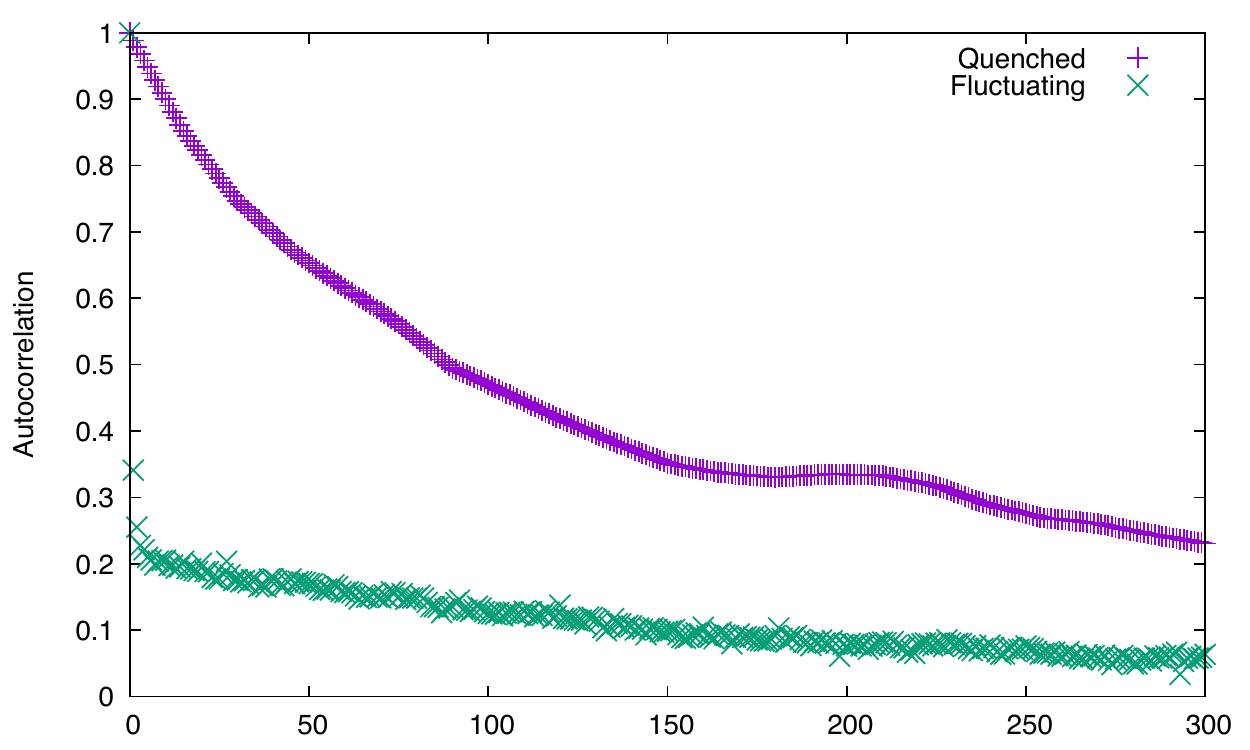}
\includegraphics[scale=0.65]{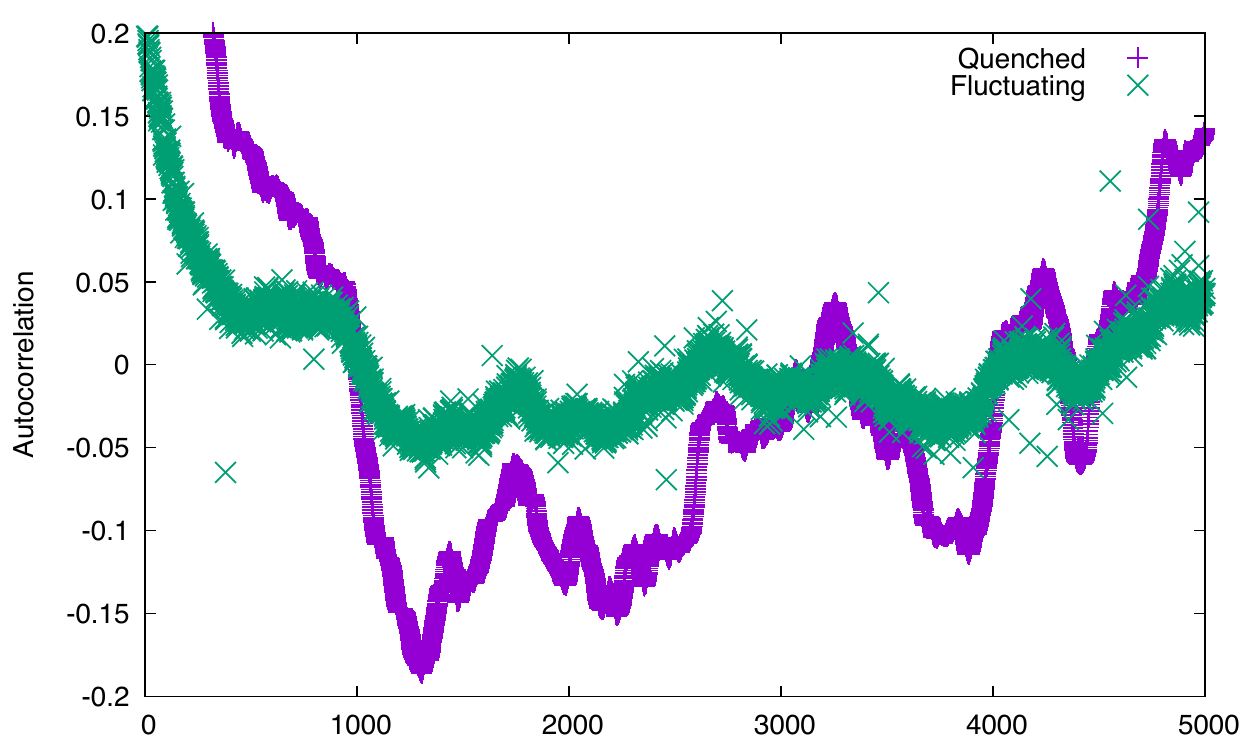}
\caption{
Quenched and fluctuating autocorrelation functions for the topological
charge for a representative example at $\beta = 6.290$ for $20^4$.
Both functions are normalised.
Top: over short times where 
for the fluctuating weight autocorrelator
the first few points show the rapid 
decay component.
Bottom: over long times showing typical fluctuations. 
}
\label{fig-staticdynamicQcorr}
\end{figure}

The autocorrelation time determines the size of errors and so is used
as a way of evaluating the efficiency of competing algorithms. 
The particular autocorrelation time relevant to computing errors is
the integrated form, $\tau_{int}$ as distinct from $\tau_{exp}$
which characterises the long time decay of the correlation function.
This distinction is emphasised by Madras and Sokal \cite{Madras:1988ei},
but a connection appeared in more recent work by Wolff \cite{Wolff:2003sm}
(also see \cite{Schaefer:2010hu}) as an estimate for $\tau_{exp}$ helps determine
an optimal value for the upper limit of the sum defining $\tau_{int}$.
It turns out that the present work illuminates the different definitions as
it allows us to give an estimate for $\tau_{exp}$ in a case where 
$\tau_{int}$ and $\tau_{exp}$ are very different.

\begin{figure}
\includegraphics[scale=0.65]{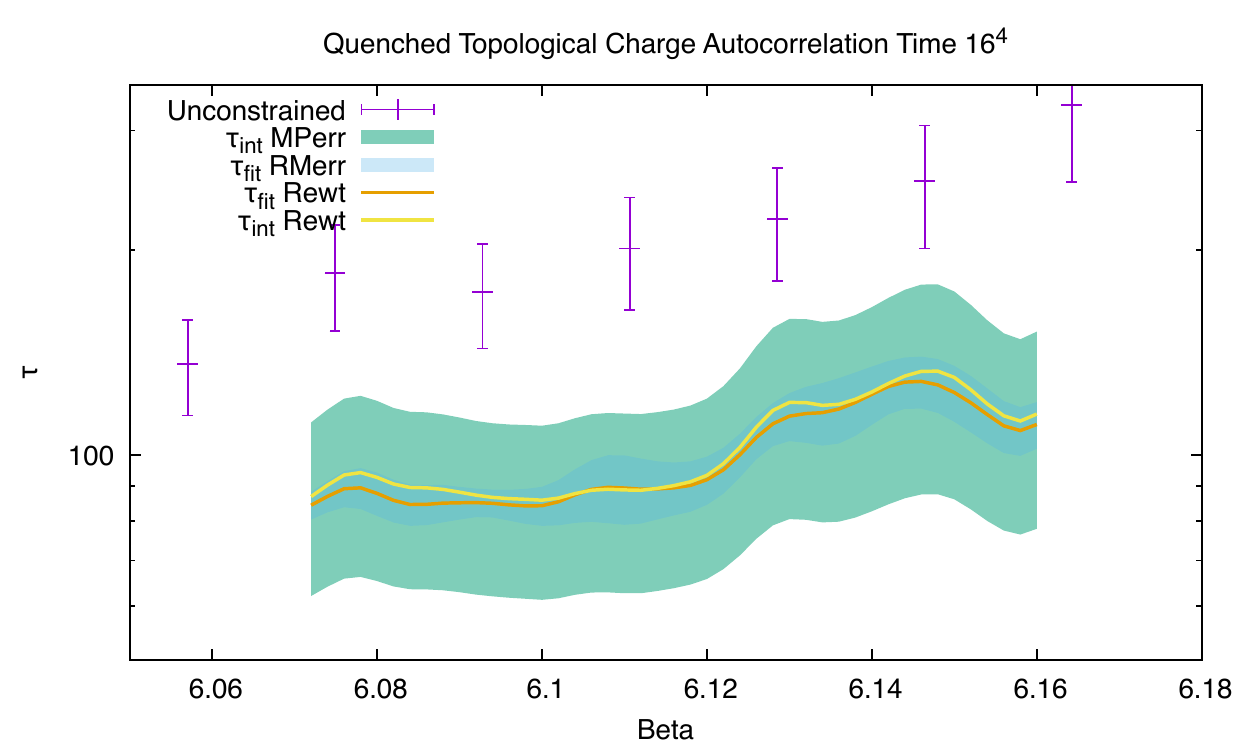}
\includegraphics[scale=0.65]{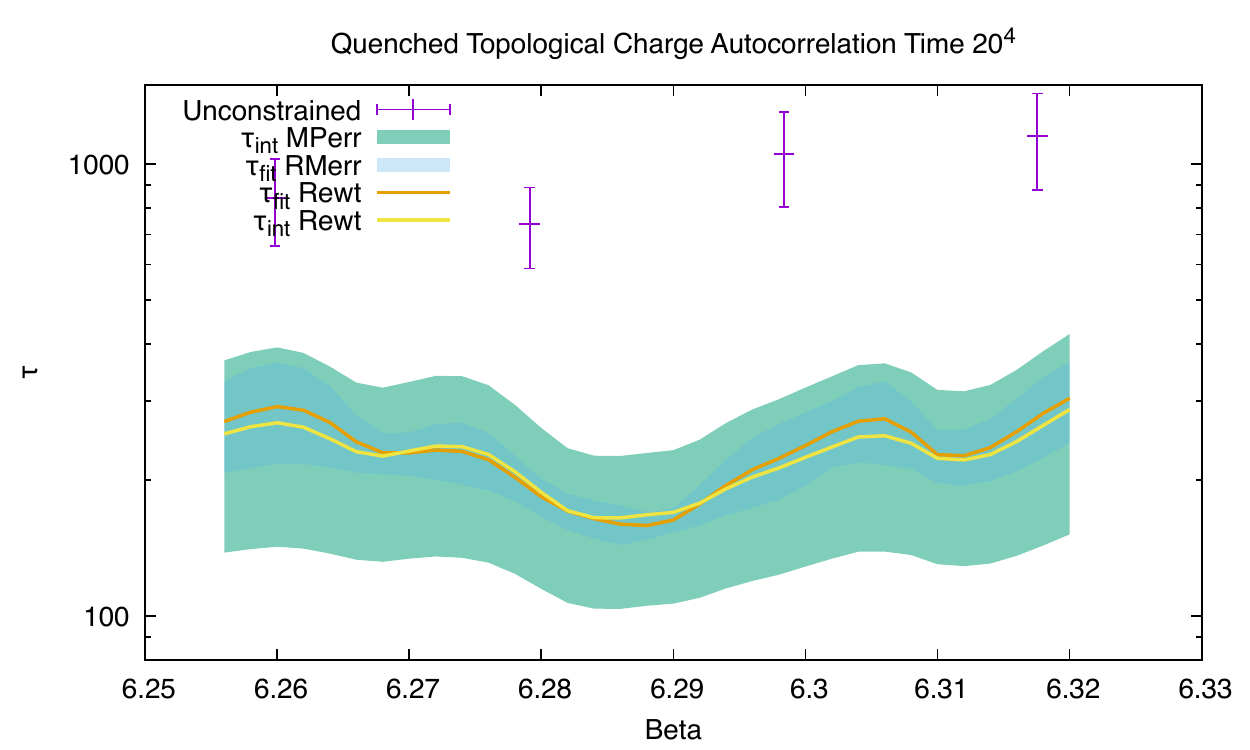}
\includegraphics[scale=0.65]{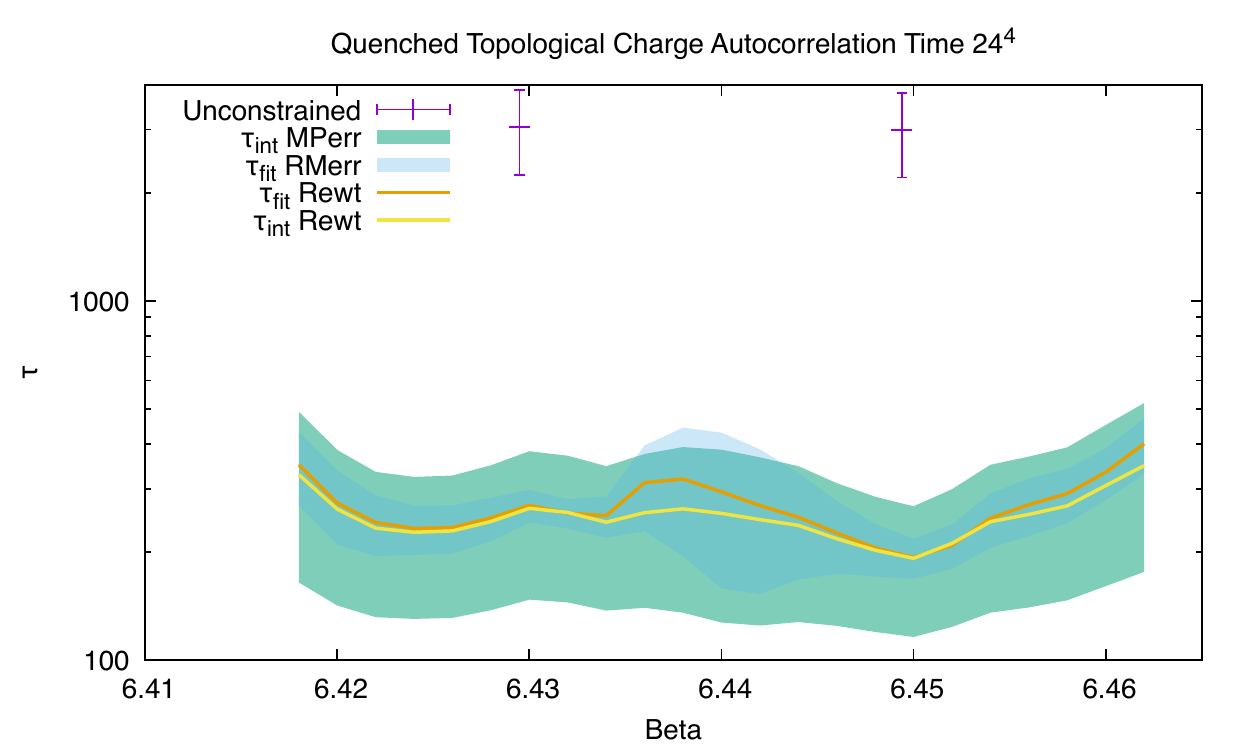}
\caption{
Autocorrelation times obtained from the 
autocorrelation function for topological charge
based on reweighting measurements with fixed weights.
Showing the 
integrated $\tau_{int}$ and the fitted $\tau_{fit}(A=1)$ as
described in the text for, from top to bottom, $16^4$, $20^4$ and $24^4$.
Also shown for comparison
are results from unconstrained simulations.
For each $\tau$ the largest source of error is 
shown with a shaded region. 
The RM errors of each $\tau$ are of similar size 
but the MP error for $\tau_{int}$ is larger. 
}
\label{fig-Statictau}
\end{figure}

First observe that the autocorrelation functions for the quenched and
fluctuating weight approaches have distinct character. 
Typical autocorrelation functions, $\Gamma(t)$, are shown in 
figure  \ref{fig-staticdynamicQcorr}. Each autocorrelation function is
normalised to $\Gamma(0)=1$ and it is clear that 
while the quenched approach leads to the usual decay, 
the fluctuating weight approach suffers from a rapid decay over a few HMC steps followed
by a component with a long timescale. The techniques used to
analyse these cases are different.

\subsubsection{Quenched Autocorrelation}
\label{sec-quenched}
The top curve of the upper plot of figure 
\ref{fig-staticdynamicQcorr}, shows that the 
quenched autocorrelation function 
for the topological charge
has the usual long timescale decay. 
We first review the standard technique for computing $\tau_{int}$ 
by analysing this usual case.

The 
expression for the
integrated autocorrelation time is:
\begin{equation}
\tau_{int}(W) = {1\over 2} + \sum_{t=1}^{W} {\Gamma(t)\over \Gamma(0)}  \ .
\end{equation}

We follow the prescription of Wolff \cite{Wolff:2003sm} to 
determine the window, $W$, by minimising the combination of
truncation and statistical errors, estimated as:

\begin{equation}
e^{-W/\tau_{exp}} + 2\sqrt{W\over T}  \ .
\label{Wolffmin}
\end{equation}

Where $T$ is the duration of the measurement simulations, 20000 steps.
For smooth autocorrelation functions
without timescales much longer than $\tau_{int}$,
the exponential autocorrelation time, $\tau_{exp}$ appearing in this expression, is
estimated as $S\, \tau_{int}$ with the parameter $S$ in the range $1\sim 2$.
All quenched reweighted autocorrelation functions 
behave like the example shown in figure \ref{fig-staticdynamicQcorr}
without any indication of timescales longer that 
$\tau_{int}$ which is of the order of 100's of HMC steps.
So we take the conventional value $S = 1.5$.

As a rough justification for this approach and
the one we will follow for fluctuating weights,
we consider a direct fit of 
the normalised autocorrelation function to
a single exponential decay: 

\begin{equation}
A  e^{-t/\tau_{fit}}   \ .
\label{singleexp}
\end{equation}

We first make a fit in which we set $A=1$, thus guaranteeing
the normalisation and the short
time behaviour that appears in $\tau_{int}$.
Fits are made over the range 1 to a few 1000's and
we  have checked that the value of the right hand cut has negligible effect on the
fit parameters. 
The resulting averaged values for $\tau_{fit}$ are shown,
in figure \ref{fig-Statictau}. 
The similarity between $\tau_{fit}(A=1)$ and  $\tau_{int}$ is striking.

An alternative fit in which $A$ is also a fitting parameter leads to a
rather longer $\tau_{fit}(A)$ which resembles $\tau_{exp}$
according to the hypothesis  $\tau_{exp} = S\,\tau_{int}$
and provides reassurance that the choice $S=1.5$ is reasonable. 
We do not
attempt to tune $S$ using this fitting information as this would have little
effect on the resulting value of $W$ or $\tau_{int}(W)$.


Figure \ref{fig-Statictau} also shows error estimates for 
$\tau_{int}$ and $\tau_{fit}(A=1)$. For a single replica, 
the error of the integrated autocorrelation time is:

\begin{equation}
\Delta\tau_{int} = 2\sqrt{W\over T}\tau_{int}  \ .
\label{Deltatauint}
\end{equation}

We regard this as the MP error and observe that it
is far larger than the error for the fitted version,
$\Delta\tau_{fit}$, which arises from uncertainty in the fitting procedure.
The RM errors
are similar for  each. 
It is  unsatisfactory that the measurement phase statistical errors 
of two ways of computing the autocorrelation time 
are so different. The fitting error of $\tau_{fit}$ appears to
be  underestimated.
Indeed, when the
measurement phase is repeated for a replica with identical RM parameters
but a different sequence of random numbers,
there is a significant difference in the results of the fit which
are well beyond the error estimate $\Delta\tau_{fit}$.
\begin{figure}
\includegraphics[scale=0.65]{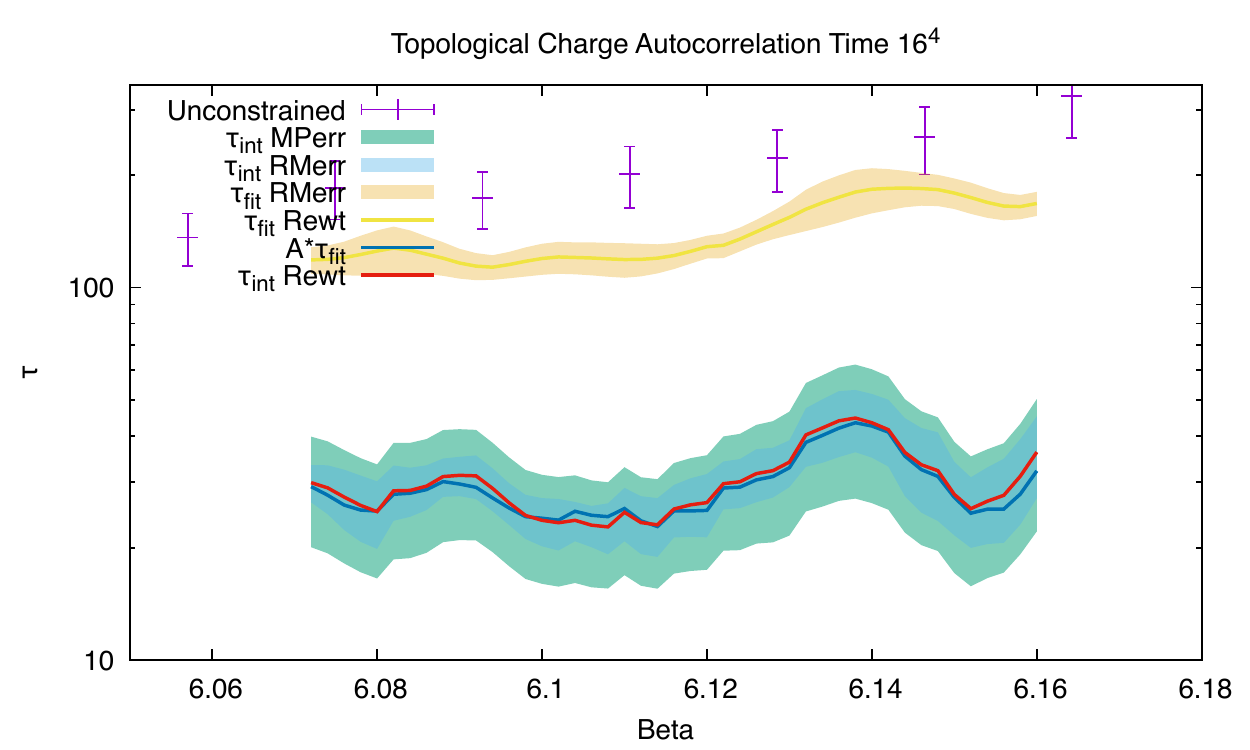}
\includegraphics[scale=0.65]{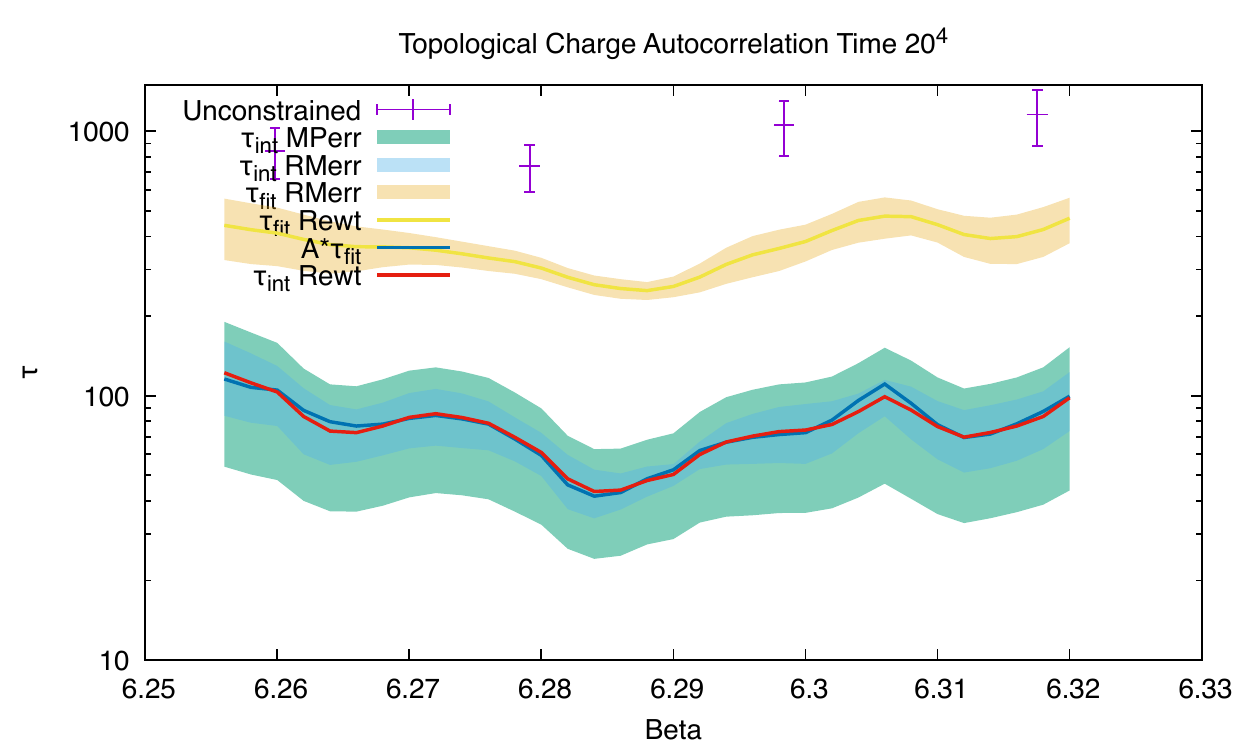}
\includegraphics[scale=0.65]{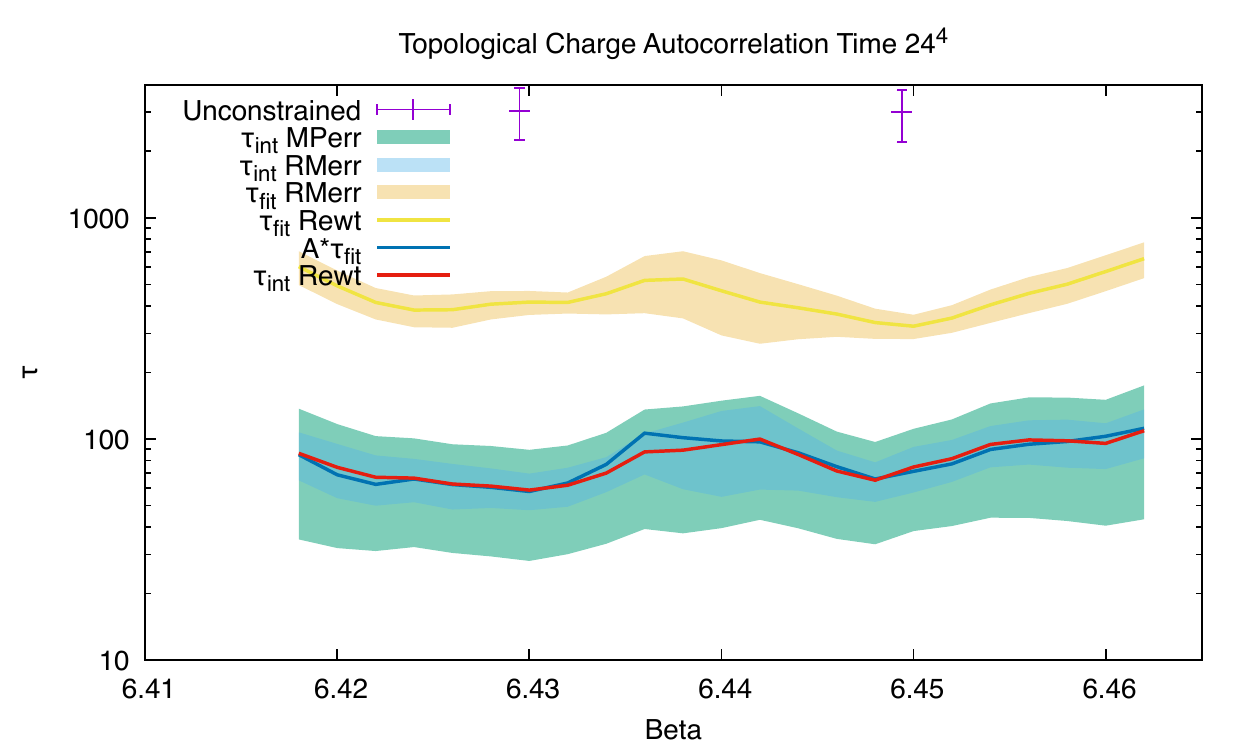}
\caption{
Autocorrelation times obtained from the 
autocorrelation function for topological charge
based on reweighting measurements with fluctuating weights.
Showing the 
integrated $\tau_{int}$, the fitted $\tau_{fitL}$ 
and the combination $A\,\tau_{fitL}$
as described in the text for, from top to bottom, $16^4$, $20^4$ and $24^4$.
Also shown for comparison
are results from traditional simulations.
For each $\tau$ the RM error and the MP error
are shown with shaded regions. The MP error 
for $\tau_{int}$ is dominant, the RM errors of each
$\tau$ are of similar size and the statistical fitting error of
 $\tau_{fitL}$ is smallest and omitted from the plot.
}
\label{fig-Dynamictau}
\end{figure}

\subsubsection{Fluctuating Weight Autocorrelation}
\label{flutwtautocorr}

The action or plaquette observable has a short
autocorrelation time  never more than a few
HMC steps irrespective of exactly which $\tau$ is computed.
This short timescale appears in the varying
weights (\ref{wt-wit})
and pollutes any long timescale that may be present
in any observable computed by reweighting with these weights.

In the case of the topological charge the fluctuations of the
weights do not entirely hide the long timescale component. 
Figure \ref{fig-staticdynamicQcorr}
shows an example of the fluctuating weight autocorrelation function for the topological charge over both short and long times.
This autocorrelation function
suffers a rapid decay over a few HMC steps, but 
clearly leaves a well distinguished long timescale component. 
This feature is observed for all RM replicas 
and all lattice sizes, though with differing amplitude for the long
component. Over the longer timescales
shown in the lower plot of figure 
\ref{fig-staticdynamicQcorr}, the varying weight autocorrelation
function continues to have short time fluctuations that manifest as the
greater thickness of the line compared with the quenched version. 
It is also apparent from this plot that  long timescale
fluctuations persist for both autocorrelation functions but 
with larger amplitude for the quenched than
the fluctuating weight case.

The Wolff procedure described in the previous subsection for computing 
the integrated autocorrelation time for fixed weights
is not appropriate for the fluctuating weight results with mixed decay times.
Indeed, reference \cite{Wolff:2003sm} suggests that the choice of parameter $S$ 
must be reexamined in this situation.
We have considered many alternative approaches with the aim of finding a 
robust prescription.
Our final approach is to directly fit
the following normalised double exponential decay according to the two stages 
described below.

\begin{equation}
(1-A) e^{-t/\tau_{fitS}} + A  e^{-t/\tau_{fitL}}   \ .
\label{doubleexp}
\end{equation}

Fitting multiple exponentials has a reputation of being fraught with problems, 
but in this case the
two $\tau$ parameters take such different values, 
$\tau_{fitL}/\tau_{fitS} > 100$,
that the technique is
robust for the variety of autocorrelation functions arising from different
RM replicas, $\beta$ and lattice sizes. Nonetheless, the procedure for the
fit is in two stages.

Firstly, all three parameters of the function are fitted over a short range,
up to a cut at about 300 steps. 
Secondly, with short timescale parameters $A$ and
$\tau_{fitS}$ frozen, the single parameter $\tau_{fitL}$ is fitted over a much longer
time range of  more than 1000 steps. 
We find that the dependence of $\tau_{fitL}$ 
on the short range cut  is smaller than the statistical errors, and
that there is negligible dependence on the long range cut
when it is taken to be sufficiently large. The same cuts are used
for all RM replicas  but they take different values according to lattice size
($16^4: 1500, 20^4: 3000, 24^4: 4000$).

The integrated autocorrelation time 
for a double exponential decay with such widely separated timescales 
will, for sufficiently large window, be dominated by the long timescale
and we expect,

\begin{equation}
\tau_{int} \approx A\, \tau_{fitL}\ .
\label{dyntauint-fit}
\end{equation}

A more accurate equation involving logs along the lines of the one discussed
in \cite{Wolff:2003sm,Schaefer:2010hu} can be derived, but
since our approach relies on widely separated scales, the simpler
equation (\ref{dyntauint-fit}) is adequate.

In view of equation (\ref{dyntauint-fit}), we set the value of $S$ used 
to estimate $\tau_{exp}$ appearing in (\ref{Wolffmin}) 
to $S=1/A$ and follow the Wolff procedure.
Figure \ref{fig-Dynamictau} shows $\tau_{int}$ computed according to
this technique along with $A \tau_{fitL}$ from the fit. 
The MP error for $\tau_{int}$ is computed using (\ref{Deltatauint})
and 
as for the quenched case, 
it is large, making the agreement all the more remarkable.
This value of $\tau_{int}$ is used to estimate errors in the topological
charge measurements and these appear in
tables  \ref{tab-TCerrs16}, \ref{tab-TCerrs20}, \ref{tab-TCerrs24}, and Figure 
\ref{fig-TC}.

The proportionality factor $A$ introduced in (\ref{doubleexp})
 can be interpreted as the coupling to the long
timescale.
Its value of around $0.2\pm 0.05$ 
does not appear to be very sensitive to the value of the reweighting
coupling $\beta$ or even to the size of the lattice. 
We expect it to depend on the parameters of the density of states method,
in particular the ratio $\delta / \sigma$.

\subsubsection{Comments}

While the shape of the autocorrelation function is very different for
the quenched and the fluctuating weights, it is interesting to see if the long
time behaviour matches. In other words, is $\tau_{exp}$ the same?
Our best estimates of this quantity are based on fits: 
$\tau_{fit}(A)$ in the quenched case and
$\tau_{fitL}$ for fluctuating weights. 
A comparison of these quantities is not shown, but
although there are large fluctuations, 
they are compatible within the errors.

To summarise this discussion of the autocorrelation time of
the topological charge: 
we have used information from the fit of a double exponential
to the fluctuating weight
autocorrelation function in order to tune the parameter $S$, 
and obtain a better value for the Wolff window leading to a
more accurate prediction for  $\tau_{int}$.
We find  relations between the integrated and fitted autocorrelation times. 
For the quenched case: $\tau_{int} \approx \tau_{fit}(A=1)$, and
for varying weights: $\tau_{int} \approx A \tau_{fitL}$. Finally, we
observe that our estimates for the long time $\tau_{exp}$ behaviour of either
quenched or fluctuating weights  is similar. 

We conclude this section by clarifying 
how these insights carry over to the computation of $\tau_{int}$
for other observables.
The quenched case is straightforward and the standard Wolff
procedure outlined  at the start of section \ref{sec-quenched}
for the topological charge is used for all observables.
For the fluctuating weight case,
even for quantities such as $E_{sym}$ after Wilson flow that have 
$\tau_{int} \approx 50\sim 60$ for unconstrained simulations, 
there is no clear long
timescale remaining that would
allow any improved estimate of $\tau_{exp}$. Hence $S$
can take the conventional value, $S=1.5$ and we follow the ordinary
Wolff procedure outlined at the start of section \ref{sec-quenched}.

\section{Scaling}
\label{scaling}

The analysis reported so far has involved
figures presenting densities against $\beta$ and
with the exception of Figure \ref{fig-overall}, have been for
a single lattice size. In this section, scaling properties are
discussed and the density of states method is used to 
illustrate the approach to the continuum
 at fixed box size.

As was discussed in section \ref{observables},
Wilson flow  introduces a new scale into the system; the
parameter $t$ specifying the endpoint of the flow
is dimensional and 
equation (\ref{eqn-WFt}) shows how it
represents the scale over
which the configuration is smoothed. 
Our simulations, both constrained and un\-con\-strain\-ed
choose $t$ in such a way that
that the scale of smoothing is a fixed fraction of the
size of the lattice. 
Then, by comparing results for couplings, $\beta$,
corresponding to the same physical box size,
the Wilson flow scale is physically identical and quantities
on different lattice sizes can be compared. 

\begin{figure}
\includegraphics[scale=0.65]{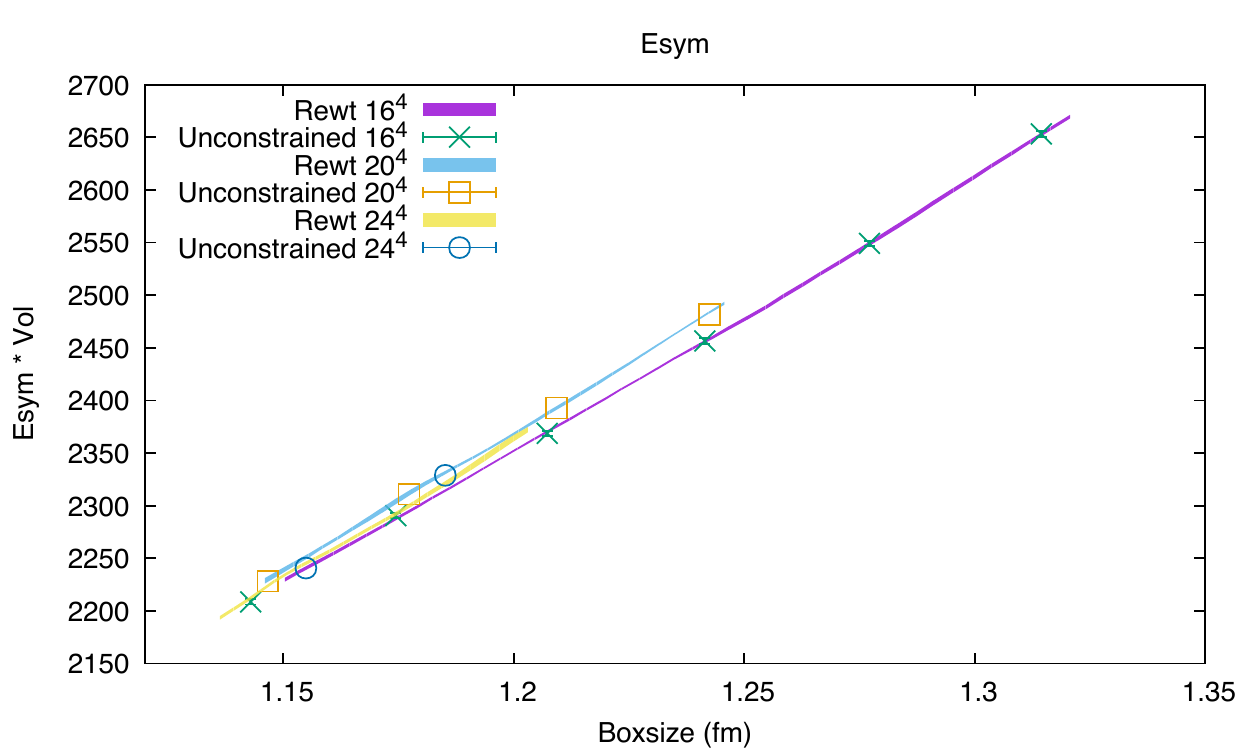}
\includegraphics[scale=0.65]{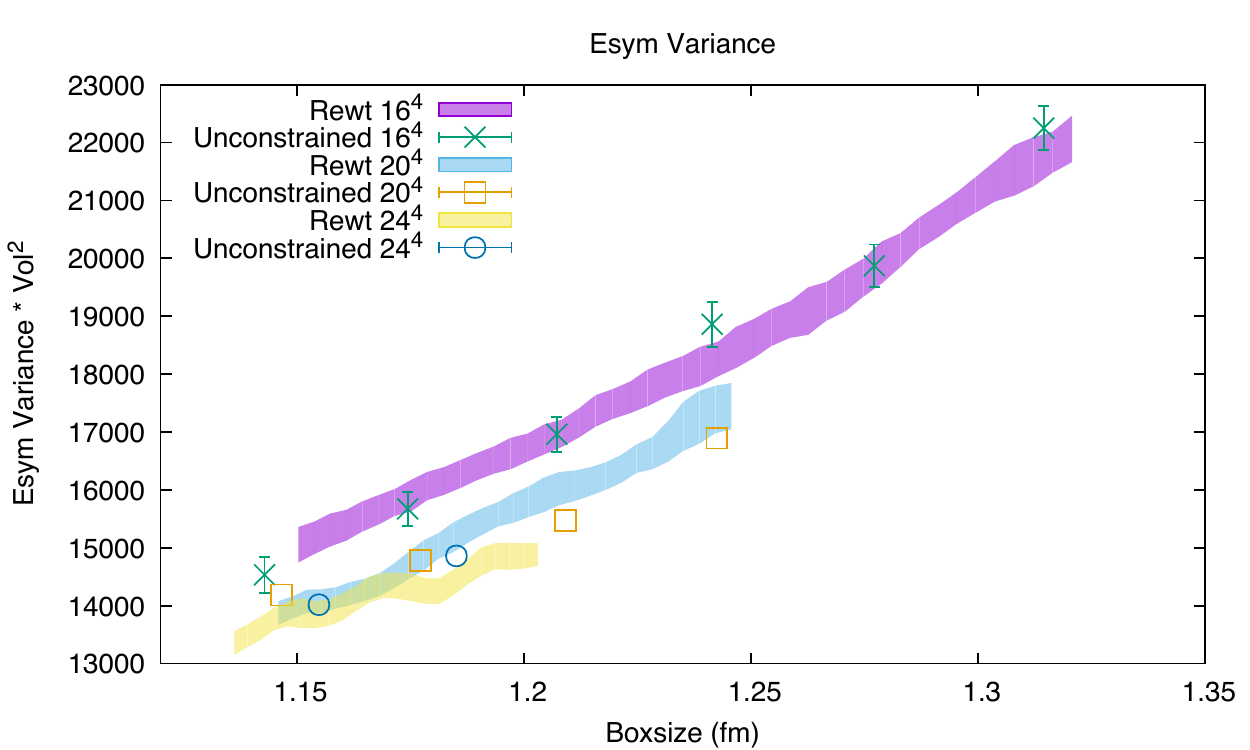}
\caption{
Scaling plot of the symmetric clover link energy
based on reweighted measurements and also 
showing traditional results. Three lines correspond to 
data from the three lattice sizes. 
The shaded areas represent the MP errors which are 
slightly larger than the RM errors (not shown).
}
\label{fig-scale-Esym}
\end{figure}

\begin{figure}
\includegraphics[scale=0.65]{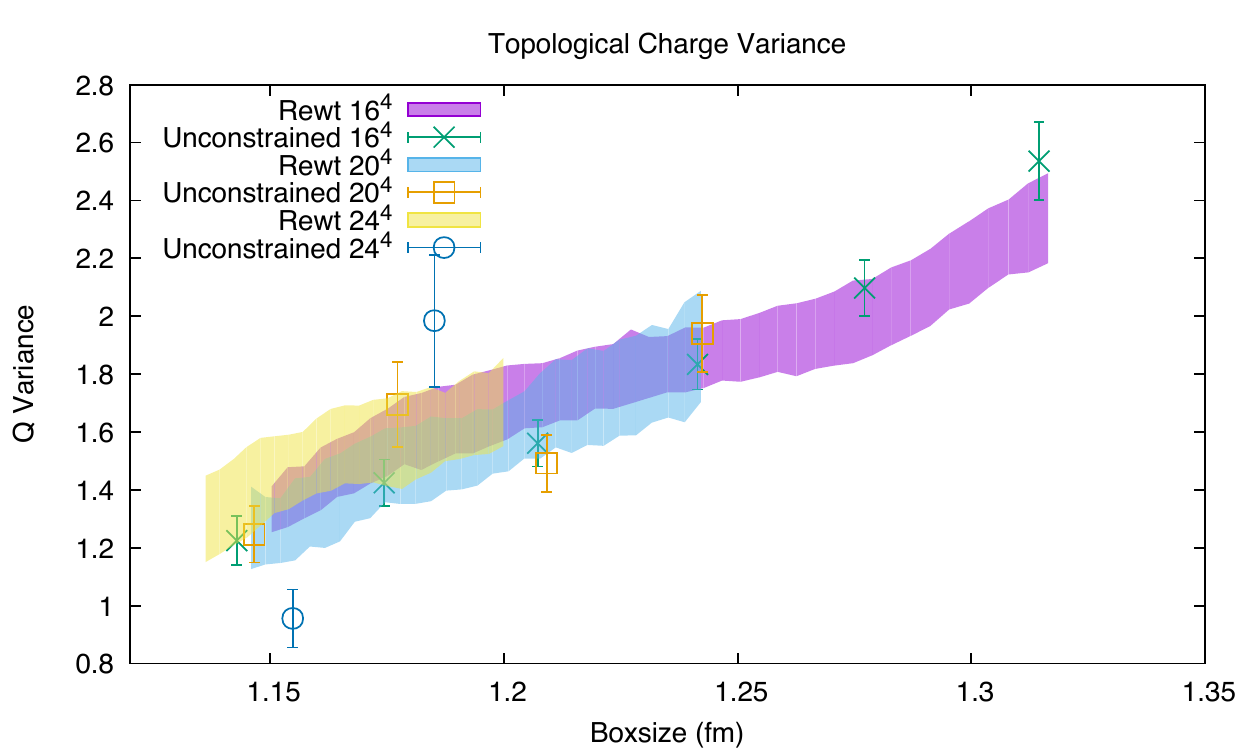}
\caption{
Scaling plot of the Topological charge susceptibility
based on reweighted measurements and also 
showing traditional results. Three lines correspond to 
data from the three lattice sizes. 
The shaded areas represent the MP errors which are 
considerably larger than the RM errors (not shown).
}
\label{fig-scale-TC}
\end{figure}

Figures \ref{fig-scale-Esym} and \ref{fig-scale-TC} 
plot global quantities for the box, versus the boxsize and display this
for different lattice resolutions.
Namely, the figures show the global observables 
$V E_{sym}$, $V^2 {\rm Var}(E_{sym})$ and $Q^2$.

The scaling of the symmetric clover link energy at the top of 
figure \ref{fig-scale-Esym} 
is almost linear and all data from different size lattices and from
both traditional and density of states simulations coincide.
The variance of this quantity displayed in the lower part of figure \ref{fig-scale-Esym} 
shows noisier data and the
lines of reweighted data from different size lattices do not quite coincide.

Figure \ref{fig-scale-TC} shows the topological charge variance scaling.
As was the case in figure \ref{fig-TC}, the data points from the traditional
simulations are noisy and especially for larger lattices,
 indicate worse ergodicity than the density of states data.

\section{Conclusions}\label{conclusions}

\begin{figure}
\includegraphics[scale=0.65]{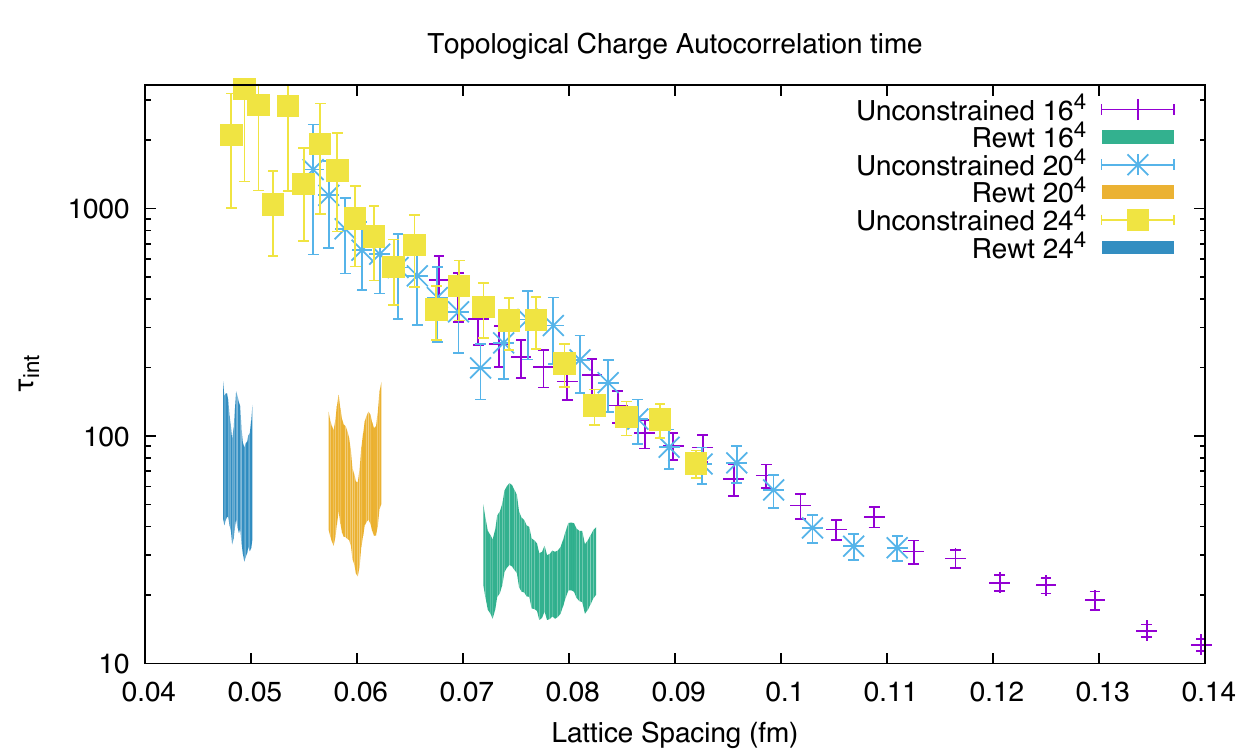}
\caption{
Overall plot of autocorrelation times for topological charge
both for unconstrained
simulations and density of states.
The band of data points shows $\tau_{int}$ for traditional simulations for
each lattice size. The shaded regions display $ \tau_{int}$
computed using reweighting
and indicate the 
statistical error which dominates the RM error.
}
\label{fig-all-TC}
\end{figure}

We have presented a comprehensive and detailed study of the use
of the density of states method for SU(3) Yang Mills theory. To do this
we derived the reweighting formalism appropriate for a smooth Gaussian
constraint and 
emphasised that the expression for a canonical expectation value
is a ratio of weighted sums and that the weights fluctuate because 
they contain a term that depends on the action.
We explored the  quenched approximation 
in which this variation is neglected and found that
when the weights vary  there is a dramatic effect on autocorrelation
times which is absent for the quenched approximation. 
However we want to emphasise that as there is no theoretical basis for  
the quenched approximation and even though it appears to be reasonably accurate
for the expectation values of most observables, there is a notable  
exception, the plaquette variance, identifying
the approximation  as uncontrollable.

We compared predictions for the plaquette and its variance (which are
both simple functions of the action) from
the LLR approach based solely on reconstructing the density of
states obtained from RM phase, and the reweighting
approach based on information from the measurement phase.
We found that these were always compatible, though reweighting led to
greater uncertainty associated with the finite duration of the measurement phase.
We noted that the
plaquette variance is a delicate test of the accuracy of the technique
and indeed it is also a good check for multi-his\-to\-gram reweighting
of traditional unconstrained simulations.

For quantities that are defined after Wilson Flow, and for which
only the reweighting approach is possible, our method allows 
families of continuum limit trajectories to be defined.
These are characterised by the
boxsize associated with the value of the coupling constant for which we
are performing the reweighting.
Scaling of the topological charge susceptibility 
at finite temperature is an interesting area
of investigation \cite{Berkowitz:2015aua} and
we have demonstrated the possibility of going to reasonably small
lattice spacing with only modest size lattices and fairly short runs.

The main conclusion of the preliminary work \cite{Cossu:2017sfu}  
concerned the reduction in autocorrelation time for the topological
charge. The conclusions presented here are more nuanced.
Figure \ref{fig-all-TC} shows the growth of the autocorrelation time
for traditional simulations along with $\tau_{int}$ for
reweighted results in each of the regions where we have
computed it. The rate of growth 
of $\tau_{int}$ for traditional unconstrained simulations
is compatible with that reported in 
the high statistics study \cite{Schaefer:2010hu}; namely $z\sim 5$ for the model
$\tau_{int}\propto a^{-z}$. 
The density of states method yields a much smaller value for $\tau_{int}$
and more importantly, the rate of growth appears to be much less. 
The size of the  errors evident in the reweighting data shown in
figure \ref{fig-all-TC}  prohibit anything better than a crude estimate.  
Using three values corresponding to the same box size gives
$z\sim 1\pm 2.5$. While the uncertainty in this estimate makes it almost meaningless,
it is clearly smaller than the exponent for unconstrained simulations.
This is an exciting result as it offers the potential to compute 
accurate values of the topological charge with less effort.
However, the small value of $\tau_{int}$ is  mainly the consequence of
the reduction in the coupling of the
topological charge to the long timescale modes of the problem.
These modes have not gone away, and indeed long timescale
modes themselves are often the
primary subject of interest in studies of the relative performance of 
different algorithms rather than the topological charge which is
frequently used simply as a proxy for studying these modes. 
In the text we discussed
our estimate of $\tau_{exp}$
which is a better measure of the long timescales of the problem
(though, as always, there may be even longer timescales beyond
the duration of the simulations).
This does show a decrease from the
equivalent parameter of standard HMC simulations
We highlight that the changes to the HMC algorithm
necessary for the density of states approach,
namely the Gaussian constraint and tempering swap, do not affect the scaling
behaviour of the timing of our implementation with the number of lattice points.

A notable aspect of this study is that we explored the errors of the
method by tracking 8  replicas for each lattice size.
We noticed that there was considerable variation in the 
convergence during the RM phase. Our study chose a simple
common time duration for this phase, but 
a more sophisticated temper dependent criterion for terminating the RM phase
that could put more effort into controlling the scaling convergence of the RM
process, might provide better overall accuracy.

We distinguished two sources of error: the RM errors and the 
MP errors.
For all results using reweighting
the relative contribution of RM and MP errors 
depend on the particular observable under study.  
The RM errors can be reduced by increasing the number of replicas
in the RM phase and the MP errors can be reduced
with a longer measurement phase.
For our choice of $N_{REPLICA}$  and duration of the measurement phase
the largest contribution to the error for 
most observables was from measurement. 

This work has kept $\delta/\delta_S$ fixed at 1.2 and scaled 
$\delta$ for different size lattices according to a fixed fraction of 
the standard deviation of the constrained action. 
These parameters affect the accuracy 
and efficiency of the method and a more detailed study would be
welcome.
For example, it would be interesting to investigate
the dependence of the topological charge susceptibility
on $\delta/\sigma$ while preserving the ratio $\delta/\delta_S$
(along the lines of figure 6 of reference
\cite{Langfeld:2015fua} which studies the dependence of 
$C_V$ in a U(1) theory),
and the coupling of the topological charge to the long timescale modes.
The number of tempers used in the 
tempering method was 24 for all lattice sizes. A wider range of tempers
would be expected to allow better mixing of regimes with
faster and slower Monte Carlo dynamics and thus improve the autocorrelation
time. As the lattice size increases, the spacing $\delta_S$ is expected
to grow as $\sqrt{V}$ in order to preserve the same level of accuracy.
The corresponding spacing in $\beta$ will therefore decrease in the same way.
If the efficiency of the tempering depends on the span of $\beta$, then
more tempers will be needed (with $N_{TEMPER} \sim \sqrt{V}$).
This scaling prediction and the 
effects of the length of the RM and measurement phases and the number of
replicas on the accuracy require further study.

\section*{Note Added}

 While this manuscript was in preparation, two studies appeared that
report on investigations of the topological charge. Ref.~\cite{Bonanno:2020hht}
performs a study in SU(6) with a novel algorithm where a tempering
procedure is used to move across systems with boundary conditions
interpolating between partially open and periodic; this algorithm is
proved to be successful at reducing the correlation time of the
topological charge. Ref.~\cite{Borsanyi:2021gqg} studies topological properties at
finite temperature using a density of states method that allows it to
sample rare events. The relative merits of our algorithm and those
new proposals would need to be assessed by dedicated studies. 

\begin{acknowledgements}
The work of BL is supported in part by the Royal Society Wolfson Research Merit Award WM170010 and by the STFC Consolidated Grant ST/P00055X/1. AR is supported by the STFC Consolidated Grant ST/P000479/1. Numerical simulations have been performed on the Swansea SUNBIRD system, provided by the Supercomputing Wales project, which is part-funded by the European Regional Development Fund (ERDF) via the Welsh Government, and on the HPC facilities at the HPCC centre of the University of Plymouth.
\end{acknowledgements}

\bibliographystyle{h-physrev}     
\bibliography{topology.bib}   

\appendix

\section{Multi-histogram}
\label{appendixa}

The density of states method reconstructs canonical expectation
values from a set of constrained expectation values at fixed
central action. It is natural to contrast this approach with
traditional multi-histogram reweighting  \cite{Ferrenberg:1989ui}
which reconstructs canonical expectation values 
at  intermediate couplings
from a set of unconstrained expectation values at fixed coupling.
The multi-histogram approach does not benefit from
the tempering update step and hence a timing comparison would be
meaningless.

Rather than make a detailed comparison, requiring additional
unconstrained runs at values of the coupling commensurate with
the set of central actions discussed in the main paper, 
this appendix studies multi-histogram based on
 the unconstrained reference simulations.
To recap, these simulations were at 24 widely spaced $\beta$ values
for each lattice size, covering the ranges listed in table \ref{ReferenceSim}.
In the following, so as to directly compare with the 
density of states results, 
we only consider the  narrower ranges 
that appear in the figures;
so only 7, 6, 4 
(for  $16^4$, $20^4$ and  $24^4$) of the 24 runs
were necessary.
Errors are obtained using a bootstrap approach of 
resampling
the original data taking into account the autocorrelation time for the
observable in question. These errors are reported in the column 
``RM-err'' of tables \ref{tab-errs16}-\ref{tab-errs24} and 
\ref{tab-TCerrs16}-\ref{tab-TCerrs24}.

\begin{figure}
\includegraphics[scale=0.65]{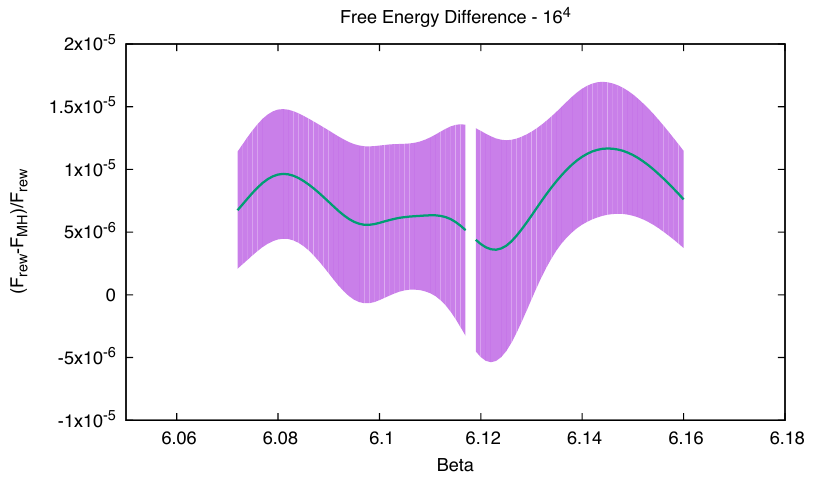}\par\medskip
\caption{
Free energy difference between Multihistogram and 
density of states approaches. Only shown for $16^4$.
The gap is related to normalisation as discussed in the text.
Errors are shown as shaded regions.
}
\label{fig-MHF}
\end{figure}

Multihistogram provides access to the partition function or the
free energy of the system and 
allows a direct comparison with the
density of states approach via equation (\ref{zeefulldos}).
A plot of $F= -\ln ({\cal Z})/\beta$ 
appears linear over the ranges in question
and no useful graphical comparison is possible.
Figure \ref{fig-MHF} shows the 
difference between $F$ computed with multihistogram 
and with LLR. Averages are taken over the bootstrap
samples in the case of multihistogram, and over RM
replicas for LLR, and the results are normalised by
setting $F(\beta=6.118) =0$.
This difference between very different
approaches is compatible with zero at two sigmas.

\begin{figure}
\includegraphics[scale=0.65]{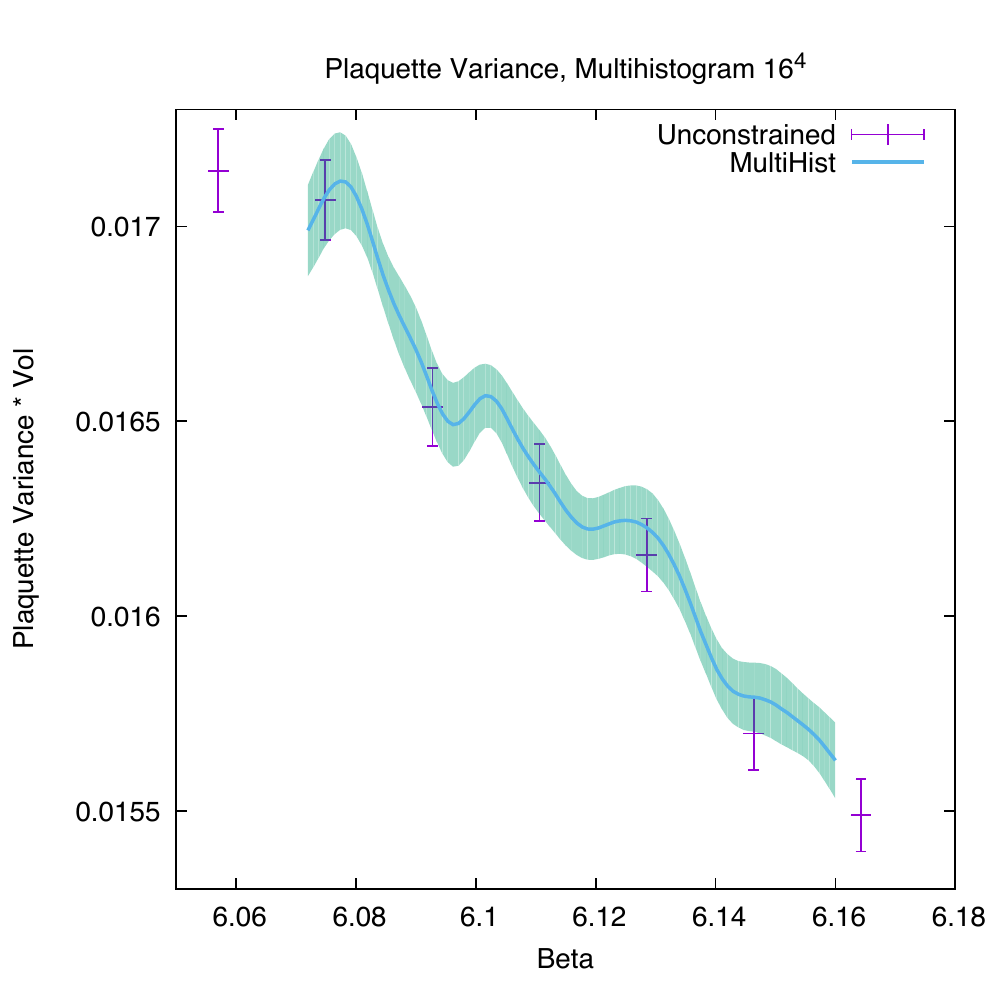}\par\medskip
\includegraphics[scale=0.65]{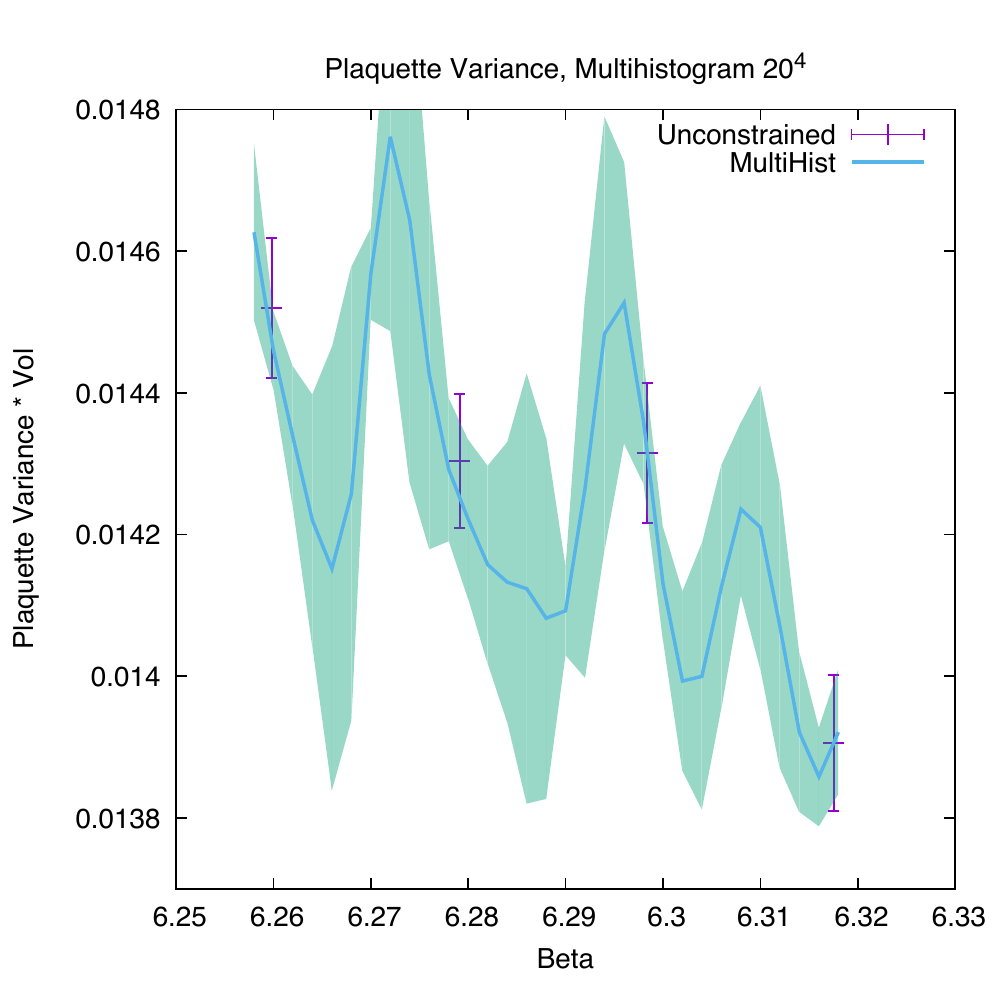}\par\medskip
\caption{
Plaquette variance from Multihistogram. 
Also shown for comparison
are results from unconstrained simulations with additional runs.
From top to bottom: $16^4$, $20^4$.  The plot for $24^4$ is not shown as
plagued by huge fluctuations.
Errors are shown as shaded regions.
}
\label{fig-MHplaq}
\end{figure}

Despite the accuracy suggested by figure \ref{fig-MHF}, the
free energy does not encode much more information than the
first moments of the plaquette expectation values. 
The plaquette variance shown in figure  \ref{fig-MHplaq} 
should be compared with the LLR predictions in the middle column
of figure \ref{fig-llr} (though note a small shift of scale at $20^4$). The LLR prediction has smaller error and is smoother. 
While the quality of the LLR prediction does decrease
for larger lattice sizes, the multihistogram results deteriorate much more
markedly and indeed are useless at $24^4$.
As already emphasised, the reason for poor performance of the multi-histogram
results is not the method itself, but our choice of data to illustrate it. 

Multihistogram can also be used for quantities such as the
symmetric cloverleaf or topological charge, both  after Wilson flow.
These behave better than the plaquette variance, but again the
excessive spacing of the coupling in the reference simulations
 leads to declining quality for
the larger lattice sizes.  

\section{Run details}
\label{appendixb}

This appendix records details of the main simulations.

Table \ref{tab-NoLLRparams} shows the couplings used for
the reference simulations. The last column 
uses the fitting function from Gattringer and Lang \cite{Gattringer:2010zz}
to compute the box size.
The duration of these reference simulations was between $6\times 10^4$ and $12\times 10^4$
steps at $16^4$,  between $5\times 10^4$ and  $11\times 10^4$ at $20^4$ and 
fixed at $9.6\times 10^4$ at $24^4$. 
Extended simulations to improve data points over a restricted range at  $20^4$ and $24^4$ were mentioned in the text.
These simulations were for four $\beta$ values and two replicas at $20^4$ and for 
 two $\beta$ values and four replicas at $24^4$. For all extended simulations the duration was $10^5$ steps with 
 Wilson Flow measurements only every 10 steps. 

The list of central action for the density of states runs is given in
Table \ref{tab-params}. Spacing, $\delta_S$, is variable and can be read off from the differences
in $S_i$.
The width of
the Gaussian constraint $\delta_i$ is also shown and the ratio 
$\delta_i/\delta_S$ is kept fixed at 1.2. The equivalent ranges
in terms of the coupling $\beta$ are given in table \ref{TemperSim}.

Table  \ref{tab-meanai} is the result of the RM phase and lists
the mean $a_i$ across replicas. The RM error of this quantity
varies slightly between tempers and lattice size, 
but  it is bounded by $4 \times10^{-5}$ in all cases.
Averaged over tempers, the error is
$2.7\times 10^{-5}$,
$1.7\times 10^{-5}$,
$1.3\times 10^{-5}$
for $16^4$, $20^4$ and $24^4$ respectively.\\

\begin{table}[!htb]
\begin{center}
\caption{
Parameter ranges for the canonical reference simulations. 
}
\begin{tabular}{|l|c|c|c|}
\hline
$16^4$  & $20^4$  & $24^4$ & Box size (fm) \\
\hline
5.78900 &       5.90359 &       6.00745 &       2.23408 \\
5.80687 &       5.92367 &       6.02941 &       2.15162 \\
5.82474 &       5.94373 &       6.05132 &       2.07358 \\
5.84261 &       5.96377 &       6.07317 &       1.99967 \\
5.86048 &       5.98380 &       6.09495 &       1.92962 \\
5.87835 &       6.00380 &       6.11665 &       1.86318 \\
5.89622 &       6.02376 &       6.13827 &       1.80012 \\
5.91409 &       6.04370 &       6.15980 &       1.74021 \\
5.93196 &       6.06360 &       6.18124 &       1.68326 \\
5.94983 &       6.08345 &       6.20257 &       1.62909 \\
5.96770 &       6.10327 &       6.22379 &       1.57752 \\
5.98557 &       6.12303 &       6.24490 &       1.52839 \\
6.00344 &       6.14274 &       6.26590 &       1.48155 \\
6.02130 &       6.16240 &       6.28677 &       1.43689 \\
6.03917 &       6.18201 &       6.30753 &       1.39422 \\
6.05704 &       6.20156 &       6.32816 &       1.35345 \\
6.07491 &       6.22104 &       6.34867 &       1.31447 \\
6.09278 &       6.24047 &       6.36906 &       1.27717 \\
6.11065 &       6.25983 &       6.38932 &       1.24145 \\
6.12852 &       6.27913 &       6.40946 &       1.20723 \\
6.14639 &       6.29837 &       6.42949 &       1.17441 \\
6.16426 &       6.31754 &       6.44939 &       1.14291 \\
6.18213 &       6.33665 &       6.46918 &       1.11266 \\
6.20000 &       6.35570 &       6.48887 &       1.08359 \\
\hline
\end{tabular}
\label{tab-NoLLRparams}
\end{center}
\end{table}

\begin{table}[!htb]
\begin{center}
\caption{
Parameters $S_i$ and $\delta_i$ used for density of states runs.}
\begin{tabular}{|cc|cc|cc|}
\hline
\multicolumn{2}{c}{$16^4$}&\multicolumn{2}{c}{$20^4$}&\multicolumn{2}{c}{$24^4$}\\
\hline
$S_i$ & $\delta_i$ & $S_i$ & $\delta_i$ & $S_i$ & $\delta_i$ \\
\hline
152685 & 228.4 & 359926 & 338.2 & 724396 & 467.0 \\ 
152891 & 228.8 & 360232 & 338.6 & 724815 & 467.5 \\ 
153098 & 229.3 & 360537 & 338.9 & 725234 & 468.0 \\ 
153305 & 229.9 & 360844 & 339.4 & 725653 & 468.6 \\ 
153513 & 230.5 & 361150 & 339.8 & 726074 & 469.1 \\ 
153722 & 231.1 & 361458 & 340.3 & 726494 & 469.6 \\ 
153931 & 231.7 & 361765 & 340.9 & 726915 & 470.0 \\ 
154140 & 232.3 & 362073 & 341.4 & 727337 & 470.5 \\ 
154350 & 233.0 & 362382 & 341.9 & 727759 & 471.0 \\ 
154561 & 233.6 & 362691 & 342.4 & 728181 & 471.5 \\ 
154772 & 234.3 & 363000 & 343.0 & 728604 & 471.9 \\ 
154984 & 234.9 & 363310 & 343.5 & 729027 & 472.4 \\ 
155196 & 235.6 & 363620 & 344.0 & 729451 & 472.8 \\ 
155409 & 236.2 & 363931 & 344.5 & 729874 & 473.2 \\ 
155623 & 236.8 & 364242 & 345.0 & 730299 & 473.7 \\ 
155837 & 237.4 & 364554 & 345.4 & 730724 & 474.1 \\ 
156052 & 238.0 & 364866 & 345.9 & 731149 & 474.5 \\ 
156267 & 238.6 & 365179 & 346.4 & 731574 & 474.9 \\ 
156483 & 239.2 & 365492 & 346.8 & 732000 & 475.4 \\ 
156699 & 239.8 & 365805 & 347.2 & 732426 & 475.8 \\ 
156916 & 240.5 & 366119 & 347.6 & 732853 & 476.2 \\ 
157133 & 241.1 & 366433 & 348.1 & 733280 & 476.6 \\ 
157351 & 241.7 & 366747 & 348.4 & 733707 & 477.0 \\ 
157570 & 242.3 & 367062 & 348.8 & 734135 & 477.4 \\ 
\hline
\end{tabular}
\label{tab-params}
\end{center}
\end{table}

\begin{table}[!htb]
  \begin{center}
\caption{
Results of RM phase: mean value of $a_i$ for each temper. 
}
\begin{tabular}{|c|c|c|}
\hline
$16^4$  & $20^4$  & $24^4$ \\
\hline
6.17897 &	6.33166 &	6.47020 \\
6.17337 &	6.32783 &	6.46743 \\
6.16766 &	6.32402 &	6.46468 \\
6.16202 &	6.32019 &	6.46195 \\
6.15644 &	6.31634 &	6.45920 \\
6.15083 &	6.31257 &	6.45644 \\
6.14527 &	6.30874 &	6.45371 \\
6.13967 &	6.30494 &	6.45098 \\
6.13411 &	6.30117 &	6.44823 \\
6.12858 &	6.29734 &	6.44547 \\
6.12304 &	6.29358 &	6.44272 \\
6.11747 &	6.28981 &	6.43999 \\
6.11199 &	6.28601 &	6.43725 \\
6.10648 &	6.28225 &	6.43452 \\
6.10099 &	6.27847 &	6.43177 \\
6.09551 &	6.27468 &	6.42907 \\
6.09002 &	6.27094 &	6.42630 \\
6.08460 &	6.26720 &	6.42361 \\
6.07918 &	6.26343 &	6.42085 \\
6.07375 &	6.25970 &	6.41815 \\
6.06833 &	6.25594 &	6.41542 \\
6.06295 &	6.25221 &	6.41272 \\
6.05763 &	6.24847 &	6.40998 \\
6.05226 &	6.24475 &	6.40731 \\
\hline
\end{tabular}
\label{tab-meanai}
\end{center}
\end{table}

\end{document}